\documentclass[11pt]{article}
\usepackage[T1]{fontenc}
\usepackage[utf8]{inputenc}
\usepackage{lmodern}
\usepackage[margin=1in]{geometry}
\usepackage{amsmath,amssymb,mathtools,mathrsfs}
\usepackage{microtype}
\usepackage{array,booktabs,tabularx,longtable,enumitem}
\usepackage{graphicx}
\usepackage{caption}
\usepackage{xurl}
\usepackage[hidelinks,unicode,bookmarksnumbered]{hyperref}
\usepackage{pgfplots}
\usepgfplotslibrary{groupplots}
\pgfplotsset{compat=1.18}
\allowdisplaybreaks[1]

\usepackage{tikz}
\usetikzlibrary{arrows.meta,positioning,calc,fit,backgrounds}
\definecolor{ideaInk}{HTML}{24364B}
\definecolor{ideaBlue}{HTML}{315B7B}
\definecolor{ideaLightBlue}{HTML}{EEF4F8}
\definecolor{ideaGreen}{HTML}{2E6C59}
\definecolor{ideaLightGreen}{HTML}{EEF6F1}
\definecolor{ideaAmber}{HTML}{946520}
\definecolor{ideaLightAmber}{HTML}{FBF4E8}
\definecolor{ideaLine}{HTML}{8C9BA7}
\tikzset{
  idea box/.style={draw=ideaLine,rounded corners=2pt,fill=white,
    align=center,inner sep=6pt,line width=0.55pt,
    font=\footnotesize,text=ideaInk},
  idea arrow/.style={-{Stealth[length=2mm]},draw=ideaInk,line width=0.65pt},
  idea note/.style={font=\footnotesize,align=center,text=ideaInk},
  idea hidden/.style={idea box,draw=ideaAmber,fill=ideaLightAmber},
  idea public/.style={idea box,draw=ideaGreen,fill=ideaLightGreen},
  idea relation/.style={idea box,draw=ideaBlue,fill=ideaLightBlue}
}

\title{Exact Compilation and Resource Laws for Hidden Integer Feedback}
\author{Ganxuan Yang$^{1,2}$\thanks{These authors contributed equally to this work.} \quad Siqi Wen$^{1,2}$\footnotemark[1] \quad Shuyan Yu$^{1,2}$ \quad Jidong Yuan$^{1}$\thanks{Corresponding author: \href{mailto:yuanjd@bjtu.edu.cn}{\texttt{yuanjd@bjtu.edu.cn}}.}\\[0.5em]{\normalsize $^{1}$Beijing Jiaotong University}\\{\normalsize $^{2}$Lancaster University}}
\hypersetup{pdfauthor={Ganxuan Yang, Siqi Wen, Shuyan Yu, Jidong Yuan}}
\date{}
\begin{document}
\maketitle

\section*{Abstract}
\phantomsection\addcontentsline{toc}{section}{Abstract}

Hidden rounding carries dependencies through integer memory. We study finite-memory feedback with predictable inputs that differ by one on at most \(C\) releases and with every output public. A shared preparation constrains future branches until a sequence of common-input releases synchronizes the conditional registers. Preserving those constraints yields exact finite-law compilation, polynomial in support width for fixed memory and participation budgets, with constructive discrete-Gaussian transfer and calibration. A commitment hierarchy quantifies the loss from releasing dependencies early and becomes exact at depth \(mC\) for memory \(m\). Strict examples include three levels with one participation and a discrete-Gaussian calibration gap exceeding 0.23. Charging actual preparation time gives a privacy resource law with a square-root remainder and a leading coefficient computable in polynomial bit cost at fixed parameters. Lean refinements cover the first-order privacy analysis and binary value algorithm; separate lemmas check finite-memory algebra and paths. Certificates establish multi-memory calibration and resource effects; workload comparisons include a finite-memory prefix-error lower bound.

\section{Introduction}

Correlated noise supports differential privacy (DP) in continual counting [6] and learning with repeated participation [7]. In an integer implementation, rounding becomes part of the stochastic transition and can carry privacy-relevant information across outputs. We study a mechanism that adds fresh integer noise and subtracts a randomly rounded weighted sum of recent noise balances. The fractional remainder sets the probability of rounding up. Upward rounding subtracts one extra unit, like a borrow in arithmetic, changing a balance that remains in memory and affects later releases.

Privacy compares two input rules along the same observed history. Their conditional noise registers can differ even when their aligned current output laws agree. Figure 1a illustrates this in the first-order case. A common-input release gives both worlds the same new noise balance, obtained by subtracting that common input from the observed output. With memory \(m\), at most \(m\) consecutive such releases flush every previous difference from the registers. We call this sequence an \emph{echo} and include all its outputs in the privacy calculation before continuing from the synchronized state.

Until synchronization, one preparation state constrains every continuation (Figure 1b). Optimizing its unused digits separately after different outputs grants extra choices. Our question is how to preserve these shared dependencies in a finite computation, how much calibration is lost when they are relaxed, and how physical time limits the preparations that an execution can reach. We answer these questions for stable rational feedback with exact past-noise registers, hidden independent rounding, predictable signed-unit differences and a participation cap on every history.

Exact calibration can change a privacy decision or the selected noise scale. Workload analysis measures the resulting accuracy: Table 2 quantifies a finite-source storage--error trade-off, while Section 7.5 locates linear prefix-error growth in the autoregressive response. This separates calibration gains from noise-architecture choices.

\begin{figure}[!htbp]
\centering
\begingroup
\def\nfM#1#2{\pgfpathmoveto{\pgfpoint{#1bp}{#2bp}}}
\def\nfL#1#2{\pgfpathlineto{\pgfpoint{#1bp}{#2bp}}}
\def\nfQ#1#2#3#4{\pgfpathquadraticcurveto{\pgfpoint{#1bp}{#2bp}}{\pgfpoint{#3bp}{#4bp}}}
\def\nfC#1#2#3#4#5#6{\pgfpathcurveto{\pgfpoint{#1bp}{#2bp}}{\pgfpoint{#3bp}{#4bp}}{\pgfpoint{#5bp}{#6bp}}}
\def\nfT#1#2#3#4#5#6{\pgftransformcm{#1}{#2}{#3}{#4}{\pgfpoint{#5bp}{#6bp}}}
\def\nfB{\begin{pgfscope}}
\def\nfE{\end{pgfscope}}
\def\nfZ{\pgfpathclose}
\def\nfFC{\pgfsetfillcolor}
\def\nfFO{\pgfsetfillopacity}
\def\nfSC{\pgfsetstrokecolor}
\def\nfSO{\pgfsetstrokeopacity}
\def\nfW{\pgfsetlinewidth}
\def\nfU{\pgfusepath}
\def\nfNZ{\pgfsetnonzerorule}
\def\nfEO{\pgfseteorule}
\def\nfBC{\pgfsetbuttcap}
\def\nfRC{\pgfsetroundcap}
\def\nfXC{\pgfsetrectcap}
\def\nfRJ{\pgfsetroundjoin}
\def\nfMJ{\pgfsetmiterjoin}
\def\nfBJ{\pgfsetbeveljoin}
\def\nfDA{\pgfsetdash}
\definecolor{nf007f82}{RGB}{0,127,130}
\definecolor{nf202b33}{RGB}{32,43,51}
\definecolor{nf586570}{RGB}{88,101,112}
\definecolor{nf667782}{RGB}{102,119,130}
\definecolor{nf8064a2}{RGB}{128,100,162}
\definecolor{nfbc845d}{RGB}{188,132,93}
\definecolor{nfdce3e6}{RGB}{220,227,230}
\definecolor{nfedf7f6}{RGB}{237,247,246}
\definecolor{nff4f0f8}{RGB}{244,240,248}
\definecolor{nff4f7f8}{RGB}{244,247,248}
\definecolor{nfffffff}{RGB}{255,255,255}
\expandafter\def\csname nfpath0\endcsname{\nfM{0}{2.236068}
\nfC{0.593012}{2.236068}{1.161816}{2.000462}{1.581139}{1.581139}
\nfC{2.000462}{1.161816}{2.236068}{0.593012}{2.236068}{0}
\nfC{2.236068}{-0.593012}{2.000462}{-1.161816}{1.581139}{-1.581139}
\nfC{1.161816}{-2.000462}{0.593012}{-2.236068}{0}{-2.236068}
\nfC{-0.593012}{-2.236068}{-1.161816}{-2.000462}{-1.581139}{-1.581139}
\nfC{-2.000462}{-1.161816}{-2.236068}{-0.593012}{-2.236068}{0}
\nfC{-2.236068}{0.593012}{-2.000462}{1.161816}{-1.581139}{1.581139}
\nfC{-1.161816}{2.000462}{-0.593012}{2.236068}{0}{2.236068}
\nfZ}
\expandafter\def\csname nfpath1\endcsname{\nfM{0}{2.236068}
\nfC{0.593012}{2.236068}{1.161816}{2.000462}{1.581139}{1.581139}
\nfC{2.000462}{1.161816}{2.236068}{0.593012}{2.236068}{0}
\nfC{2.236068}{-0.593012}{2.000462}{-1.161816}{1.581139}{-1.581139}
\nfC{1.161816}{-2.000462}{0.593012}{-2.236068}{0}{-2.236068}
\nfC{-0.593012}{-2.236068}{-1.161816}{-2.000462}{-1.581139}{-1.581139}
\nfC{-2.000462}{-1.161816}{-2.236068}{-0.593012}{-2.236068}{0}
\nfC{-2.236068}{0.593012}{-2.000462}{1.161816}{-1.581139}{1.581139}
\nfC{-1.161816}{2.000462}{-0.593012}{2.236068}{0}{2.236068}
\nfZ}
\expandafter\def\csname nfpath2\endcsname{\nfM{0}{2.236068}
\nfC{0.593012}{2.236068}{1.161816}{2.000462}{1.581139}{1.581139}
\nfC{2.000462}{1.161816}{2.236068}{0.593012}{2.236068}{0}
\nfC{2.236068}{-0.593012}{2.000462}{-1.161816}{1.581139}{-1.581139}
\nfC{1.161816}{-2.000462}{0.593012}{-2.236068}{0}{-2.236068}
\nfC{-0.593012}{-2.236068}{-1.161816}{-2.000462}{-1.581139}{-1.581139}
\nfC{-2.000462}{-1.161816}{-2.236068}{-0.593012}{-2.236068}{0}
\nfC{-2.236068}{0.593012}{-2.000462}{1.161816}{-1.581139}{1.581139}
\nfC{-1.161816}{2.000462}{-0.593012}{2.236068}{0}{2.236068}
\nfZ}
\expandafter\def\csname nfpath3\endcsname{\nfM{1116}{2306}
\nfL{319}{2450}
\nfQ{453}{2931}{781}{3162}
\nfQ{1109}{3394}{1756}{3394}
\nfQ{2344}{3394}{2631}{3255}
\nfQ{2919}{3116}{3036}{2902}
\nfQ{3153}{2688}{3153}{2116}
\nfL{3144}{1091}
\nfQ{3144}{653}{3186}{445}
\nfQ{3228}{238}{3344}{0}
\nfL{2475}{0}
\nfQ{2441}{88}{2391}{259}
\nfQ{2369}{338}{2359}{363}
\nfQ{2134}{144}{1878}{34}
\nfQ{1622}{-75}{1331}{-75}
\nfQ{819}{-75}{523}{203}
\nfQ{228}{481}{228}{906}
\nfQ{228}{1188}{362}{1408}
\nfQ{497}{1628}{739}{1745}
\nfQ{981}{1863}{1438}{1950}
\nfQ{2053}{2066}{2291}{2166}
\nfL{2291}{2253}
\nfQ{2291}{2506}{2166}{2614}
\nfQ{2041}{2722}{1694}{2722}
\nfQ{1459}{2722}{1328}{2630}
\nfQ{1197}{2538}{1116}{2306}
\nfZ
\nfM{2291}{1594}
\nfQ{2122}{1538}{1756}{1459}
\nfQ{1391}{1381}{1278}{1306}
\nfQ{1106}{1184}{1106}{997}
\nfQ{1106}{813}{1243}{678}
\nfQ{1381}{544}{1594}{544}
\nfQ{1831}{544}{2047}{700}
\nfQ{2206}{819}{2256}{991}
\nfQ{2291}{1103}{2291}{1419}
\nfL{2291}{1594}
\nfZ}
\expandafter\def\csname nfpath4\endcsname{\nfM{278}{2263}
\nfQ{278}{2963}{488}{3438}
\nfQ{644}{3788}{914}{4066}
\nfQ{1184}{4344}{1506}{4478}
\nfQ{1934}{4659}{2494}{4659}
\nfQ{3506}{4659}{4114}{4031}
\nfQ{4722}{3403}{4722}{2284}
\nfQ{4722}{1175}{4119}{548}
\nfQ{3516}{-78}{2506}{-78}
\nfQ{1484}{-78}{881}{545}
\nfQ{278}{1169}{278}{2263}
\nfZ
\nfM{1231}{2294}
\nfQ{1231}{1516}{1590}{1114}
\nfQ{1950}{713}{2503}{713}
\nfQ{3056}{713}{3411}{1111}
\nfQ{3766}{1509}{3766}{2306}
\nfQ{3766}{3094}{3420}{3481}
\nfQ{3075}{3869}{2503}{3869}
\nfQ{1931}{3869}{1581}{3476}
\nfQ{1231}{3084}{1231}{2294}
\nfZ}
\expandafter\def\csname nfpath5\endcsname{\nfM{3478}{0}
\nfL{2600}{0}
\nfL{2600}{1694}
\nfQ{2600}{2231}{2544}{2389}
\nfQ{2488}{2547}{2361}{2634}
\nfQ{2234}{2722}{2056}{2722}
\nfQ{1828}{2722}{1647}{2597}
\nfQ{1466}{2472}{1398}{2265}
\nfQ{1331}{2059}{1331}{1503}
\nfL{1331}{0}
\nfL{453}{0}
\nfL{453}{3319}
\nfL{1269}{3319}
\nfL{1269}{2831}
\nfQ{1703}{3394}{2363}{3394}
\nfQ{2653}{3394}{2893}{3289}
\nfQ{3134}{3184}{3257}{3021}
\nfQ{3381}{2859}{3429}{2653}
\nfQ{3478}{2447}{3478}{2063}
\nfL{3478}{0}
\nfZ}
\expandafter\def\csname nfpath6\endcsname{\nfM{2381}{1056}
\nfL{3256}{909}
\nfQ{3088}{428}{2723}{176}
\nfQ{2359}{-75}{1813}{-75}
\nfQ{947}{-75}{531}{491}
\nfQ{203}{944}{203}{1634}
\nfQ{203}{2459}{634}{2926}
\nfQ{1066}{3394}{1725}{3394}
\nfQ{2466}{3394}{2894}{2905}
\nfQ{3322}{2416}{3303}{1406}
\nfL{1103}{1406}
\nfQ{1113}{1016}{1316}{798}
\nfQ{1519}{581}{1822}{581}
\nfQ{2028}{581}{2168}{693}
\nfQ{2309}{806}{2381}{1056}
\nfZ
\nfM{2431}{1944}
\nfQ{2422}{2325}{2234}{2523}
\nfQ{2047}{2722}{1778}{2722}
\nfQ{1491}{2722}{1303}{2513}
\nfQ{1116}{2303}{1119}{1944}
\nfL{2431}{1944}
\nfZ}
\expandafter\def\csname nfpath7\endcsname{}
\expandafter\def\csname nfpath8\endcsname{\nfM{434}{3319}
\nfL{1253}{3319}
\nfL{1253}{2831}
\nfQ{1413}{3081}{1684}{3237}
\nfQ{1956}{3394}{2288}{3394}
\nfQ{2866}{3394}{3269}{2941}
\nfQ{3672}{2488}{3672}{1678}
\nfQ{3672}{847}{3265}{386}
\nfQ{2859}{-75}{2281}{-75}
\nfQ{2006}{-75}{1782}{34}
\nfQ{1559}{144}{1313}{409}
\nfL{1313}{-1263}
\nfL{434}{-1263}
\nfL{434}{3319}
\nfZ
\nfM{1303}{1716}
\nfQ{1303}{1156}{1525}{889}
\nfQ{1747}{622}{2066}{622}
\nfQ{2372}{622}{2575}{867}
\nfQ{2778}{1113}{2778}{1672}
\nfQ{2778}{2194}{2568}{2447}
\nfQ{2359}{2700}{2050}{2700}
\nfQ{1728}{2700}{1515}{2451}
\nfQ{1303}{2203}{1303}{1716}
\nfZ}
\expandafter\def\csname nfpath9\endcsname{\nfM{1334}{4581}
\nfL{1334}{2897}
\nfQ{1759}{3394}{2350}{3394}
\nfQ{2653}{3394}{2897}{3281}
\nfQ{3141}{3169}{3264}{2994}
\nfQ{3388}{2819}{3433}{2606}
\nfQ{3478}{2394}{3478}{1947}
\nfL{3478}{0}
\nfL{2600}{0}
\nfL{2600}{1753}
\nfQ{2600}{2275}{2550}{2415}
\nfQ{2500}{2556}{2373}{2639}
\nfQ{2247}{2722}{2056}{2722}
\nfQ{1838}{2722}{1666}{2615}
\nfQ{1494}{2509}{1414}{2295}
\nfQ{1334}{2081}{1334}{1663}
\nfL{1334}{0}
\nfL{456}{0}
\nfL{456}{4581}
\nfL{1334}{4581}
\nfZ}
\expandafter\def\csname nfpath10\endcsname{\nfM{150}{947}
\nfL{1031}{1081}
\nfQ{1088}{825}{1259}{692}
\nfQ{1431}{559}{1741}{559}
\nfQ{2081}{559}{2253}{684}
\nfQ{2369}{772}{2369}{919}
\nfQ{2369}{1019}{2306}{1084}
\nfQ{2241}{1147}{2013}{1200}
\nfQ{950}{1434}{666}{1628}
\nfQ{272}{1897}{272}{2375}
\nfQ{272}{2806}{612}{3100}
\nfQ{953}{3394}{1669}{3394}
\nfQ{2350}{3394}{2681}{3172}
\nfQ{3013}{2950}{3138}{2516}
\nfL{2309}{2363}
\nfQ{2256}{2556}{2107}{2659}
\nfQ{1959}{2763}{1684}{2763}
\nfQ{1338}{2763}{1188}{2666}
\nfQ{1088}{2597}{1088}{2488}
\nfQ{1088}{2394}{1175}{2328}
\nfQ{1294}{2241}{1995}{2081}
\nfQ{2697}{1922}{2975}{1691}
\nfQ{3250}{1456}{3250}{1038}
\nfQ{3250}{581}{2869}{253}
\nfQ{2488}{-75}{1741}{-75}
\nfQ{1063}{-75}{667}{200}
\nfQ{272}{475}{150}{947}
\nfZ}
\expandafter\def\csname nfpath11\endcsname{\nfM{438}{878}
\nfL{1316}{878}
\nfL{1316}{250}
\nfQ{1316}{-131}{1250}{-351}
\nfQ{1184}{-572}{1001}{-747}
\nfQ{819}{-922}{538}{-1022}
\nfL{366}{-659}
\nfQ{631}{-572}{743}{-419}
\nfQ{856}{-266}{863}{0}
\nfL{438}{0}
\nfL{438}{878}
\nfZ}
\expandafter\def\csname nfpath12\endcsname{\nfM{1981}{3319}
\nfL{1981}{2619}
\nfL{1381}{2619}
\nfL{1381}{1281}
\nfQ{1381}{875}{1398}{808}
\nfQ{1416}{741}{1477}{697}
\nfQ{1538}{653}{1625}{653}
\nfQ{1747}{653}{1978}{738}
\nfL{2053}{56}
\nfQ{1747}{-75}{1359}{-75}
\nfQ{1122}{-75}{931}{4}
\nfQ{741}{84}{652}{211}
\nfQ{563}{338}{528}{553}
\nfQ{500}{706}{500}{1172}
\nfL{500}{2619}
\nfL{97}{2619}
\nfL{97}{3319}
\nfL{500}{3319}
\nfL{500}{3978}
\nfL{1381}{4491}
\nfL{1381}{3319}
\nfL{1981}{3319}
\nfZ}
\expandafter\def\csname nfpath13\endcsname{\nfM{1078}{0}
\nfL{28}{3319}
\nfL{881}{3319}
\nfL{1503}{1144}
\nfL{2075}{3319}
\nfL{2922}{3319}
\nfL{3475}{1144}
\nfL{4109}{3319}
\nfL{4975}{3319}
\nfL{3909}{0}
\nfL{3066}{0}
\nfL{2494}{2134}
\nfL{1931}{0}
\nfL{1078}{0}
\nfZ}
\expandafter\def\csname nfpath14\endcsname{\nfM{256}{1706}
\nfQ{256}{2144}{472}{2553}
\nfQ{688}{2963}{1083}{3178}
\nfQ{1478}{3394}{1966}{3394}
\nfQ{2719}{3394}{3200}{2905}
\nfQ{3681}{2416}{3681}{1669}
\nfQ{3681}{916}{3195}{420}
\nfQ{2709}{-75}{1972}{-75}
\nfQ{1516}{-75}{1102}{131}
\nfQ{688}{338}{472}{736}
\nfQ{256}{1134}{256}{1706}
\nfZ
\nfM{1156}{1659}
\nfQ{1156}{1166}{1390}{903}
\nfQ{1625}{641}{1969}{641}
\nfQ{2313}{641}{2545}{903}
\nfQ{2778}{1166}{2778}{1666}
\nfQ{2778}{2153}{2545}{2415}
\nfQ{2313}{2678}{1969}{2678}
\nfQ{1625}{2678}{1390}{2415}
\nfQ{1156}{2153}{1156}{1659}
\nfZ}
\expandafter\def\csname nfpath15\endcsname{\nfM{2644}{0}
\nfL{2644}{497}
\nfQ{2463}{231}{2167}{78}
\nfQ{1872}{-75}{1544}{-75}
\nfQ{1209}{-75}{943}{72}
\nfQ{678}{219}{559}{484}
\nfQ{441}{750}{441}{1219}
\nfL{441}{3319}
\nfL{1319}{3319}
\nfL{1319}{1794}
\nfQ{1319}{1094}{1367}{936}
\nfQ{1416}{778}{1544}{686}
\nfQ{1672}{594}{1869}{594}
\nfQ{2094}{594}{2272}{717}
\nfQ{2450}{841}{2515}{1023}
\nfQ{2581}{1206}{2581}{1919}
\nfL{2581}{3319}
\nfL{3459}{3319}
\nfL{3459}{0}
\nfL{2644}{0}
\nfZ}
\expandafter\def\csname nfpath16\endcsname{\nfM{3353}{2338}
\nfL{2488}{2181}
\nfQ{2444}{2441}{2289}{2572}
\nfQ{2134}{2703}{1888}{2703}
\nfQ{1559}{2703}{1364}{2476}
\nfQ{1169}{2250}{1169}{1719}
\nfQ{1169}{1128}{1367}{884}
\nfQ{1566}{641}{1900}{641}
\nfQ{2150}{641}{2309}{783}
\nfQ{2469}{925}{2534}{1272}
\nfL{3397}{1125}
\nfQ{3263}{531}{2881}{228}
\nfQ{2500}{-75}{1859}{-75}
\nfQ{1131}{-75}{698}{384}
\nfQ{266}{844}{266}{1656}
\nfQ{266}{2478}{700}{2936}
\nfQ{1134}{3394}{1875}{3394}
\nfQ{2481}{3394}{2839}{3133}
\nfQ{3197}{2872}{3353}{2338}
\nfZ}
\expandafter\def\csname nfpath17\endcsname{\nfM{1300}{0}
\nfL{422}{0}
\nfL{422}{3319}
\nfL{1238}{3319}
\nfL{1238}{2847}
\nfQ{1447}{3181}{1614}{3287}
\nfQ{1781}{3394}{1994}{3394}
\nfQ{2294}{3394}{2572}{3228}
\nfL{2300}{2463}
\nfQ{2078}{2606}{1888}{2606}
\nfQ{1703}{2606}{1575}{2504}
\nfQ{1447}{2403}{1373}{2137}
\nfQ{1300}{1872}{1300}{1025}
\nfL{1300}{0}
\nfZ}
\expandafter\def\csname nfpath18\endcsname{\nfM{422}{0}
\nfL{422}{4581}
\nfL{1300}{4581}
\nfL{1300}{2931}
\nfQ{1706}{3394}{2263}{3394}
\nfQ{2869}{3394}{3266}{2955}
\nfQ{3663}{2516}{3663}{1694}
\nfQ{3663}{844}{3258}{384}
\nfQ{2853}{-75}{2275}{-75}
\nfQ{1991}{-75}{1714}{67}
\nfQ{1438}{209}{1238}{488}
\nfL{1238}{0}
\nfL{422}{0}
\nfZ
\nfM{1294}{1731}
\nfQ{1294}{1216}{1456}{969}
\nfQ{1684}{619}{2063}{619}
\nfQ{2353}{619}{2558}{867}
\nfQ{2763}{1116}{2763}{1650}
\nfQ{2763}{2219}{2556}{2470}
\nfQ{2350}{2722}{2028}{2722}
\nfQ{1713}{2722}{1503}{2476}
\nfQ{1294}{2231}{1294}{1731}
\nfZ}
\expandafter\def\csname nfpath19\endcsname{\nfM{1116}{0}
\nfL{22}{4581}
\nfL{969}{4581}
\nfL{1659}{1434}
\nfL{2497}{4581}
\nfL{3597}{4581}
\nfL{4400}{1381}
\nfL{5103}{4581}
\nfL{6034}{4581}
\nfL{4922}{0}
\nfL{3941}{0}
\nfL{3028}{3425}
\nfL{2119}{0}
\nfL{1116}{0}
\nfZ}
\expandafter\def\csname nfpath20\endcsname{\nfM{394}{3319}
\nfL{1203}{3319}
\nfL{1203}{2866}
\nfQ{1638}{3394}{2238}{3394}
\nfQ{2556}{3394}{2790}{3262}
\nfQ{3025}{3131}{3175}{2866}
\nfQ{3394}{3131}{3647}{3262}
\nfQ{3900}{3394}{4188}{3394}
\nfQ{4553}{3394}{4806}{3245}
\nfQ{5059}{3097}{5184}{2809}
\nfQ{5275}{2597}{5275}{2122}
\nfL{5275}{0}
\nfL{4397}{0}
\nfL{4397}{1897}
\nfQ{4397}{2391}{4306}{2534}
\nfQ{4184}{2722}{3931}{2722}
\nfQ{3747}{2722}{3584}{2609}
\nfQ{3422}{2497}{3350}{2280}
\nfQ{3278}{2063}{3278}{1594}
\nfL{3278}{0}
\nfL{2400}{0}
\nfL{2400}{1819}
\nfQ{2400}{2303}{2353}{2443}
\nfQ{2306}{2584}{2207}{2653}
\nfQ{2109}{2722}{1941}{2722}
\nfQ{1738}{2722}{1575}{2612}
\nfQ{1413}{2503}{1342}{2297}
\nfQ{1272}{2091}{1272}{1613}
\nfL{1272}{0}
\nfL{394}{0}
\nfL{394}{3319}
\nfZ}
\expandafter\def\csname nfpath21\endcsname{\nfM{459}{3769}
\nfL{459}{4581}
\nfL{1338}{4581}
\nfL{1338}{3769}
\nfL{459}{3769}
\nfZ
\nfM{459}{0}
\nfL{459}{3319}
\nfL{1338}{3319}
\nfL{1338}{0}
\nfL{459}{0}
\nfZ}
\expandafter\def\csname nfpath22\endcsname{\nfM{3503}{0}
\nfL{2688}{0}
\nfL{2688}{488}
\nfQ{2484}{203}{2207}{64}
\nfQ{1931}{-75}{1650}{-75}
\nfQ{1078}{-75}{670}{386}
\nfQ{263}{847}{263}{1672}
\nfQ{263}{2516}{659}{2955}
\nfQ{1056}{3394}{1663}{3394}
\nfQ{2219}{3394}{2625}{2931}
\nfL{2625}{4581}
\nfL{3503}{4581}
\nfL{3503}{0}
\nfZ
\nfM{1159}{1731}
\nfQ{1159}{1200}{1306}{963}
\nfQ{1519}{619}{1900}{619}
\nfQ{2203}{619}{2415}{876}
\nfQ{2628}{1134}{2628}{1647}
\nfQ{2628}{2219}{2422}{2470}
\nfQ{2216}{2722}{1894}{2722}
\nfQ{1581}{2722}{1370}{2473}
\nfQ{1159}{2225}{1159}{1731}
\nfZ}
\expandafter\def\csname nfpath23\endcsname{\nfM{2347}{1181}
\nfL{1550}{1181}
\nfQ{1547}{1353}{1547}{1391}
\nfQ{1547}{1778}{1675}{2028}
\nfQ{1803}{2278}{2187}{2590}
\nfQ{2572}{2903}{2647}{3000}
\nfQ{2763}{3153}{2763}{3338}
\nfQ{2763}{3594}{2558}{3776}
\nfQ{2353}{3959}{2006}{3959}
\nfQ{1672}{3959}{1447}{3768}
\nfQ{1222}{3578}{1138}{3188}
\nfL{331}{3288}
\nfQ{366}{3847}{808}{4237}
\nfQ{1250}{4628}{1969}{4628}
\nfQ{2725}{4628}{3172}{4233}
\nfQ{3619}{3838}{3619}{3313}
\nfQ{3619}{3022}{3455}{2762}
\nfQ{3291}{2503}{2753}{2056}
\nfQ{2475}{1825}{2408}{1684}
\nfQ{2341}{1544}{2347}{1181}
\nfZ
\nfM{1550}{0}
\nfL{1550}{878}
\nfL{2428}{878}
\nfL{2428}{0}
\nfL{1550}{0}
\nfZ}
\expandafter\def\csname nfpath24\endcsname{\nfM{1491}{2950}
\nfL{1838}{788}
\nfL{3456}{3500}
\nfL{4066}{3500}
\nfL{1972}{0}
\nfL{1713}{-1331}
\nfL{1138}{-1331}
\nfL{1397}{0}
\nfL{988}{2613}
\nfQ{925}{3006}{628}{3006}
\nfL{475}{3006}
\nfL{569}{3500}
\nfL{788}{3500}
\nfQ{1403}{3500}{1491}{2950}
\nfZ}
\expandafter\def\csname nfpath25\endcsname{\nfM{678}{2906}
\nfL{4684}{2906}
\nfL{4684}{2381}
\nfL{678}{2381}
\nfL{678}{2906}
\nfZ
\nfM{678}{1631}
\nfL{4684}{1631}
\nfL{4684}{1100}
\nfL{678}{1100}
\nfL{678}{1631}
\nfZ}
\expandafter\def\csname nfpath26\endcsname{\nfM{794}{531}
\nfL{1825}{531}
\nfL{1825}{4091}
\nfL{703}{3866}
\nfL{703}{4441}
\nfL{1819}{4666}
\nfL{2450}{4666}
\nfL{2450}{531}
\nfL{3481}{531}
\nfL{3481}{0}
\nfL{794}{0}
\nfL{794}{531}
\nfZ}
\expandafter\def\csname nfpath27\endcsname{\nfM{1625}{4666}
\nfL{2156}{4666}
\nfL{531}{-594}
\nfL{0}{-594}
\nfL{1625}{4666}
\nfZ}
\expandafter\def\csname nfpath28\endcsname{\nfM{2597}{2516}
\nfQ{3050}{2419}{3304}{2112}
\nfQ{3559}{1806}{3559}{1356}
\nfQ{3559}{666}{3084}{287}
\nfQ{2609}{-91}{1734}{-91}
\nfQ{1441}{-91}{1130}{-33}
\nfQ{819}{25}{488}{141}
\nfL{488}{750}
\nfQ{750}{597}{1062}{519}
\nfQ{1375}{441}{1716}{441}
\nfQ{2309}{441}{2620}{675}
\nfQ{2931}{909}{2931}{1356}
\nfQ{2931}{1769}{2642}{2001}
\nfQ{2353}{2234}{1838}{2234}
\nfL{1294}{2234}
\nfL{1294}{2753}
\nfL{1863}{2753}
\nfQ{2328}{2753}{2575}{2939}
\nfQ{2822}{3125}{2822}{3475}
\nfQ{2822}{3834}{2567}{4026}
\nfQ{2313}{4219}{1838}{4219}
\nfQ{1578}{4219}{1281}{4162}
\nfQ{984}{4106}{628}{3988}
\nfL{628}{4550}
\nfQ{988}{4650}{1302}{4700}
\nfQ{1616}{4750}{1894}{4750}
\nfQ{2613}{4750}{3031}{4423}
\nfQ{3450}{4097}{3450}{3541}
\nfQ{3450}{3153}{3228}{2886}
\nfQ{3006}{2619}{2597}{2516}
\nfZ}
\expandafter\def\csname nfpath29\endcsname{\nfM{288}{1472}
\nfL{859}{1522}
\nfQ{900}{1178}{1048}{958}
\nfQ{1197}{738}{1509}{602}
\nfQ{1822}{466}{2213}{466}
\nfQ{2559}{466}{2825}{569}
\nfQ{3091}{672}{3220}{851}
\nfQ{3350}{1031}{3350}{1244}
\nfQ{3350}{1459}{3225}{1620}
\nfQ{3100}{1781}{2813}{1891}
\nfQ{2628}{1963}{1997}{2114}
\nfQ{1366}{2266}{1113}{2400}
\nfQ{784}{2572}{623}{2826}
\nfQ{463}{3081}{463}{3397}
\nfQ{463}{3744}{659}{4045}
\nfQ{856}{4347}{1234}{4503}
\nfQ{1613}{4659}{2075}{4659}
\nfQ{2584}{4659}{2973}{4495}
\nfQ{3363}{4331}{3572}{4012}
\nfQ{3781}{3694}{3797}{3291}
\nfL{3216}{3247}
\nfQ{3169}{3681}{2898}{3903}
\nfQ{2628}{4125}{2100}{4125}
\nfQ{1550}{4125}{1298}{3923}
\nfQ{1047}{3722}{1047}{3438}
\nfQ{1047}{3191}{1225}{3031}
\nfQ{1400}{2872}{2139}{2705}
\nfQ{2878}{2538}{3153}{2413}
\nfQ{3553}{2228}{3743}{1945}
\nfQ{3934}{1663}{3934}{1294}
\nfQ{3934}{928}{3725}{604}
\nfQ{3516}{281}{3123}{101}
\nfQ{2731}{-78}{2241}{-78}
\nfQ{1619}{-78}{1198}{103}
\nfQ{778}{284}{539}{648}
\nfQ{300}{1013}{288}{1472}
\nfZ}
\expandafter\def\csname nfpath30\endcsname{\nfM{2588}{409}
\nfQ{2275}{144}{1986}{34}
\nfQ{1697}{-75}{1366}{-75}
\nfQ{819}{-75}{525}{192}
\nfQ{231}{459}{231}{875}
\nfQ{231}{1119}{342}{1320}
\nfQ{453}{1522}{633}{1644}
\nfQ{813}{1766}{1038}{1828}
\nfQ{1203}{1872}{1538}{1913}
\nfQ{2219}{1994}{2541}{2106}
\nfQ{2544}{2222}{2544}{2253}
\nfQ{2544}{2597}{2384}{2738}
\nfQ{2169}{2928}{1744}{2928}
\nfQ{1347}{2928}{1158}{2789}
\nfQ{969}{2650}{878}{2297}
\nfL{328}{2372}
\nfQ{403}{2725}{575}{2942}
\nfQ{747}{3159}{1072}{3276}
\nfQ{1397}{3394}{1825}{3394}
\nfQ{2250}{3394}{2515}{3294}
\nfQ{2781}{3194}{2906}{3042}
\nfQ{3031}{2891}{3081}{2659}
\nfQ{3109}{2516}{3109}{2141}
\nfL{3109}{1391}
\nfQ{3109}{606}{3145}{398}
\nfQ{3181}{191}{3288}{0}
\nfL{2700}{0}
\nfQ{2613}{175}{2588}{409}
\nfZ
\nfM{2541}{1666}
\nfQ{2234}{1541}{1622}{1453}
\nfQ{1275}{1403}{1131}{1340}
\nfQ{988}{1278}{909}{1158}
\nfQ{831}{1038}{831}{891}
\nfQ{831}{666}{1001}{516}
\nfQ{1172}{366}{1500}{366}
\nfQ{1825}{366}{2078}{508}
\nfQ{2331}{650}{2450}{897}
\nfQ{2541}{1088}{2541}{1459}
\nfL{2541}{1666}
\nfZ}
\expandafter\def\csname nfpath31\endcsname{\nfM{422}{0}
\nfL{422}{3319}
\nfL{925}{3319}
\nfL{925}{2853}
\nfQ{1081}{3097}{1340}{3245}
\nfQ{1600}{3394}{1931}{3394}
\nfQ{2300}{3394}{2536}{3241}
\nfQ{2772}{3088}{2869}{2813}
\nfQ{3263}{3394}{3894}{3394}
\nfQ{4388}{3394}{4653}{3120}
\nfQ{4919}{2847}{4919}{2278}
\nfL{4919}{0}
\nfL{4359}{0}
\nfL{4359}{2091}
\nfQ{4359}{2428}{4304}{2576}
\nfQ{4250}{2725}{4106}{2815}
\nfQ{3963}{2906}{3769}{2906}
\nfQ{3419}{2906}{3187}{2673}
\nfQ{2956}{2441}{2956}{1928}
\nfL{2956}{0}
\nfL{2394}{0}
\nfL{2394}{2156}
\nfQ{2394}{2531}{2256}{2718}
\nfQ{2119}{2906}{1806}{2906}
\nfQ{1569}{2906}{1367}{2781}
\nfQ{1166}{2656}{1075}{2415}
\nfQ{984}{2175}{984}{1722}
\nfL{984}{0}
\nfL{422}{0}
\nfZ}
\expandafter\def\csname nfpath32\endcsname{\nfM{2694}{1069}
\nfL{3275}{997}
\nfQ{3138}{488}{2766}{206}
\nfQ{2394}{-75}{1816}{-75}
\nfQ{1088}{-75}{661}{373}
\nfQ{234}{822}{234}{1631}
\nfQ{234}{2469}{665}{2931}
\nfQ{1097}{3394}{1784}{3394}
\nfQ{2450}{3394}{2872}{2941}
\nfQ{3294}{2488}{3294}{1666}
\nfQ{3294}{1616}{3291}{1516}
\nfL{816}{1516}
\nfQ{847}{969}{1125}{678}
\nfQ{1403}{388}{1819}{388}
\nfQ{2128}{388}{2347}{550}
\nfQ{2566}{713}{2694}{1069}
\nfZ
\nfM{847}{1978}
\nfL{2700}{1978}
\nfQ{2663}{2397}{2488}{2606}
\nfQ{2219}{2931}{1791}{2931}
\nfQ{1403}{2931}{1139}{2672}
\nfQ{875}{2413}{847}{1978}
\nfZ}
\expandafter\def\csname nfpath33\endcsname{}
\expandafter\def\csname nfpath34\endcsname{\nfM{2588}{1216}
\nfL{3141}{1144}
\nfQ{3050}{572}{2676}{248}
\nfQ{2303}{-75}{1759}{-75}
\nfQ{1078}{-75}{664}{370}
\nfQ{250}{816}{250}{1647}
\nfQ{250}{2184}{428}{2587}
\nfQ{606}{2991}{970}{3192}
\nfQ{1334}{3394}{1763}{3394}
\nfQ{2303}{3394}{2647}{3120}
\nfQ{2991}{2847}{3088}{2344}
\nfL{2541}{2259}
\nfQ{2463}{2594}{2264}{2762}
\nfQ{2066}{2931}{1784}{2931}
\nfQ{1359}{2931}{1093}{2626}
\nfQ{828}{2322}{828}{1663}
\nfQ{828}{994}{1084}{691}
\nfQ{1341}{388}{1753}{388}
\nfQ{2084}{388}{2306}{591}
\nfQ{2528}{794}{2588}{1216}
\nfZ}
\expandafter\def\csname nfpath35\endcsname{\nfM{2597}{0}
\nfL{2597}{488}
\nfQ{2209}{-75}{1544}{-75}
\nfQ{1250}{-75}{995}{37}
\nfQ{741}{150}{617}{320}
\nfQ{494}{491}{444}{738}
\nfQ{409}{903}{409}{1263}
\nfL{409}{3319}
\nfL{972}{3319}
\nfL{972}{1478}
\nfQ{972}{1038}{1006}{884}
\nfQ{1059}{663}{1231}{536}
\nfQ{1403}{409}{1656}{409}
\nfQ{1909}{409}{2131}{539}
\nfQ{2353}{669}{2445}{892}
\nfQ{2538}{1116}{2538}{1541}
\nfL{2538}{3319}
\nfL{3100}{3319}
\nfL{3100}{0}
\nfL{2597}{0}
\nfZ}
\expandafter\def\csname nfpath36\endcsname{\nfM{416}{0}
\nfL{416}{3319}
\nfL{922}{3319}
\nfL{922}{2816}
\nfQ{1116}{3169}{1280}{3281}
\nfQ{1444}{3394}{1641}{3394}
\nfQ{1925}{3394}{2219}{3213}
\nfL{2025}{2691}
\nfQ{1819}{2813}{1613}{2813}
\nfQ{1428}{2813}{1281}{2702}
\nfQ{1134}{2591}{1072}{2394}
\nfQ{978}{2094}{978}{1738}
\nfL{978}{0}
\nfL{416}{0}
\nfZ}
\expandafter\def\csname nfpath37\endcsname{\nfM{422}{0}
\nfL{422}{3319}
\nfL{928}{3319}
\nfL{928}{2847}
\nfQ{1294}{3394}{1984}{3394}
\nfQ{2284}{3394}{2536}{3286}
\nfQ{2788}{3178}{2913}{3003}
\nfQ{3038}{2828}{3088}{2588}
\nfQ{3119}{2431}{3119}{2041}
\nfL{3119}{0}
\nfL{2556}{0}
\nfL{2556}{2019}
\nfQ{2556}{2363}{2490}{2533}
\nfQ{2425}{2703}{2258}{2804}
\nfQ{2091}{2906}{1866}{2906}
\nfQ{1506}{2906}{1245}{2678}
\nfQ{984}{2450}{984}{1813}
\nfL{984}{0}
\nfL{422}{0}
\nfZ}
\expandafter\def\csname nfpath38\endcsname{\nfM{1650}{503}
\nfL{1731}{6}
\nfQ{1494}{-44}{1306}{-44}
\nfQ{1000}{-44}{831}{53}
\nfQ{663}{150}{594}{308}
\nfQ{525}{466}{525}{972}
\nfL{525}{2881}
\nfL{113}{2881}
\nfL{113}{3319}
\nfL{525}{3319}
\nfL{525}{4141}
\nfL{1084}{4478}
\nfL{1084}{3319}
\nfL{1650}{3319}
\nfL{1650}{2881}
\nfL{1084}{2881}
\nfL{1084}{941}
\nfQ{1084}{700}{1114}{631}
\nfQ{1144}{563}{1211}{522}
\nfQ{1278}{481}{1403}{481}
\nfQ{1497}{481}{1650}{503}
\nfZ}
\expandafter\def\csname nfpath39\endcsname{\nfM{213}{1659}
\nfQ{213}{2581}{725}{3025}
\nfQ{1153}{3394}{1769}{3394}
\nfQ{2453}{3394}{2887}{2945}
\nfQ{3322}{2497}{3322}{1706}
\nfQ{3322}{1066}{3130}{698}
\nfQ{2938}{331}{2570}{128}
\nfQ{2203}{-75}{1769}{-75}
\nfQ{1072}{-75}{642}{372}
\nfQ{213}{819}{213}{1659}
\nfZ
\nfM{791}{1659}
\nfQ{791}{1022}{1069}{705}
\nfQ{1347}{388}{1769}{388}
\nfQ{2188}{388}{2466}{706}
\nfQ{2744}{1025}{2744}{1678}
\nfQ{2744}{2294}{2464}{2611}
\nfQ{2184}{2928}{1769}{2928}
\nfQ{1347}{2928}{1069}{2612}
\nfQ{791}{2297}{791}{1659}
\nfZ}
\expandafter\def\csname nfpath40\endcsname{\nfM{1034}{0}
\nfL{19}{3319}
\nfL{600}{3319}
\nfL{1128}{1403}
\nfL{1325}{691}
\nfQ{1338}{744}{1497}{1375}
\nfL{2025}{3319}
\nfL{2603}{3319}
\nfL{3100}{1394}
\nfL{3266}{759}
\nfL{3456}{1400}
\nfL{4025}{3319}
\nfL{4572}{3319}
\nfL{3534}{0}
\nfL{2950}{0}
\nfL{2422}{1988}
\nfL{2294}{2553}
\nfL{1622}{0}
\nfL{1034}{0}
\nfZ}
\expandafter\def\csname nfpath41\endcsname{\nfM{197}{991}
\nfL{753}{1078}
\nfQ{800}{744}{1014}{566}
\nfQ{1228}{388}{1613}{388}
\nfQ{2000}{388}{2187}{545}
\nfQ{2375}{703}{2375}{916}
\nfQ{2375}{1106}{2209}{1216}
\nfQ{2094}{1291}{1634}{1406}
\nfQ{1016}{1563}{777}{1677}
\nfQ{538}{1791}{414}{1992}
\nfQ{291}{2194}{291}{2438}
\nfQ{291}{2659}{392}{2848}
\nfQ{494}{3038}{669}{3163}
\nfQ{800}{3259}{1026}{3326}
\nfQ{1253}{3394}{1513}{3394}
\nfQ{1903}{3394}{2198}{3281}
\nfQ{2494}{3169}{2634}{2976}
\nfQ{2775}{2784}{2828}{2463}
\nfL{2278}{2388}
\nfQ{2241}{2644}{2061}{2787}
\nfQ{1881}{2931}{1553}{2931}
\nfQ{1166}{2931}{1000}{2803}
\nfQ{834}{2675}{834}{2503}
\nfQ{834}{2394}{903}{2306}
\nfQ{972}{2216}{1119}{2156}
\nfQ{1203}{2125}{1616}{2013}
\nfQ{2213}{1853}{2448}{1751}
\nfQ{2684}{1650}{2818}{1456}
\nfQ{2953}{1263}{2953}{975}
\nfQ{2953}{694}{2789}{445}
\nfQ{2625}{197}{2315}{61}
\nfQ{2006}{-75}{1616}{-75}
\nfQ{969}{-75}{630}{194}
\nfQ{291}{463}{197}{991}
\nfZ}
\expandafter\def\csname nfpath42\endcsname{\nfM{1984}{4856}
\nfQ{1566}{4138}{1362}{3434}
\nfQ{1159}{2731}{1159}{2009}
\nfQ{1159}{1288}{1364}{580}
\nfQ{1569}{-128}{1984}{-844}
\nfL{1484}{-844}
\nfQ{1016}{-109}{783}{600}
\nfQ{550}{1309}{550}{2009}
\nfQ{550}{2706}{781}{3412}
\nfQ{1013}{4119}{1484}{4856}
\nfL{1984}{4856}
\nfZ}
\expandafter\def\csname nfpath43\endcsname{\nfM{750}{794}
\nfL{1409}{794}
\nfL{1409}{256}
\nfL{897}{-744}
\nfL{494}{-744}
\nfL{750}{256}
\nfL{750}{794}
\nfZ}
\expandafter\def\csname nfpath44\endcsname{\nfM{2034}{4250}
\nfQ{1547}{4250}{1301}{3770}
\nfQ{1056}{3291}{1056}{2328}
\nfQ{1056}{1369}{1301}{889}
\nfQ{1547}{409}{2034}{409}
\nfQ{2525}{409}{2770}{889}
\nfQ{3016}{1369}{3016}{2328}
\nfQ{3016}{3291}{2770}{3770}
\nfQ{2525}{4250}{2034}{4250}
\nfZ
\nfM{2034}{4750}
\nfQ{2819}{4750}{3233}{4129}
\nfQ{3647}{3509}{3647}{2328}
\nfQ{3647}{1150}{3233}{529}
\nfQ{2819}{-91}{2034}{-91}
\nfQ{1250}{-91}{836}{529}
\nfQ{422}{1150}{422}{2328}
\nfQ{422}{3509}{836}{4129}
\nfQ{1250}{4750}{2034}{4750}
\nfZ}
\expandafter\def\csname nfpath45\endcsname{\nfM{513}{4856}
\nfL{1013}{4856}
\nfQ{1481}{4119}{1714}{3412}
\nfQ{1947}{2706}{1947}{2009}
\nfQ{1947}{1309}{1714}{600}
\nfQ{1481}{-109}{1013}{-844}
\nfL{513}{-844}
\nfQ{928}{-128}{1133}{580}
\nfQ{1338}{1288}{1338}{2009}
\nfQ{1338}{2731}{1133}{3434}
\nfQ{928}{4138}{513}{4856}
\nfZ}
\expandafter\def\csname nfpath46\endcsname{\nfM{2853}{2969}
\nfQ{2766}{3016}{2653}{3041}
\nfQ{2541}{3066}{2413}{3066}
\nfQ{1953}{3066}{1609}{2717}
\nfQ{1266}{2369}{1153}{1784}
\nfL{800}{0}
\nfL{225}{0}
\nfL{909}{3500}
\nfL{1484}{3500}
\nfL{1375}{2956}
\nfQ{1603}{3259}{1920}{3421}
\nfQ{2238}{3584}{2597}{3584}
\nfQ{2691}{3584}{2781}{3573}
\nfQ{2872}{3563}{2963}{3538}
\nfL{2853}{2969}
\nfZ}
\expandafter\def\csname nfpath47\endcsname{\nfM{3059}{4863}
\nfL{2969}{4384}
\nfL{2419}{4384}
\nfQ{2106}{4384}{1964}{4261}
\nfQ{1822}{4138}{1753}{3809}
\nfL{1691}{3500}
\nfL{2638}{3500}
\nfL{2553}{3053}
\nfL{1606}{3053}
\nfL{1013}{0}
\nfL{434}{0}
\nfL{1031}{3053}
\nfL{481}{3053}
\nfL{563}{3500}
\nfL{1113}{3500}
\nfL{1159}{3744}
\nfQ{1278}{4363}{1576}{4613}
\nfQ{1875}{4863}{2516}{4863}
\nfL{3059}{4863}
\nfZ}
\expandafter\def\csname nfpath48\endcsname{\nfM{266}{2259}
\nfQ{266}{3072}{433}{3567}
\nfQ{600}{4063}{929}{4331}
\nfQ{1259}{4600}{1759}{4600}
\nfQ{2128}{4600}{2406}{4451}
\nfQ{2684}{4303}{2865}{4023}
\nfQ{3047}{3744}{3150}{3342}
\nfQ{3253}{2941}{3253}{2259}
\nfQ{3253}{1453}{3087}{958}
\nfQ{2922}{463}{2592}{192}
\nfQ{2263}{-78}{1759}{-78}
\nfQ{1097}{-78}{719}{397}
\nfQ{266}{969}{266}{2259}
\nfZ
\nfM{844}{2259}
\nfQ{844}{1131}{1108}{757}
\nfQ{1372}{384}{1759}{384}
\nfQ{2147}{384}{2411}{759}
\nfQ{2675}{1134}{2675}{2259}
\nfQ{2675}{3391}{2411}{3762}
\nfQ{2147}{4134}{1753}{4134}
\nfQ{1366}{4134}{1134}{3806}
\nfQ{844}{3388}{844}{2259}
\nfZ}
\expandafter\def\csname nfpath49\endcsname{\nfM{488}{0}
\nfL{488}{4581}
\nfL{1109}{4581}
\nfL{3516}{984}
\nfL{3516}{4581}
\nfL{4097}{4581}
\nfL{4097}{0}
\nfL{3475}{0}
\nfL{1069}{3600}
\nfL{1069}{0}
\nfL{488}{0}
\nfZ}
\expandafter\def\csname nfpath50\endcsname{\nfM{47}{0}
\nfL{1259}{1725}
\nfL{138}{3319}
\nfL{841}{3319}
\nfL{1350}{2541}
\nfQ{1494}{2319}{1581}{2169}
\nfQ{1719}{2375}{1834}{2534}
\nfL{2394}{3319}
\nfL{3066}{3319}
\nfL{1919}{1756}
\nfL{3153}{0}
\nfL{2463}{0}
\nfL{1781}{1031}
\nfL{1600}{1309}
\nfL{728}{0}
\nfL{47}{0}
\nfZ}
\expandafter\def\csname nfpath51\endcsname{\nfM{2419}{4116}
\nfL{825}{1625}
\nfL{2419}{1625}
\nfL{2419}{4116}
\nfZ
\nfM{2253}{4666}
\nfL{3047}{4666}
\nfL{3047}{1625}
\nfL{3713}{1625}
\nfL{3713}{1100}
\nfL{3047}{1100}
\nfL{3047}{0}
\nfL{2419}{0}
\nfL{2419}{1100}
\nfL{313}{1100}
\nfL{313}{1709}
\nfL{2253}{4666}
\nfZ}
\expandafter\def\csname nfpath52\endcsname{\nfM{1228}{531}
\nfL{3431}{531}
\nfL{3431}{0}
\nfL{469}{0}
\nfL{469}{531}
\nfQ{828}{903}{1448}{1529}
\nfQ{2069}{2156}{2228}{2338}
\nfQ{2531}{2678}{2651}{2914}
\nfQ{2772}{3150}{2772}{3378}
\nfQ{2772}{3750}{2511}{3984}
\nfQ{2250}{4219}{1831}{4219}
\nfQ{1534}{4219}{1204}{4116}
\nfQ{875}{4013}{500}{3803}
\nfL{500}{4441}
\nfQ{881}{4594}{1212}{4672}
\nfQ{1544}{4750}{1819}{4750}
\nfQ{2544}{4750}{2975}{4387}
\nfQ{3406}{4025}{3406}{3419}
\nfQ{3406}{3131}{3298}{2873}
\nfQ{3191}{2616}{2906}{2266}
\nfQ{2828}{2175}{2409}{1742}
\nfQ{1991}{1309}{1228}{531}
\nfZ}
\expandafter\def\csname nfpath53\endcsname{\nfM{678}{2272}
\nfL{4684}{2272}
\nfL{4684}{1741}
\nfL{678}{1741}
\nfL{678}{2272}
\nfZ}
\expandafter\def\csname nfpath54\endcsname{\nfM{494}{0}
\nfL{494}{4581}
\nfL{2222}{4581}
\nfQ{2678}{4581}{2919}{4538}
\nfQ{3256}{4481}{3484}{4323}
\nfQ{3713}{4166}{3852}{3881}
\nfQ{3991}{3597}{3991}{3256}
\nfQ{3991}{2672}{3619}{2267}
\nfQ{3247}{1863}{2275}{1863}
\nfL{1100}{1863}
\nfL{1100}{0}
\nfL{494}{0}
\nfZ
\nfM{1100}{2403}
\nfL{2284}{2403}
\nfQ{2872}{2403}{3119}{2622}
\nfQ{3366}{2841}{3366}{3238}
\nfQ{3366}{3525}{3220}{3729}
\nfQ{3075}{3934}{2838}{4000}
\nfQ{2684}{4041}{2272}{4041}
\nfL{1100}{4041}
\nfL{1100}{2403}
\nfZ}
\expandafter\def\csname nfpath55\endcsname{\nfM{941}{0}
\nfL{419}{0}
\nfL{419}{4581}
\nfL{981}{4581}
\nfL{981}{2947}
\nfQ{1338}{3394}{1891}{3394}
\nfQ{2197}{3394}{2470}{3270}
\nfQ{2744}{3147}{2920}{2923}
\nfQ{3097}{2700}{3197}{2384}
\nfQ{3297}{2069}{3297}{1709}
\nfQ{3297}{856}{2875}{390}
\nfQ{2453}{-75}{1863}{-75}
\nfQ{1275}{-75}{941}{416}
\nfL{941}{0}
\nfZ
\nfM{934}{1684}
\nfQ{934}{1088}{1097}{822}
\nfQ{1363}{388}{1816}{388}
\nfQ{2184}{388}{2453}{708}
\nfQ{2722}{1028}{2722}{1663}
\nfQ{2722}{2313}{2464}{2622}
\nfQ{2206}{2931}{1841}{2931}
\nfQ{1472}{2931}{1203}{2611}
\nfQ{934}{2291}{934}{1684}
\nfZ}
\expandafter\def\csname nfpath56\endcsname{\nfM{409}{0}
\nfL{409}{4581}
\nfL{972}{4581}
\nfL{972}{0}
\nfL{409}{0}
\nfZ}
\expandafter\def\csname nfpath57\endcsname{\nfM{425}{3934}
\nfL{425}{4581}
\nfL{988}{4581}
\nfL{988}{3934}
\nfL{425}{3934}
\nfZ
\nfM{425}{0}
\nfL{425}{3319}
\nfL{988}{3319}
\nfL{988}{0}
\nfL{425}{0}
\nfZ}
\expandafter\def\csname nfpath58\endcsname{\nfM{422}{-1272}
\nfL{422}{3319}
\nfL{934}{3319}
\nfL{934}{2888}
\nfQ{1116}{3141}{1344}{3267}
\nfQ{1572}{3394}{1897}{3394}
\nfQ{2322}{3394}{2647}{3175}
\nfQ{2972}{2956}{3137}{2557}
\nfQ{3303}{2159}{3303}{1684}
\nfQ{3303}{1175}{3120}{767}
\nfQ{2938}{359}{2589}{142}
\nfQ{2241}{-75}{1856}{-75}
\nfQ{1575}{-75}{1351}{44}
\nfQ{1128}{163}{984}{344}
\nfL{984}{-1272}
\nfL{422}{-1272}
\nfZ
\nfM{931}{1641}
\nfQ{931}{1000}{1190}{694}
\nfQ{1450}{388}{1819}{388}
\nfQ{2194}{388}{2461}{705}
\nfQ{2728}{1022}{2728}{1688}
\nfQ{2728}{2322}{2467}{2637}
\nfQ{2206}{2953}{1844}{2953}
\nfQ{1484}{2953}{1207}{2617}
\nfQ{931}{2281}{931}{1641}
\nfZ}
\expandafter\def\csname nfpath59\endcsname{\nfM{469}{0}
\nfL{469}{4581}
\nfL{2188}{4581}
\nfQ{2713}{4581}{3030}{4442}
\nfQ{3347}{4303}{3526}{4014}
\nfQ{3706}{3725}{3706}{3409}
\nfQ{3706}{3116}{3547}{2856}
\nfQ{3388}{2597}{3066}{2438}
\nfQ{3481}{2316}{3704}{2022}
\nfQ{3928}{1728}{3928}{1328}
\nfQ{3928}{1006}{3792}{729}
\nfQ{3656}{453}{3456}{303}
\nfQ{3256}{153}{2954}{76}
\nfQ{2653}{0}{2216}{0}
\nfL{469}{0}
\nfZ
\nfM{1075}{2656}
\nfL{2066}{2656}
\nfQ{2469}{2656}{2644}{2709}
\nfQ{2875}{2778}{2992}{2937}
\nfQ{3109}{3097}{3109}{3338}
\nfQ{3109}{3566}{3000}{3739}
\nfQ{2891}{3913}{2687}{3977}
\nfQ{2484}{4041}{1991}{4041}
\nfL{1075}{4041}
\nfL{1075}{2656}
\nfZ
\nfM{1075}{541}
\nfL{2216}{541}
\nfQ{2509}{541}{2628}{563}
\nfQ{2838}{600}{2978}{687}
\nfQ{3119}{775}{3209}{942}
\nfQ{3300}{1109}{3300}{1328}
\nfQ{3300}{1584}{3169}{1773}
\nfQ{3038}{1963}{2805}{2039}
\nfQ{2572}{2116}{2134}{2116}
\nfL{1075}{2116}
\nfL{1075}{541}
\nfZ}
\expandafter\def\csname nfpath60\endcsname{\nfM{2575}{0}
\nfL{2575}{419}
\nfQ{2259}{-75}{1647}{-75}
\nfQ{1250}{-75}{917}{144}
\nfQ{584}{363}{401}{755}
\nfQ{219}{1147}{219}{1656}
\nfQ{219}{2153}{384}{2558}
\nfQ{550}{2963}{881}{3178}
\nfQ{1213}{3394}{1622}{3394}
\nfQ{1922}{3394}{2156}{3267}
\nfQ{2391}{3141}{2538}{2938}
\nfL{2538}{4581}
\nfL{3097}{4581}
\nfL{3097}{0}
\nfL{2575}{0}
\nfZ
\nfM{797}{1656}
\nfQ{797}{1019}{1065}{703}
\nfQ{1334}{388}{1700}{388}
\nfQ{2069}{388}{2326}{689}
\nfQ{2584}{991}{2584}{1609}
\nfQ{2584}{2291}{2321}{2609}
\nfQ{2059}{2928}{1675}{2928}
\nfQ{1300}{2928}{1048}{2622}
\nfQ{797}{2316}{797}{1656}
\nfZ}
\expandafter\def\csname nfpath61\endcsname{\nfM{503}{0}
\nfL{503}{4581}
\nfL{2534}{4581}
\nfQ{3147}{4581}{3465}{4457}
\nfQ{3784}{4334}{3975}{4021}
\nfQ{4166}{3709}{4166}{3331}
\nfQ{4166}{2844}{3850}{2509}
\nfQ{3534}{2175}{2875}{2084}
\nfQ{3116}{1969}{3241}{1856}
\nfQ{3506}{1613}{3744}{1247}
\nfL{4541}{0}
\nfL{3778}{0}
\nfL{3172}{953}
\nfQ{2906}{1366}{2734}{1584}
\nfQ{2563}{1803}{2427}{1890}
\nfQ{2291}{1978}{2150}{2013}
\nfQ{2047}{2034}{1813}{2034}
\nfL{1109}{2034}
\nfL{1109}{0}
\nfL{503}{0}
\nfZ
\nfM{1109}{2559}
\nfL{2413}{2559}
\nfQ{2828}{2559}{3062}{2645}
\nfQ{3297}{2731}{3419}{2920}
\nfQ{3541}{3109}{3541}{3331}
\nfQ{3541}{3656}{3305}{3865}
\nfQ{3069}{4075}{2559}{4075}
\nfL{1109}{4075}
\nfL{1109}{2559}
\nfZ}
\expandafter\def\csname nfpath62\endcsname{\nfM{422}{0}
\nfL{422}{4581}
\nfL{984}{4581}
\nfL{984}{2938}
\nfQ{1378}{3394}{1978}{3394}
\nfQ{2347}{3394}{2619}{3248}
\nfQ{2891}{3103}{3008}{2847}
\nfQ{3125}{2591}{3125}{2103}
\nfL{3125}{0}
\nfL{2563}{0}
\nfL{2563}{2103}
\nfQ{2563}{2525}{2380}{2717}
\nfQ{2197}{2909}{1863}{2909}
\nfQ{1613}{2909}{1392}{2779}
\nfQ{1172}{2650}{1078}{2428}
\nfQ{984}{2206}{984}{1816}
\nfL{984}{0}
\nfL{422}{0}
\nfZ}
\expandafter\def\csname nfpath63\endcsname{\nfM{319}{-275}
\nfL{866}{-356}
\nfQ{900}{-609}{1056}{-725}
\nfQ{1266}{-881}{1628}{-881}
\nfQ{2019}{-881}{2231}{-725}
\nfQ{2444}{-569}{2519}{-288}
\nfQ{2563}{-116}{2559}{434}
\nfQ{2191}{0}{1641}{0}
\nfQ{956}{0}{581}{494}
\nfQ{206}{988}{206}{1678}
\nfQ{206}{2153}{378}{2554}
\nfQ{550}{2956}{876}{3175}
\nfQ{1203}{3394}{1644}{3394}
\nfQ{2231}{3394}{2613}{2919}
\nfL{2613}{3319}
\nfL{3131}{3319}
\nfL{3131}{450}
\nfQ{3131}{-325}{2973}{-648}
\nfQ{2816}{-972}{2473}{-1159}
\nfQ{2131}{-1347}{1631}{-1347}
\nfQ{1038}{-1347}{672}{-1080}
\nfQ{306}{-813}{319}{-275}
\nfZ
\nfM{784}{1719}
\nfQ{784}{1066}{1043}{766}
\nfQ{1303}{466}{1694}{466}
\nfQ{2081}{466}{2343}{764}
\nfQ{2606}{1063}{2606}{1700}
\nfQ{2606}{2309}{2336}{2618}
\nfQ{2066}{2928}{1684}{2928}
\nfQ{1309}{2928}{1046}{2623}
\nfQ{784}{2319}{784}{1719}
\nfZ}
\expandafter\def\csname nfpath64\endcsname{\nfM{744}{3500}
\nfL{3475}{3500}
\nfL{3372}{2975}
\nfL{738}{459}
\nfL{2913}{459}
\nfL{2822}{0}
\nfL{-19}{0}
\nfL{84}{525}
\nfL{2719}{3041}
\nfL{653}{3041}
\nfL{744}{3500}
\nfZ}
\expandafter\def\csname nfpath65\endcsname{\nfM{691}{4666}
\nfL{3169}{4666}
\nfL{3169}{4134}
\nfL{1269}{4134}
\nfL{1269}{2991}
\nfQ{1406}{3038}{1543}{3061}
\nfQ{1681}{3084}{1819}{3084}
\nfQ{2600}{3084}{3056}{2656}
\nfQ{3513}{2228}{3513}{1497}
\nfQ{3513}{744}{3044}{326}
\nfQ{2575}{-91}{1722}{-91}
\nfQ{1428}{-91}{1123}{-41}
\nfQ{819}{9}{494}{109}
\nfL{494}{744}
\nfQ{775}{591}{1075}{516}
\nfQ{1375}{441}{1709}{441}
\nfQ{2250}{441}{2565}{725}
\nfQ{2881}{1009}{2881}{1497}
\nfQ{2881}{1984}{2565}{2268}
\nfQ{2250}{2553}{1709}{2553}
\nfQ{1456}{2553}{1204}{2497}
\nfQ{953}{2441}{691}{2322}
\nfL{691}{4666}
\nfZ}
\expandafter\def\csname nfpath66\endcsname{\nfM{2113}{2584}
\nfQ{1688}{2584}{1439}{2293}
\nfQ{1191}{2003}{1191}{1497}
\nfQ{1191}{994}{1439}{701}
\nfQ{1688}{409}{2113}{409}
\nfQ{2538}{409}{2786}{701}
\nfQ{3034}{994}{3034}{1497}
\nfQ{3034}{2003}{2786}{2293}
\nfQ{2538}{2584}{2113}{2584}
\nfZ
\nfM{3366}{4563}
\nfL{3366}{3988}
\nfQ{3128}{4100}{2886}{4159}
\nfQ{2644}{4219}{2406}{4219}
\nfQ{1781}{4219}{1451}{3797}
\nfQ{1122}{3375}{1075}{2522}
\nfQ{1259}{2794}{1537}{2939}
\nfQ{1816}{3084}{2150}{3084}
\nfQ{2853}{3084}{3261}{2657}
\nfQ{3669}{2231}{3669}{1497}
\nfQ{3669}{778}{3244}{343}
\nfQ{2819}{-91}{2113}{-91}
\nfQ{1303}{-91}{875}{529}
\nfQ{447}{1150}{447}{2328}
\nfQ{447}{3434}{972}{4092}
\nfQ{1497}{4750}{2381}{4750}
\nfQ{2619}{4750}{2861}{4703}
\nfQ{3103}{4656}{3366}{4563}
\nfZ}
\expandafter\def\csname nfpath67\endcsname{\nfM{525}{4666}
\nfL{3525}{4666}
\nfL{3525}{4397}
\nfL{1831}{0}
\nfL{1172}{0}
\nfL{2766}{4134}
\nfL{525}{4134}
\nfL{525}{4666}
\nfZ}
\expandafter\def\csname nfpath68\endcsname{\nfM{472}{0}
\nfL{472}{4581}
\nfL{3613}{4581}
\nfL{3613}{3806}
\nfL{1397}{3806}
\nfL{1397}{2722}
\nfL{3309}{2722}
\nfL{3309}{1947}
\nfL{1397}{1947}
\nfL{1397}{0}
\nfL{472}{0}
\nfZ}
\expandafter\def\csname nfpath69\endcsname{\nfM{459}{0}
\nfL{459}{4581}
\nfL{1338}{4581}
\nfL{1338}{0}
\nfL{459}{0}
\nfZ}
\expandafter\def\csname nfpath70\endcsname{\nfM{3763}{1606}
\nfL{4369}{1453}
\nfQ{4178}{706}{3683}{314}
\nfQ{3188}{-78}{2472}{-78}
\nfQ{1731}{-78}{1267}{223}
\nfQ{803}{525}{561}{1097}
\nfQ{319}{1669}{319}{2325}
\nfQ{319}{3041}{592}{3573}
\nfQ{866}{4106}{1370}{4382}
\nfQ{1875}{4659}{2481}{4659}
\nfQ{3169}{4659}{3637}{4309}
\nfQ{4106}{3959}{4291}{3325}
\nfL{3694}{3184}
\nfQ{3534}{3684}{3231}{3912}
\nfQ{2928}{4141}{2469}{4141}
\nfQ{1941}{4141}{1586}{3887}
\nfQ{1231}{3634}{1087}{3207}
\nfQ{944}{2781}{944}{2328}
\nfQ{944}{1744}{1114}{1308}
\nfQ{1284}{872}{1643}{656}
\nfQ{2003}{441}{2422}{441}
\nfQ{2931}{441}{3284}{734}
\nfQ{3638}{1028}{3763}{1606}
\nfZ}
\expandafter\def\csname nfpath71\endcsname{\nfM{556}{0}
\nfL{556}{2881}
\nfL{59}{2881}
\nfL{59}{3319}
\nfL{556}{3319}
\nfL{556}{3672}
\nfQ{556}{4006}{616}{4169}
\nfQ{697}{4388}{901}{4523}
\nfQ{1106}{4659}{1475}{4659}
\nfQ{1713}{4659}{2000}{4603}
\nfL{1916}{4113}
\nfQ{1741}{4144}{1584}{4144}
\nfQ{1328}{4144}{1222}{4034}
\nfQ{1116}{3925}{1116}{3625}
\nfL{1116}{3319}
\nfL{1763}{3319}
\nfL{1763}{2881}
\nfL{1116}{2881}
\nfL{1116}{0}
\nfL{556}{0}
\nfZ}
\expandafter\def\csname nfpath72\endcsname{\nfM{231}{1491}
\nfL{1131}{1578}
\nfQ{1213}{1125}{1461}{912}
\nfQ{1709}{700}{2131}{700}
\nfQ{2578}{700}{2804}{889}
\nfQ{3031}{1078}{3031}{1331}
\nfQ{3031}{1494}{2936}{1608}
\nfQ{2841}{1722}{2603}{1806}
\nfQ{2441}{1863}{1863}{2006}
\nfQ{1119}{2191}{819}{2459}
\nfQ{397}{2838}{397}{3381}
\nfQ{397}{3731}{595}{4036}
\nfQ{794}{4341}{1167}{4500}
\nfQ{1541}{4659}{2069}{4659}
\nfQ{2931}{4659}{3367}{4281}
\nfQ{3803}{3903}{3825}{3272}
\nfL{2900}{3231}
\nfQ{2841}{3584}{2645}{3739}
\nfQ{2450}{3894}{2059}{3894}
\nfQ{1656}{3894}{1428}{3728}
\nfQ{1281}{3622}{1281}{3444}
\nfQ{1281}{3281}{1419}{3166}
\nfQ{1594}{3019}{2269}{2859}
\nfQ{2944}{2700}{3267}{2529}
\nfQ{3591}{2359}{3773}{2064}
\nfQ{3956}{1769}{3956}{1334}
\nfQ{3956}{941}{3737}{597}
\nfQ{3519}{253}{3119}{86}
\nfQ{2719}{-81}{2122}{-81}
\nfQ{1253}{-81}{787}{320}
\nfQ{322}{722}{231}{1491}
\nfZ}
\expandafter\def\csname nfpath73\endcsname{\nfM{1372}{0}
\nfL{34}{3319}
\nfL{956}{3319}
\nfL{1581}{1625}
\nfL{1763}{1059}
\nfQ{1834}{1275}{1853}{1344}
\nfQ{1897}{1484}{1947}{1625}
\nfL{2578}{3319}
\nfL{3481}{3319}
\nfL{2163}{0}
\nfL{1372}{0}
\nfZ}
\expandafter\def\csname nfpath74\endcsname{\nfM{184}{1300}
\nfL{731}{1375}
\nfQ{753}{850}{928}{656}
\nfQ{1103}{463}{1413}{463}
\nfQ{1641}{463}{1806}{567}
\nfQ{1972}{672}{2034}{851}
\nfQ{2097}{1031}{2097}{1425}
\nfL{2097}{4581}
\nfL{2703}{4581}
\nfL{2703}{1459}
\nfQ{2703}{884}{2564}{568}
\nfQ{2425}{253}{2123}{87}
\nfQ{1822}{-78}{1416}{-78}
\nfQ{813}{-78}{492}{269}
\nfQ{172}{616}{184}{1300}
\nfZ}
\expandafter\def\csname nfpath75\endcsname{\nfM{3500}{4581}
\nfL{4106}{4581}
\nfL{4106}{1934}
\nfQ{4106}{1244}{3950}{837}
\nfQ{3794}{431}{3386}{176}
\nfQ{2978}{-78}{2316}{-78}
\nfQ{1672}{-78}{1262}{144}
\nfQ{853}{366}{678}{786}
\nfQ{503}{1206}{503}{1934}
\nfL{503}{4581}
\nfL{1109}{4581}
\nfL{1109}{1938}
\nfQ{1109}{1341}{1220}{1058}
\nfQ{1331}{775}{1601}{622}
\nfQ{1872}{469}{2263}{469}
\nfQ{2931}{469}{3215}{772}
\nfQ{3500}{1075}{3500}{1938}
\nfL{3500}{4581}
\nfZ}
\expandafter\def\csname nfpath76\endcsname{\nfM{1081}{4666}
\nfL{2694}{4666}
\nfQ{3350}{4666}{3675}{4422}
\nfQ{4000}{4178}{4000}{3688}
\nfQ{4000}{3238}{3720}{2911}
\nfQ{3441}{2584}{2988}{2516}
\nfQ{3375}{2428}{3569}{2181}
\nfQ{3763}{1934}{3763}{1522}
\nfQ{3763}{819}{3242}{409}
\nfQ{2722}{0}{1819}{0}
\nfL{172}{0}
\nfL{1081}{4666}
\nfZ
\nfM{1234}{2228}
\nfL{903}{519}
\nfL{1919}{519}
\nfQ{2491}{519}{2800}{781}
\nfQ{3109}{1044}{3109}{1522}
\nfQ{3109}{1891}{2904}{2059}
\nfQ{2700}{2228}{2247}{2228}
\nfL{1234}{2228}
\nfZ
\nfM{1606}{4147}
\nfL{1331}{2741}
\nfL{2272}{2741}
\nfQ{2775}{2741}{3058}{2959}
\nfQ{3341}{3178}{3341}{3566}
\nfQ{3341}{3869}{3150}{4008}
\nfQ{2959}{4147}{2541}{4147}
\nfL{1606}{4147}
\nfZ}
\expandafter\def\csname nfpath77\endcsname{\nfM{4447}{4306}
\nfL{4319}{3641}
\nfQ{4019}{3944}{3683}{4091}
\nfQ{3347}{4238}{2956}{4238}
\nfQ{2422}{4238}{2017}{3981}
\nfQ{1613}{3725}{1319}{3200}
\nfQ{1131}{2863}{1032}{2486}
\nfQ{934}{2109}{934}{1728}
\nfQ{934}{1091}{1264}{756}
\nfQ{1594}{422}{2222}{422}
\nfQ{2656}{422}{3056}{561}
\nfQ{3456}{700}{3834}{978}
\nfL{3688}{231}
\nfQ{3316}{72}{2936}{-9}
\nfQ{2556}{-91}{2175}{-91}
\nfQ{1278}{-91}{773}{396}
\nfQ{269}{884}{269}{1753}
\nfQ{269}{2309}{461}{2846}
\nfQ{653}{3384}{1013}{3828}
\nfQ{1394}{4300}{1883}{4525}
\nfQ{2372}{4750}{3022}{4750}
\nfQ{3422}{4750}{3780}{4639}
\nfQ{4138}{4528}{4447}{4306}
\nfZ}
\expandafter\def\csname nfpath78\endcsname{\nfM{309}{2231}
\nfQ{309}{3372}{921}{4017}
\nfQ{1534}{4663}{2503}{4663}
\nfQ{3138}{4663}{3647}{4359}
\nfQ{4156}{4056}{4423}{3514}
\nfQ{4691}{2972}{4691}{2284}
\nfQ{4691}{1588}{4409}{1038}
\nfQ{4128}{488}{3612}{205}
\nfQ{3097}{-78}{2500}{-78}
\nfQ{1853}{-78}{1343}{234}
\nfQ{834}{547}{571}{1087}
\nfQ{309}{1628}{309}{2231}
\nfZ
\nfM{934}{2222}
\nfQ{934}{1394}{1379}{917}
\nfQ{1825}{441}{2497}{441}
\nfQ{3181}{441}{3623}{922}
\nfQ{4066}{1403}{4066}{2288}
\nfQ{4066}{2847}{3877}{3264}
\nfQ{3688}{3681}{3323}{3911}
\nfQ{2959}{4141}{2506}{4141}
\nfQ{1863}{4141}{1398}{3698}
\nfQ{934}{3256}{934}{2222}
\nfZ}
\expandafter\def\csname nfpath79\endcsname{\nfM{3381}{2694}
\nfL{356}{2694}
\nfL{356}{3219}
\nfL{3381}{3219}
\nfL{3381}{2694}
\nfZ
\nfM{3381}{1303}
\nfL{356}{1303}
\nfL{356}{1828}
\nfL{3381}{1828}
\nfL{3381}{1303}
\nfZ}
\def\nfGA#1#2#3{\nfB
\nfFC{nf667782}
\nfFO{1}
\nfSC{nfffffff}
\nfSO{1}
\nfW{0.5bp}
\nfBC
\nfRJ
\nfNZ
\pgftransformcm{1}{0}{0}{-1}{\pgfpoint{#1bp}{#2bp}}
\csname nfpath#3\endcsname
\nfU{fill,stroke}
\nfE}
\def\nfGB#1#2#3{\nfB
\nfFC{nf007f82}
\nfFO{1}
\nfSC{nfffffff}
\nfSO{1}
\nfW{0.5bp}
\nfBC
\nfRJ
\nfNZ
\pgftransformcm{1}{0}{0}{-1}{\pgfpoint{#1bp}{#2bp}}
\csname nfpath#3\endcsname
\nfU{fill,stroke}
\nfE}
\def\nfGC#1#2#3{\nfB
\nfFC{nfbc845d}
\nfFO{1}
\nfSC{nfffffff}
\nfSO{1}
\nfW{0.5bp}
\nfBC
\nfRJ
\nfNZ
\pgftransformcm{1}{0}{0}{-1}{\pgfpoint{#1bp}{#2bp}}
\csname nfpath#3\endcsname
\nfU{fill,stroke}
\nfE}
\def\nfGD#1#2#3{\nfB
\nfFC{nf202b33}
\nfFO{1}
\nfNZ
\pgftransformcm{0.00148437}{0}{0}{0.00148437}{\pgfpoint{#1bp}{#2bp}}
\csname nfpath#3\endcsname
\nfU{fill}
\nfE}
\def\nfGE#1#2#3{\nfB
\nfFC{nf202b33}
\nfFO{1}
\nfNZ
\pgftransformcm{0.00134375}{0}{0}{0.00134375}{\pgfpoint{#1bp}{#2bp}}
\csname nfpath#3\endcsname
\nfU{fill}
\nfE}
\def\nfGF#1#2#3{\nfB
\nfFC{nf202b33}
\nfFO{1}
\nfNZ
\pgftransformcm{0.00125}{0}{0}{0.00125}{\pgfpoint{#1bp}{#2bp}}
\csname nfpath#3\endcsname
\nfU{fill}
\nfE}
\def\nfGG#1#2#3{\nfB
\nfFC{nf586570}
\nfFO{1}
\nfNZ
\pgftransformcm{0.00115625}{0}{0}{0.00115625}{\pgfpoint{#1bp}{#2bp}}
\csname nfpath#3\endcsname
\nfU{fill}
\nfE}
\def\nfGH#1#2#3{\nfB
\nfFC{nf202b33}
\nfFO{1}
\nfNZ
\pgftransformcm{0.00126562}{0}{0}{0.00126562}{\pgfpoint{#1bp}{#2bp}}
\csname nfpath#3\endcsname
\nfU{fill}
\nfE}
\def\nfGI#1#2#3{\nfB
\nfFC{nf202b33}
\nfFO{1}
\nfNZ
\pgftransformcm{0.00114062}{0}{0}{0.00114062}{\pgfpoint{#1bp}{#2bp}}
\csname nfpath#3\endcsname
\nfU{fill}
\nfE}
\def\nfGJ#1#2#3{\nfB
\nfFC{nf202b33}
\nfFO{1}
\nfNZ
\pgftransformcm{0.0015625}{0}{0}{0.0015625}{\pgfpoint{#1bp}{#2bp}}
\csname nfpath#3\endcsname
\nfU{fill}
\nfE}
\def\nfGK#1#2#3{\nfB
\nfFC{nf202b33}
\nfFO{1}
\nfNZ
\pgftransformcm{0.00109375}{0}{0}{0.00109375}{\pgfpoint{#1bp}{#2bp}}
\csname nfpath#3\endcsname
\nfU{fill}
\nfE}
\def\nfGL#1#2#3{\nfB
\nfFC{nf586570}
\nfFO{1}
\nfNZ
\pgftransformcm{0.001125}{0}{0}{0.001125}{\pgfpoint{#1bp}{#2bp}}
\csname nfpath#3\endcsname
\nfU{fill}
\nfE}
\def\nfGM#1#2#3{\nfB
\nfFC{nf8064a2}
\nfFO{1}
\nfNZ
\pgftransformcm{0.00129687}{0}{0}{0.00129687}{\pgfpoint{#1bp}{#2bp}}
\csname nfpath#3\endcsname
\nfU{fill}
\nfE}
\def\nfGN#1#2#3{\nfB
\nfFC{nf586570}
\nfFO{1}
\nfNZ
\pgftransformcm{0.00109375}{0}{0}{0.00109375}{\pgfpoint{#1bp}{#2bp}}
\csname nfpath#3\endcsname
\nfU{fill}
\nfE}
\def\nfGO#1#2#3{\nfB
\nfFC{nf202b33}
\nfFO{1}
\nfNZ
\pgftransformcm{0.00117187}{0}{0}{0.00117187}{\pgfpoint{#1bp}{#2bp}}
\csname nfpath#3\endcsname
\nfU{fill}
\nfE}
\def\nfGP#1#2#3{\nfB
\nfFC{nf202b33}
\nfFO{1}
\nfNZ
\pgftransformcm{0.00082031}{0}{0}{0.00082031}{\pgfpoint{#1bp}{#2bp}}
\csname nfpath#3\endcsname
\nfU{fill}
\nfE}
\def\nfGQ#1#2#3{\nfB
\nfFC{nf667782}
\nfFO{1}
\nfNZ
\pgftransformcm{0.00129687}{0}{0}{0.00129687}{\pgfpoint{#1bp}{#2bp}}
\csname nfpath#3\endcsname
\nfU{fill}
\nfE}
\def\nfGR#1#2#3{\nfB
\nfFC{nf007f82}
\nfFO{1}
\nfNZ
\pgftransformcm{0.00129687}{0}{0}{0.00129687}{\pgfpoint{#1bp}{#2bp}}
\csname nfpath#3\endcsname
\nfU{fill}
\nfE}
\def\nfGS#1#2#3{\nfB
\nfFC{nf8064a2}
\nfFO{1}
\nfNZ
\pgftransformcm{0.00110937}{0}{0}{0.00110937}{\pgfpoint{#1bp}{#2bp}}
\csname nfpath#3\endcsname
\nfU{fill}
\nfE}
\def\nfGT#1#2#3{\nfB
\nfFC{nf202b33}
\nfFO{1}
\nfNZ
\pgftransformcm{0.00121875}{0}{0}{0.00121875}{\pgfpoint{#1bp}{#2bp}}
\csname nfpath#3\endcsname
\nfU{fill}
\nfE}
\def\nfGU#1#2#3{\nfB
\nfFC{nf202b33}
\nfFO{1}
\nfNZ
\pgftransformcm{0.00085312}{0}{0}{0.00085312}{\pgfpoint{#1bp}{#2bp}}
\csname nfpath#3\endcsname
\nfU{fill}
\nfE}
\def\nfGV#1#2#3{\nfB
\nfFC{nf007f82}
\nfFO{1}
\nfNZ
\pgftransformcm{0.00110937}{0}{0}{0.00110937}{\pgfpoint{#1bp}{#2bp}}
\csname nfpath#3\endcsname
\nfU{fill}
\nfE}

\endgroup
\caption{\textbf{The same current probabilities can lead to different futures.} \textbf{a}, At $\gamma=1/3$, state pairs $(1,0)$ and $(4,3)$ have identical output probabilities after alignment. Output $f=0$ leaves reference noise balances $0$ and $-1$, whose next rounding remainders are $0$ and $2/3$. \textbf{b}, In this first-order schematic, reading left to right, the one-phase analysis renews its choice after each participation; a full word retains one of $B^C$ starting choices; intermediate blocks keep selected dependencies together. Each retained block is transformed as a whole when the integer state shifts. The eight-digit rows are schematic.}
\label{fig:intro-motivation}
\end{figure}

\textbf{Exact privacy computation for finite memory.} Complete preparation words, vector carries and a synchronizing echo yield an exact finite testing object. For fixed \(m\ge1,C\ge2\), its support-width exponent is \((m^2+m+1)C-m-1\) (Theorem 4). The first-order exponent \(3C-2\) and its sharper C2 reduction are recovered. Public preparation and certified source transfer connect this finite object to discrete Gaussian (DG) privacy over all deterministic finite deadlines.

\textbf{The calibration cost of shorter dependencies.} Recursive proof groups retain a chosen number of future digits, with complete carry transport and convex aggregation. The resulting bounds decrease to the exact value at depth \(mC\) (Theorems 5--7). Three-memory noise already exhibits three strict levels with one participation. For a two-memory DG example, a 0.20 epsilon tolerance requires depth two throughout a specified risk interval.

\textbf{Privacy under a physical deadline.} Under \(T=\lfloor\tau C\rfloor\), optimal privacy cost is \(C\Phi(\tau)+O(\sqrt C)\) at fixed risk and time ratio (Theorems 9--10). The leading coefficient uses the actual hidden-output Kullback--Leibler (KL) reward. Weighted contraction, closed-class entry and hard-budget realization connect its occupation description to legal executions. A finite-window algorithm computes \(\Phi\) to error \(2^{-k}\) with polynomial bit cost in \(k\) for fixed memory and rational parameters (Theorem 11).

Sections 3--6 follow the preparation constraint through exact computation, relaxation and physical-time optimization. Scalar examples expose the carry geometry; Section 6.4 sharpens finite-deadline constants. Section 7 evaluates calibration, resources and workloads; Appendix I specifies formal coverage.

\section{Related work}

The noise process belongs to the mean-preserving rounded integer autoregressive family [35], with coefficients \(-a_i/B\) and DG innovations. Weiß and Zhu study its stationarity, moments and transition probabilities. We construct finite privacy objects that preserve shared preparations, complete public observation and predictable participation constraints, then quantify the effects of relaxing dependencies and limiting physical time.

Discrete-Gaussian translation bounds [1], Rényi differential privacy (RDP) and filters [2,3], privacy profiles [34] and privacy loss distribution (PLD) accountants [19,21] supply the statistical foundations. Profiles already sharpen private selection [38]; PLD realizations support subsampling and random allocation [42]. Optimal RDP-to-testing conversion [17] applies to a bound, while an exact fixed-pair curve determines testing information under a finite-moment condition [24]. Bok [39] shows that order-specific dominating pairs can favor RDP accounting for private stochastic gradient descent with fixed-size sampling and replace adjacency. Our separation identifies the experiment identity lost by orderwise optimization over a family; the block comparisons preserve progressively more cross-time constraints.

Partially observable Markov decision processes (POMDPs) and incremental pruning [18,23] use convex value geometry. Information relaxation [36], delayed-information bounds [20] and generalized parameter lifting [15] explain enlarged decision freedom. Natural-filter analyses [40,41] establish additional conditions for composing trade-off curves selected along adaptive paths; Regehr et al. [41] also give a worst-case PLD recursion. Our fixed-cap recursion preserves legal histories. The contribution is the carry and echo construction of a finite testing object from (2), its support-sensitive complexity and the calibration cost of specified relaxations. Appendix K.3 relates the resulting object to parametric Markov chains.

Controlled sensing [31], state-action frequencies [48] and constrained Markov decision process (MDP) approximation [25,30,32] supply information-rate precedents. Poisson bounds [49] connect stationary rewards to finite trajectories. Lan and Wang [37] constrain sample count and elapsed time under independent source arrivals, expected budgets and sequential stopping. Here caps hold pathwise and deadlines are deterministic. Weighted contraction, actual closed-class entry and likelihood compensation [29,33] preserve the hidden-output reward while connecting occupations to the privacy inverse.

Probabilistic abstraction [4,14] supplies probability bounds; CheckDP [8], relational symbolic execution [9] and Gaussian-program verification [12,13] automate privacy reasoning. SampCert [10] and Clutch-DP [28] verify foundational or modular implementations. SuperDP [43] synthesizes expectation-martingale witnesses for pure-DP refutation. Our general proofs cover complete integer transcripts, legal compilation, commitments and resource laws; the first-order specialization also has Lean refinements of these objects and a binary value algorithm. Implementation sensitivity [5] and public dither [11] motivate explicit sampling and observation contracts.

Gaussian matrix mechanisms already protect complete adaptive streams through uniform strategy sensitivity [44]. Alongside workload optimization [26] and analytic calibration [27], correlation designs include buffered Toeplitz mechanisms [7], one-step innovation cancellation [45] and banded inverse factorizations [46,47]. They study prefix error, repeated participation and storage or regeneration of innovations. Our recurrence feeds back an exact integer register with hidden rounding. Section 7 compares designs under common workload and output contracts; its hard-cap comparisons use no random-allocation amplification [42].

\section{Mechanism, common coordinates, and privacy information}

\subsection{The mechanism and the analysis objects}

Fix \(m\ge1\), \(B\ge2\), \(a=(a_1,\ldots,a_m)\in\mathbb Z^m\), and \(\sigma>0\), with
\[
\gcd(a_1,B)=1,\qquad r:=\frac{\sum_{i=1}^m|a_i|}{B}<1.
\tag{1}
\]

Let \(\eta_t\) have mass \(g_\sigma(z)=Z_\sigma^{-1}\exp[-z^2/(2\sigma^2)]\) on \(\mathbb Z\), and independently draw \(U_t\) uniformly from \(\{0,\ldots,B-1\}\). All fresh draws are independent. The mechanism is
\[
K_{1-m}=\cdots=K_0=0,\qquad
K_t=\eta_t-\left\lfloor\frac{\sum_{i=1}^m a_iK_{t-i}+U_t}{B}\right\rfloor,
\qquad Y_t=x_t+K_t.
\tag{2}
\]

The register stores the past actual noise balances \(K\) exactly. The parameter \(\sigma^2\) is the squared DG scale; its actual variance is denoted separately when needed. Our first-order specialization has \(m=1\), \(0<a_1=a<B\), and \(\gamma=a/B\).

Every release is public; the rounding draws and internal states are hidden. Floors are mathematical floors, including at negative arguments. An ordered pair of deterministic input rules selects \(x_t^P,x_t^Q\) from the common public history before the fresh draws. Their difference \(j_t=x_t^Q-x_t^P\) belongs to \(\{-1,0,1\}\) and is nonzero at most \(C\) times on every history. Consecutive participations are allowed. Let \(\mathcal F_C\) include both ordered directions, and let \(\mathcal F=\mathcal F_2\) for the information separation below. For \(H_\lambda(P\Vert Q)=\sum_y(P(y)-\lambda Q(y))_+\), define
\[
\begin{aligned}
\Delta_{C,T}(\varepsilon)&=\sup_{\pi^P\sim_C\pi^Q}H_{e^\varepsilon}(P_T\Vert Q_T),
&\Delta_C^*(\varepsilon)&=\sup_{T<\infty}\Delta_{C,T}(\varepsilon),\\
E_{\mathcal F}(\alpha)&=\sup_{(P,Q)\in\mathcal F}D_\alpha(P\Vert Q).
\end{aligned}
\tag{3}
\]

The fixed-deadline profile retains all \(T\) physical outputs, including common-input waiting and echoes after the last participation. Taking its supremum over deterministic finite deadlines gives the target of exact compilation and commitment analysis. For \(0<\delta<1\), inverse calibration means \(\inf\{\varepsilon\ge0:\Delta(\varepsilon)\le\delta\}\), giving \(\varepsilon_C^*(\delta)\) and \(\varepsilon_{C,T}^*(\delta)\), respectively. Rényi order one means KL divergence. A generic experiment family's profile is denoted \(\Delta_{\mathcal A}\) in the same way.

Before release \(t\), write \(s^b=(K^b_{t-1},\ldots,K^b_{t-m})\) for world \(b\). The common public history and the fixed input rules determine both conditional registers. Their difference and its update are
\[
s^P=s^Q+d,\qquad d=(j_{t-1},\ldots,j_{t-m}),\qquad
d_j=(j,d_1,\ldots,d_{m-1}).
\tag{4}
\]

The compiler uses the Q-reference register to align output rows; the KL analysis uses the P-reference register to take expectations along an autonomous noise chain. Equation (4) converts between these coordinates of the same conditional state. In the first-order examples, the integer shift and two successive normalized remainders are
\[
b(k)=-\lfloor\gamma k\rfloor,\qquad
r(k)=\{\gamma k\},\qquad s(k)=\{\gamma b(k)\}.
\tag{5}
\]

To evaluate how well an observer can distinguish the two inputs, weight their output laws by \(x,y\). For innovation law \(\nu\), let \(G_c^\nu(x,y)\) be the largest weighted correct-testing score after synchronization with \(c\) participations remaining. Normalizing the weights gives \(\mathcal G_c^\nu\); subtracting the second weight gives the privacy payoff \(D_c^\nu\):
\[
\begin{aligned}
\mathcal G_c^\nu(t)&=G_c^\nu(t,1-t),\qquad
G_c^\nu(x,y)=(x+y)\mathcal G_c^\nu\!\left(\frac{x}{x+y}\right),\\
D_c^\nu(\lambda)&=G_c^\nu(1,\lambda)-\lambda.
\end{aligned}
\tag{6}
\]

Here \(x,y\ge0\), \(x+y>0\), \(t\in[0,1]\), and \(G_0^\nu(x,y)=\max(x,y)\). We suppress \(\nu\) when fixed. At a synchronized node of the finite-law game, a \emph{preparation} is one freely chosen common integer register. Its subsequent choices remain shared across the affected output branches. For a full-support innovation law with finite first moment, public common-input waiting reaches each selected preparation; deterministic truncation retains every timeout branch. Appendix A.1 proves that this realizes the all-finite-horizon value. Under a fixed deadline, the preparation time remains part of the optimization.

\subsection{Transcript moment bounds}

The difference register in (4) clears after at most \(m\) consecutive zero actions. Each participation can therefore affect at most \(m+1\) informative rows. Put \(D=(m+1)C\). Temporarily retaining the current rounding coin bounds each row by a DG shift of magnitude at most two; marginalizing that coin preserves the bound [1].

\textbf{Transcript moment envelope.} For every path-legal policy pair and every finite \(T\), both ordered transcript directions satisfy
\[
\mathbb E_Q\left[\left(\frac{dP}{dQ}\right)^2\right]
\le M_C=e^{4D/\sigma^2},\qquad
H_{e^\varepsilon}(P\Vert Q)
\le\min\left\{\frac{M_C}{4e^\varepsilon},
\frac{M_C-1}{4(e^\varepsilon-1)}\right\}
\quad(\varepsilon>0).
\tag{7}
\]

A rank potential pays for remaining participations and the pending echo; synchronized waiting has moment one. Taking the supremum and inverting the centered event bound gives
\[
\varepsilon_C^*(\delta)\le
\log\left(1+\frac{e^{4D/\sigma^2}-1}{4\delta}\right),
\qquad 0<\delta<1,
\tag{8}
\]

with \(\varepsilon_0^*(\delta)=0\). This controls endpoint search and inverse conditioning. Appendix F gives the general argument; the first-order specialization has \(D=2C\).

\subsection{Information and implementation separations}

The first-order subfamily already separates statistical summaries from the legal continuation structure. The next two results fix \(m=1\) and \(\gamma=a/B\).

\textbf{Theorem 1 (family-envelope separation).} At \(a/B=1/3\), \(\sigma=3\), an experiment family \(\mathcal F'\supset\mathcal F\) satisfies
\[
E_{\mathcal F'}(\alpha)=E_{\mathcal F}(\alpha)\quad(\alpha>0),
\qquad
\Delta_{\mathcal F}(2)<10^{-4}<\Delta_{\mathcal F'}(2).
\tag{9}
\]

Adjoin the binary experiment \(P_b(1)=23/10000\), \(Q_b(1)=27/100000\), and its reverse. At every positive order, both divergences lie below those of the zero-initial three-release pair with inputs \((0,0,0)\) and \((1,1,0)\). Yet
\[
\Delta_{\mathcal F}(2)\le0.000084311988530666,
\qquad
H_{e^2}(P_b\Vert Q_b)>0.000304954853288724.
\tag{10}
\]

Appendix B proves continuum domination by interval and tail bounds. The family envelope loses experiment identity. For a fixed pair, the full exact RDP curve determines its testing information when finite at some order greater than one [24]; (7) supplies that condition here.

Now allow a common public predictable offset \(s_t(Y_{<t})\in\{0,\ldots,B-1\}\) inside the floor in (2). Choosing \(s=(r-ak)\bmod B\) realizes any desired rounding residue before the fresh draws. Thus the phase-controlled relaxation is the worst testing curve of this larger implementation class.

\textbf{Proposition 2 (offset-contract separation).} At the same slope and noise scale, take \(x^P=(0,0)\), \(x^Q=(1,1)\), and
\[
s_1=0,\qquad s_2=(1-Y_1)\bmod3.
\tag{11}
\]

An observable event has payoff at least \(0.000118891298210654>10^{-4}\). Literal hidden-rounding sums verify the event (Appendix B.4). The predictable offset changes the subsequent mixture, explaining why exact analysis must retain arithmetic semantics as well as statistical dependence.

\section{Exact compilation through legal regeneration}

A preparation remains shared across every observation it constrains. The first-order C2 geometry makes this constraint explicit; complete vector carries and multi-step synchronization then give the finite-memory compiler. Throughout this section \(\nu(z)=w_z/S\) is the explicitly supplied normalized finite law on \([-L,L]\). The C2 reduction assumes symmetry; the general point compiler permits asymmetric counts and zeros.

\subsection{First-order geometry and the sharp C2 reduction}

Fix \(m=1\), \(0<a<B\) and a symmetric finite law. Use the Q-reference state \(k\), preceding difference \(d=1\), \(r=\{\gamma k\}\), \(e=\lfloor r+\gamma\rfloor\), and \(r'=r+\gamma-e\). In aligned output coordinate \(f=n+\lfloor\gamma k\rfloor\), the complete rows are
\[
q_r(f)=(1-r)\nu(f)+r\nu(f+1),\qquad
p_{r,j}(f)=(1-r')\nu(f+j+e)+r'\nu(f+j+e+1).
\tag{12}
\]

Let \(I=[-L-3,L+1]\cap\mathbb Z\), \(J=[-L-1,L]\cap\mathbb Z\), and \(G_B=\{0,1/B,\ldots,(B-1)/B\}\). The echo, one-participation root and pending continuation are
\[
\begin{aligned}
E(r;x,y)&=\sum_{f\in I}\max\{xp_{r,0}(f),yq_r(f)\},\\
G_1(x,y)&=\max_{\beta\in G_B}\sum_{f\in J}
E(\{\beta+\gamma f\};x\nu(f+1),y\nu(f)),
\end{aligned}
\tag{13}
\]

\[
\begin{aligned}
F(r,s;x,y)=\max\bigg\{&\max(x,y),\ \sum_{f\in I}G_1(xp_{r,0}(f),yq_r(f)),\\
&\sum_{f\in I}E(\{s+\gamma f\};xp_{r,1}(f),yq_r(f)),\\
&\sum_{f\in I}E(\{-s-\gamma f\};xp_{r,-1}(f),yq_r(f))\bigg\}.
\end{aligned}
\tag{14}
\]

Each action is chosen before its complete output sum. The next reference state is \(b(k)+f\), with phase \(\{s+\gamma f\}\); reflection handles the negative action. Between rounding boundaries \(F\) is convex in each phase. Reindexing \(f'=f+1\) across the first boundary gives
\[
F(1,s;x,y)=F(0,\{s-\gamma\};x,y).
\tag{15}
\]

Thus one shared preparation \((\beta,\chi)\) gives
\[
\begin{aligned}
r_f&=\{\beta+\gamma f\},\qquad
s_f=\{\chi-\gamma\lfloor\beta+\gamma f\rfloor\},\\
G_2(x,y)&=\max_{\beta,\chi\in G_B}\sum_{f\in J}
F(r_f,s_f;x\nu(f+1),y\nu(f)).
\end{aligned}
\tag{16}
\]

The pair remains fixed across all first-output branches. Two exact evaluations realize (16): one combines convex testing curves, and the other reconstructs complete joint rows by counting hidden-rounding preimages.

\textbf{Theorem 3 (exact C2 support and bit complexity).} The root maximum in (16) can be restricted, for every nonnegative weight pair, to
\[
\begin{aligned}
K_0&=\{-m\gamma\bmod1:-L-1\le m\le L+1\},\\
K_2&=\{n\gamma\bmod1:-2L-3\le n\le2L+4\}.
\end{aligned}
\tag{17}
\]

At most \(\min(B,2L+3)\min(B,4L+8)\) legal pairs suffice. For \(\lambda=u/v\), integers \(u\ge0,v\ge1\), set \(\ell=L+1\) and
\[
\kappa=1+\lceil\log_2(B+1)\rceil+\lceil\log_2(S+1)\rceil
+\lceil\log_2(u+v+1)\rceil+\lceil\log_2(L+2)\rceil.
\tag{18}
\]

With exact coordinates and schoolbook arithmetic, either path computes \(G_2(1,\lambda)-\lambda\) in
\[
O\!\left((\ell^4\log(\ell+1)+\log B)\kappa^2\right)
\quad\text{bit operations},\qquad O(\ell^3\kappa)\quad\text{bits of storage}.
\tag{19}
\]

The cost includes normalization, initialization, curve construction and binary input/\allowbreak{}output. Between successive \(K_0\) points the objective is convex; (15) preserves the carry at a wrap. The second-coordinate boundaries then lie in \(K_2\) (Figure 2; Appendix A). Reached phase classes require \(O(\ell^3)\) lines, with intermediate integers bounded by \((u+v+1)(BS)^4\) (Appendix C). The general compiler below extends the support exponents.

\input{figures/ctwo_compilation.tex}

The shared choice is essential: for law \((1,2,1)/4\), \(\gamma=1/3\), and \(\lambda=1/2\), selecting \(\chi\) separately after each first output changes the value from \(5/6\) to \(107/128\).

\subsection{Complete words, vector carries and finite-memory compilation}

Define the deterministic baseline map \(\mathbf b(s)=(-\lfloor a^\top s/B\rfloor,s_1,\ldots,s_{m-1})\) and phase digits \(\theta_i(s)=a^\top\mathbf b^i(s)\bmod B\). Let \(M\) have first row \(-a^\top\) and subdiagonal \(B\). The identities
\[
\mathbf b(s+B^qv)=\mathbf b(s)+B^{q-1}Mv,\qquad
a^\top M^q e_1\equiv(-1)^q a_1^{q+1}\pmod B
\tag{20}
\]

construct every length-\(H\) word by successive changes \(B^q u_qe_1\). Previously fixed digits remain unchanged and the next digit has an invertible coefficient. The representative \((k,0,\ldots,0)\), \(0\le k<B^H\), is unique in this set and has \(O(H\log B)\) bits. Appendix F.1 proves the construction and records a separate last-tap-coprime route.

For a current digit \(\theta\), Q-offset \(w\), difference \(d\) and action \(j\), let \((b_Q,r_Q)=\operatorname{divmod}(\theta+a^\top w,B)\) and \((b_P,r_P)=\operatorname{divmod}(\theta+a^\top(w+d),B)\). The complete hidden-rounding rows are
\[
\begin{aligned}
q(f)&=(1-r_Q/B)\nu(f+b_Q)+(r_Q/B)\nu(f+b_Q+1),\\
p(f)&=(1-r_P/B)\nu(f+j+b_P)+(r_P/B)\nu(f+j+b_P+1).
\end{aligned}
\tag{21}
\]

After output \(f\), the offset is \((f,w_1,\ldots,w_{m-1})\). A current offset or a closed phase endpoint is normalized by transporting the whole unread word. Starting from an integer displacement \(u_0\), set
\[
(k_i,\phi_i)=\operatorname{divmod}(\theta_i-a^\top u_i,B),\qquad
u_{i+1}=(k_i,u_{i,1},\ldots,u_{i,m-1}).
\tag{22}
\]

This carry yields a common invertible translation of outputs in both worlds and a triangular bijection on compatible suffixes. It preserves the entire conditional experiment (Appendix F.1).

Let \(h(0)=0\) and \(h(d)=m+1-\min\{i:d_i\ne0\}\) otherwise. At remaining budget \(c\),
\[
\mathcal R(c,d)=(m+1)c+h(d),\qquad
\mathcal K(c,d)=mc+h(d).
\tag{23}
\]

Every nonsynchronized informative edge reduces both quantities. At a synchronized root the first row can be made pure, \(p_0(f)=\nu(f+j),q_0(f)=\nu(f)\), by convexity in its phase and complete carry. This row reduces \(\mathcal R\) by one and leaves \(\mathcal K=mc\). The sign and the entire following \(mc\)-digit word are chosen before summing over this first output. Thus there are at most \(D\) informative outputs.

\textbf{Regeneration lemma.} Whenever an action gives \(d_j=0\), its full continuation is
\[
\sum_f G_{c-|j|}(xp(f),yq(f)).
\tag{24}
\]

After each of these outputs both conditional registers agree. Their unused preparation suffix can then be eliminated. While \(d_j\ne0\), the compiler retains the complete transported suffix. In the first-order example with \(\gamma=1/3\), law \((1,2,1)/4\) and \(d=1\), states 0 and 9 have the same echo value \(9293/3456\), but
\[
F_2^+(0;1,2)-F_2^+(9;1,2)=7/576.
\tag{25}
\]

For support width \(\ell=L+3\), every offset lies in a box of radius \(O_{m,C}(\ell)\). Each digit has \(O_{m,C}(\ell^m)\) critical values, obtained from \((-a^\top w)\bmod B\) and the endpoints. Within the resulting cells each fixed policy/\allowbreak{}event score is multiaffine, so a legal vertex cover is exact. Complete zero-budget flush curves and reuse of already consumed prefixes reduce the cost of root calls (Appendix F.2--F.4).

\textbf{Theorem 4 (finite-memory exact point compilation).} For fixed \(m\ge1,C\ge2\), finite rational \(\nu(z)=w_z/S\) on \([-L,L]\), and integer weights \(x,y\ge0\), \(x+y>0\), put \(A=m(m+1)\), \(D=(m+1)C\), \(\ell=L+3\), and
\[
\begin{aligned}
\kappa_{m,C}={}&1+D\lceil\log_2(BS+1)\rceil+\lceil\log_2(x+y+1)\rceil\\
&+(mC+1)\lceil\log_2(B+1)\rceil
+\lceil\log_2((mC+2)(L+3)+2)\rceil.
\end{aligned}
\tag{26}
\]

There is an exact point algorithm for \(G_C(x,y)\) with
\[
\begin{aligned}
\text{bit time}&=O_{m,C}\!\left((\ell^{(A+1)C-m-1}\log(\ell+1)+\log(B+1))\kappa_{m,C}^2\right),\\
\text{space}&=O_{m,C}\!\left(\ell^{\max\{A+m,A(C-2)\}}\kappa_{m,C}\right)\quad\text{bits}.
\end{aligned}
\tag{27}
\]

Rational weights are cleared first. For \(C=1\), an exact full curve has \(O_m(\ell^{A+1})\) lines and construction cost \(O_m(\ell^{A+1}\log(\ell+1))\) arithmetic operations. With \(m=1\), (27) recovers time exponent \(3C-2\) and space exponent \(\max(3,2C-4)\); Theorem 3 gives the sharper C2 constants.

Algorithm 1 summarizes the recursion; F.3 proves its cost. At fixed \(m,C\), the support exponent is 11 for \(m=C=2\) and 22 for \(m=3,C=2\); its optimality remains open. Section 5 relaxes dependency depth for this finite-budget target. Section 6 computes the deadline-normalized leading coefficient.

\par\smallskip\begingroup
\small
\setlength{\tabcolsep}{3pt}
\renewcommand{\arraystretch}{1.10}
\begin{tabularx}{\linewidth}{@{}>{\hsize=1.000000\hsize\linewidth=\hsize\raggedright\arraybackslash}X@{}}
\toprule
\textbf{Algorithm 1. Exact finite-law point compilation} \\
\midrule
\textbf{Input/\allowbreak{}output.} Given \(m,C,a,B,\nu,x,y\) as in Theorem 4, return \(G_C(x,y)\). \\
\textbf{1. Precompute.} Form critical sets (F.4), normalize endpoints by (22), and store complete flush curves (F.6) and \(G_1\). \\
\textbf{2. Start a root.} Return \(\max(x,y)\) if \(xy=0\) or \(c=0\); query \(G_1\) if \(c=1\). Otherwise enumerate signs and complete legal \(mc\)-digit words from (F.5). \\
\textbf{3. Fix preparation, then observe.} Evaluate (F.3), keeping the sign and word fixed across the entire first-output sum before maximization. \\
\textbf{4. Process pending nodes.} Carry the offset and unread suffix by (22). A one-world branch returns \(\max(x,y)\); zero budget uses its flush curve. Otherwise form each complete action row (21). \\
\textbf{5. Continue.} After each output, call \(G_{c-|j|}\) if \(d_j=0\); otherwise recurse with the transported suffix and updated state/\allowbreak{}weights. Sum complete rows before maximizing over actions and stopping. \\
\textbf{6. Cache.} Reuse complete normalized flush and consumed-prefix keys as in F.3, retaining query weights. Ranks (23) give termination. \\
\bottomrule
\end{tabularx}
\endgroup\smallskip\par

\subsection{DG transfer and constructive inverse calibration}

For \(C\ge1\), suppose the total variation (TV) distance satisfies \(\operatorname{TV}(\nu,g_\sigma)\le\tau\le1\) and \(0\le\lambda_-\le e^\varepsilon\le\lambda_+\). Let \(v_*=G_C(1,\lambda_-)-\lambda_-\), using homogeneous rational weights. Coupling the at most \(D=(m+1)C\) informative transitions gives
\[
\begin{aligned}
b_C&=(1+\lambda_+)\{1-(1-\tau)^D\},\\
\Delta_C^*(\varepsilon)&\in
\left[\max\{0,v_*-(\lambda_+-\lambda_-)-b_C\},\;
\min\{1,v_*+b_C\}\right].
\end{aligned}
\tag{28}
\]

The threshold correction uses unit Lipschitzness. Coupling follows the informative preparation game; a fixed physical program charges every release. Appendix F.4 constructs the common source and proves the fixed-parameter polynomial precision bound. Appendix C.5 retains the sharper first-order C2 error allocation.

\textbf{Constructive calibration corollary.} For fixed memory, integer coefficients satisfying (1), rational \(\sigma^2\), every finite \(C\), compact \(I\subset(0,1)\), and positive rational epsilon tolerance, a finite-source construction and a terminating sequence of exact point queries enclose \(\varepsilon_C^*(\delta)\) to that tolerance throughout \(I\).

For \(D_C(\lambda)=\Delta_C^*(\log\lambda)\) and \(1\le x<y\), the moment envelope gives
\[
D_C(y)\le M_C/y,\qquad
D_C(x)-D_C(y)\ge (y-x)D_C(y)^2/M_C.
\tag{29}
\]

Applying \(P(E)^2\le M_CQ(E)\) to near-maximizing events proves the positive secant bound. It makes both endpoint search and source-aware inverse refinement effective. Clipping the inverse at \(\lambda=1\) includes the zero-epsilon boundary; Appendix F.4 proves completion; Appendix K.1 gives the first-order uniform-risk construction.

\textbf{Physical realization.} For each fixed \(C\ge1\), there is a finite \(L_C\), depending on the mechanism and budget, such that for every finite \(T\) and \(\varepsilon\ge0\),
\[
0\le\Delta_C^*(\varepsilon)-\Delta_{C,T}(\varepsilon)
\le\frac{L_C}{T+1}.
\tag{30}
\]

Complete words are implemented by public first-hit preparation; all unfinished branches are retained. The bound is uniform in the testing threshold. Consequently,
\[
\varepsilon_C^*\!\left(\delta+\frac{L_C}{T+1}\right)
\le\varepsilon_{C,T}^*(\delta)\le\varepsilon_C^*(\delta)
\tag{31}
\]

whenever the shifted risk lies in \((0,1)\). This is a risk-axis correction; Appendix A.1 proves it from a uniform expected completion bound.

\section{Commitment depth and calibrated approximation}

The exact compiler retains a preparation until synchronization. A commitment relaxation releases selected parts earlier, producing a larger comparison problem with a smaller dependency description. We use the same finite-law value \(G_C\) and its normalized form \(\mathcal G_C\) from Section 3.1.

\subsection{Fixed cuts and recursive proof decompositions}

A fixed schedule places cuts in the future digit word of each synchronized segment. At a cut, the next whole block and current action are selected before the next output; subsequent carries transport the retained digits together. Full register synchronization ends the segment. Write \(U_{\mathcal S}\) for the resulting comparison value.

\textbf{Theorem 5 (selective commitment).} For all nonnegative weights and nested cut schedules,
\[
G_C\le U_{\mathcal S_0}\le U_{\mathcal S_1}
\quad\text{when }\mathcal S_0\subseteq\mathcal S_1,
\qquad U_{\varnothing}=G_C.
\tag{32}
\]

A finer schedule simulates a coarser one by remembering its complete chosen block and exposing the transported prefixes at successive boundaries. Both world probabilities are preserved. In the first-order implementation, cuts are labeled by remaining budget \(\mathcal S\subseteq\{0,\ldots,C-2\}\), giving
\[
\ell_{\mathcal S}(c)=c-\max\bigl(\{b\in\mathcal S:b<c\}\cup\{-1\}\bigr).
\tag{33}
\]

This labeling and its sparse one-/\allowbreak{}two-digit graph are retained in Appendix E.2.

The recursive hierarchy has synchronized roots, nonsynchronized free nodes and nodes with a committed prefix. At a free node \((c,d)\), a proof group chooses a length \(1\le\ell\le\min\{H,mc+h(d)\}\), then maximizes over all prefixes of that length and all complete actions. A synchronized root allows \(1\le\ell\le\min\{H,mc\}\), with the sign and prefix fixed before its pure first row. After each output, the entire unread prefix is carried; a free node is reached only when that prefix is exhausted. Complete group backups combine as
\[
P_s=\operatorname{Cvx}\!\left(\min_{g\in\mathfrak G_s}
\mathcal B_{s,g}(P_{\rm child})\right),
\qquad
\operatorname{Cvx}(u)(t)=
\inf_{\substack{\alpha_i\ge0,\ \sum_i\alpha_i=1\\
                 \sum_i\alpha_i t_i=t}}
\sum_i\alpha_i u(t_i).
\tag{34}
\]

The root is \(P_H\). Evaluate children before parents: sum complete output rows, maximize over allowed prefixes and actions within each group, minimize over groups, then take the convex minorant. Each group bounds the same convex concrete target, so Jensen's inequality makes this last step sound. The group minimum selects a proof decomposition; preparations and input actions remain maximized inside each group.

\textbf{Theorem 6 (finite-memory commitment hierarchy).} For a declared family \(\mathcal A_H\) of fixed schedules with blocks of length at most \(H\),
\[
\begin{aligned}
\mathcal G_C&\le P_{H+1}\le P_H
\le\operatorname{Cvx}\!\left(\min_{\mathcal S\in\mathcal A_H}U_{\mathcal S}\right)
\le\min_{\mathcal S\in\mathcal A_H}U_{\mathcal S},\\
H\ge\max(1,mC)&\quad\Longrightarrow\quad P_H=\mathcal G_C.
\end{aligned}
\tag{35}
\]

The graph has at most
\[
(C+1)+C(3^m-1)\left(1+\sum_{\ell=1}^{\min(H,mC)}B^\ell\right)
\tag{36}
\]

states, with \(O(L+1)\) output branches per action. This counts semantic states; full-curve aggregation is performed on their rational breakpoints. The proof uses the decreasing ranks (23) and the bijection of compatible completions under carry (Appendix G.1). At full depth, no renewal occurs before genuine synchronization, so the exact preparation game is recovered.

For \(m=2,C=2,B=3\), (36) gives at most 67 states at \(H=1\) and 211 at \(H=2\). Cost also depends on the number and bit length of curve breakpoints and their group aggregation.

Every normalized proof curve has sufficient supporting lines with both coordinates in \([0,1]\). A source change of TV distance \(\tau_\nu\) changes one complete backup by at most \(\tau_\nu\); maximum, minimum and convex minorant are nonexpansive. Consequently the entire hierarchy, including its convexification, satisfies
\[
\|P_H^\mu-P_H^\nu\|_\infty\le D\tau_\nu,
\qquad
|D_H^\mu(\lambda)-D_H^\nu(\lambda)|\le(1+\lambda)D\tau_\nu,
\tag{37}
\]

where \(D=(m+1)C\) and \(D_H(\lambda)=\widehat P_H(1,\lambda)-\lambda\). A fixed-cut game also admits the sharper coupling factor \(1-(1-\tau_\nu)^D\) in (28). Appendix G.2 proves the operator bound for the recursive hierarchy.

\subsection{Memory-dependent strictness and calibrated depth}

\textbf{Theorem 7 (strict finite-memory hierarchies).} Let \(\nu_0=(-1,0,1)\mapsto(1,2,1)/4\). For \(a=(1,1),B=3,C=2\), exact complete root curves give
\[
(\widehat P_1,\widehat P_2,\widehat P_3,\widehat P_4)(1,10)
=\frac{(883740,879994,878918,878641)}{82944}.
\tag{38}
\]

For \(a=(1,1,1),B=4,C=1\), they give
\[
(\widehat P_1,\widehat P_2,\widehat P_3)(1,2)
=\frac{(39669,39485,39301)}{16384}.
\tag{39}
\]

Thus a single participation can require several retained memory digits. The last level in each display is exact. The two-memory strict gaps persist for a specified full-support perturbation of \(\nu_0\). These are complete rational-curve identities, including group minimization and convex closure (Appendix G.3).

For DG, let \(\varepsilon_{P_H}\) be the clipped inverse of \(D_H\). On a risk interval \(I\subset(0,1)\), define
\[
H_{\rm cal}(C,\zeta;I)=
\min\left\{H:\sup_{\delta\in I}
[\varepsilon_{P_H}(\delta)-\varepsilon_C^*(\delta)]\le\zeta\right\}.
\tag{40}
\]

The endpoint \(P_{mC}=\mathcal G_C^{\rm DG}\) ensures existence for \(C\ge1\). Every layer's inverse is constructively computable: the moment envelope and convex supporting coordinates give a positive secant bound analogous to (29), so two interior threshold queries always resolve a refinement step (Appendix G.2).

For \(a=(1,1),B=3,C=1,\sigma^2=1\), certified source transfer and directed inverse queries give
\[
\begin{aligned}
\varepsilon_{P_1}(10^{-5})&\in(5.68479845,5.68479848],\\
\varepsilon_1^*(10^{-5})=\varepsilon_{P_2}(10^{-5})&\in(5.45309338,5.45309341],\\
H_{\rm cal}(1,0.20;[10^{-5},1.1\times10^{-5}])&=2.
\end{aligned}
\tag{41}
\]

The depth-one epsilon loss lies in \((0.23170504,0.23170510)\) at risk \(10^{-5}\), and exceeds 0.20011104 throughout the displayed risk interval. Figure 4 presents this calibration effect alongside the three-memory strict levels. Appendix K.2 retains the first-order DG depth witnesses as specialized comparisons.

\subsection{Numerical precision and structural loss}

For a fixed comparison game, feasible lower supports and upper chords form a convex-function sandwich [22]. The weight \(w(t)=\min(t,1-t)\) controls profile error near the axes: a complete observation row satisfies \(\sum_o\min(tp_o,(1-t)q_o)\le w(t)\). Whole-action maximization preserves this bound.

\textbf{Theorem 8 (weighted refinement at finite informative rank).} Consider a fixed-cut graph with complete probability rows and maximum informative rank \(D=(m+1)C\). At each rank \(h\), fix a lower oracle with integer event coefficients in \([0,(BS)^h]\). For \(0<\xi\le1\), dyadic refinement with exact axes and local weighted tolerance \(\xi\) needs \(O(\xi^{-1/2}\log((BS)^h/\xi))\) points per node. With ordinate scale \(\mathsf Q\), the root weighted width is at most \(D(\xi+1/\mathsf Q)w(t)\), giving profile width at most \(D(\xi+1/\mathsf Q)\) for every \(\lambda\ge1\). Appendix E.3 proves the dimension-independent argument; the certificates in E.5 instantiate \(m=1\).

For recursive \(P_H\), convex closure can leave the event-coefficient lattice required by Theorem 8. We therefore certify its proof functions directly: lower supports are checked against every complete group, and upper Bellman inequalities and rank induction give \(\underline P_H\le P_H\le\overline P_H\). This distinguishes numerical uncertainty from structural relaxation:
\[
\overline P_H-\mathcal G_C
=(\overline P_H-P_H)+(P_H-\mathcal G_C).
\tag{42}
\]

If the exact endpoint is also bracketed, its structural gap obeys
\[
\max\{0,\underline P_H-\overline P_{mC}\}
\le P_H-\mathcal G_C
\le\overline P_H-\underline P_{mC}.
\tag{43}
\]

A positive lower endpoint certifies a gap that persists as arithmetic is refined. Physical calibration obtains its lower bound from event coordinates of a legal strategy. Section 7 uses the first type to diagnose the chosen proof function and the second to bracket the mechanism's privacy cost.

\section{Physical time and participation}

The same mechanism has a different resource problem when the physical deadline is fixed. Use the P-reference register \(s\) and difference memory \(d\) of (4). The P noise register is autonomous; a predictable triangular transcript translation removes its input. Define
\[
\begin{aligned}
\mathsf K_s(n)&=\frac1B\sum_{u=0}^{B-1}
 g_\sigma\!\left(n+\left\lfloor\frac{a^\top s+u}{B}\right\rfloor\right),\\
g(s,d,j)&=\sum_n\mathsf K_s(n)
\log\frac{\mathsf K_s(n)}{\mathsf K_{s-d}(n-j)}.
\end{aligned}
\tag{44}
\]

The next state is \(((n,s_1,\ldots,s_{m-1}),d_j)\). The rounding mixtures are formed before taking the logarithm. With \(R=2/\sigma^2\), DG shift bounds imply \(0\le g\le R\) and \(g(s,0,0)=0\). The chain rule identifies complete-transcript KL with the expected sum of these rewards. One participation can affect the current row and at most \(m\) later rows.

\subsection{The unrestricted participation rate}

Let \(A_C=\sup_{\pi,T}D_{\rm KL}(P_{\pi,T}\Vert Q_{\pi,T})\), over the legal finite programs in (3), and let \(V_c(s,d)\) be the corresponding continuation value. Public first-hit preparation and complete flushing give
\[
V_c(s,0)=A_c,\qquad A_c\le V_c(s,d)\le A_c+mR,
\qquad A_b+A_c\le A_{b+c}\le A_b+A_c+mR.
\tag{45}
\]

For concatenation, flush the first block and return publicly to zero before starting the next. For the upper bound, cut after the \(b\)-th participation; the initial difference memory can change only the next \(m\) rewards. All waits are realized by the original process and truncated at deterministic deadlines.

\textbf{Theorem 9 (participation rate and testing inverse).} For a fixed mechanism (2), the limit \(a_* =\lim_{C\to\infty}A_C/C=\sup_{C\ge1}A_C/C\) exists, and for each fixed \(0<\delta<1\),
\[
0\le Ca_*-A_C\le mR,\qquad
\varepsilon_C^*(\delta)=Ca_*+O_{m,a,B,\sigma,\delta}(\sqrt C).
\tag{46}
\]

The proof compensates likelihood increments by the continuation value (45). Synchronized idle rows contribute zero; at most \((m+1)C\) informative rows contribute to the concentration bound. Near-optimal finite KL policies supply the matching tests. Appendix H.1 gives the two directions. The leading rate also holds along deterministic horizons with \(T_C/C\to\infty\).

\subsection{The hard-resource frontier}

For a participation price \(u\ge0\), define the zero-initial value
\[
J_t(u)=\sup_\pi\mathbb E_P\!\left[
\sum_{i=1}^t g(S_{i-1},d_{i-1},j_i)-u\sum_{i=1}^t|j_i|\right].
\tag{47}
\]

Write \(c_i=|a_i|/B\), \(\chi=(1-r)/m\), \(\alpha=1-\chi\), and
\[
w_i=(m-i+1)\chi+\sum_{k=i}^m c_k,\quad
\|s\|_w=\sum_iw_i|s_i|,\quad W=\sum_iw_i,\quad
\mathsf L=\frac{R}{w_{\min}(1-\alpha)}.
\tag{48}
\]

The common-noise coupling contracts this norm by \(\alpha<1\). Since \(w_1=1\), finite-time Bellman values are \(\mathsf L\)-Lipschitz and have difference-memory span at most \(mR\). Put \(v_0=\sigma^2+1/4\), \({V_\infty}=v_0/(1-\alpha^2)\), and \({b_\psi}=\mathsf L W\sqrt{{V_\infty}}+mR\).

\textbf{Theorem 10 (finite-memory hard-resource law).} The limit \(\psi(u)=\lim_{t\to\infty}J_t(u)/t\) exists with \(|J_t(u)-t\psi(u)|\le{b_\psi}\). For fixed mechanism parameters, \(\tau>0\) and \(\delta\in(0,1)\),
\[
\begin{gathered}
\Phi(\tau)=\min_{0\le u\le(m+1)R}\{u+\tau\psi(u)\}
=\tau\max_{z\in\mathcal Z:\,c(z)\le1/\tau}r(z),\\
\varepsilon_{C,\lfloor\tau C\rfloor}^*(\delta)
=C\Phi(\tau)+O_{m,a,B,\sigma,\tau,\delta}(\sqrt C).
\end{gathered}
\tag{49}
\]

Here \(\mathcal Z\) contains normalized stationary occupations of the original countable state/\allowbreak{}action kernel, \(c(z)=\sum |j|z\), and \(r(z)=\sum gz\). Its noise-register marginal is the common invariant distribution of the autonomous chain. The frontier is nondecreasing and concave, equals \(\tau\psi(0)\) for \(0<\tau\le1\), and tends to \(a_*\).

Multi-memory closed classes can correlate the register and difference memory. For example, with \(m=2\) and \(j_t=1\{K_{t-1}=0\}\), the next state satisfies \(d_1=1\{s_2=0\}\). The proof uses the shared invariant marginal to select at most two fixed optimal closed classes, then enters them through actual noise paths and charged actions. Poisson potentials control the selected policies' participation fluctuations. Suppressing excess actions enforces the hard cap at a KL cost of at most \((m+1)R\) per deleted participation (Appendix H.6).

For the optimal hard-budget KL \(A_{C,T}\), this gives \(C\Phi(\tau)-K_*\sqrt C-B_*\le A_{C,\lfloor\tau C\rfloor}\le C\Phi(\tau)+{b_\psi}\), with finite constants depending on the fixed parameters and chosen classes. A Bellman-compensated likelihood martingale connects these bounds to testing. For \(a_*>0\), finite saturation is characterized by
\[
\Phi(\tau)=a_*
\quad\Longleftrightarrow\quad
\exists z\in\mathcal Z:\ c(z)=1/\tau,\quad r(z)=a_*c(z).
\tag{50}
\]

If \(a_*=0\), the frontier is identically zero.

\subsection{A finite-window value algorithm}

Clamp only the computational register to \([-M,M]^m\), retaining the full-output reward (44). One common normalized rational source approximates the DG law within TV error \(\eta_p\); rational rewards approximate the original hidden-output KL within \(\eta_g\), retaining zero synchronized-idle reward. Write \(\widetilde{\mathcal P}\) for this finite transition kernel.

\textbf{Theorem 11 (finite-memory window precision).} The finite occupation linear program (LP) has value \(\widetilde\Phi_M(\tau)\) satisfying
\[
|\Phi(\tau)-\widetilde\Phi_M(\tau)|
\le\tau\left[
\mathsf L W\frac{2{V_\infty}}{M}e^{-M^2/(2{V_\infty})}
+\eta_g+2M\mathsf L\eta_p\right].
\tag{51}
\]

For fixed \(m,a,B\) satisfying (1), rational \(\sigma^2,\tau>0\), there are constants \(K,d\) and a rational algorithm returning \(v_k\) on every \(k\in\mathbb N\), with
\[
|v_k-\Phi(\tau)|\le2^{-k},\qquad
\operatorname{cost}(k)\le K(k+1)^d.
\tag{52}
\]

Algorithm 2 constructs and solves the bounded rational LP. With \(M=O(\sqrt{k+1})\), there are \(3^{m+1}(2M+1)^m\) variables and a conservative dense encoded size \(O((k+1)^{m+1})\). The constants and polynomial degree depend on the fixed memory and parameters (Appendix H.3).

For \(x=(s,d)\), occupation variables \(z_{x,j}\) obey nonnegativity, total mass one, and
\[
\begin{aligned}
\sum_jz_{y,j}&=\sum_{x,j}z_{x,j}\widetilde{\mathcal P}(x,j;y),\\
\sum_{x,j}|j|z_{x,j}&\le1/\tau,\qquad
\widetilde\Phi_M(\tau)=\max_z\ \tau\sum_{x,j}\widetilde g(x,j)z_{x,j}.
\end{aligned}
\tag{53}
\]

Extending window values by clamped lookup and coupling with the autonomous source proves (51) in both directions. Rational primal and dual witnesses also allow residuals: the verifier charges flow imbalance, excess participation, Bellman residuals and all three error terms in (51).

\par\smallskip\begingroup
\small
\setlength{\tabcolsep}{3pt}
\renewcommand{\arraystretch}{1.10}
\begin{tabularx}{\linewidth}{@{}>{\hsize=1.000000\hsize\linewidth=\hsize\raggedright\arraybackslash}X@{}}
\toprule
\textbf{Algorithm 2. Compute the resource coefficient to precision \(2^{-k}\)} \\
\midrule
\textbf{Input/\allowbreak{}output.} Given fixed parameters as in Theorem 11 and precision \(k\), return \(v_k\) with \(|v_k-\Phi(\tau)|\le2^{-k}\). \\
\textbf{1. Allocate error.} Choose integer \(M\ge1\) and source/\allowbreak{}reward tolerances making each term in (51), including \(\tau\), at most \(2^{-k}/4\), using outward rational comparisons. \\
\textbf{2. Construct coefficients.} Use one normalized rational DG source for all transitions. Evaluate KL after forming full-output rounding mixtures, enclosing omitted signed reward tails. \\
\textbf{3. Form the LP.} Enumerate \(s\in[-M,M]^m,d\in\{-1,0,1\}^m,j\in\{-1,0,1\}\). Clamp next registers, retain full-output rewards, and impose (53) and \(0\le z\le1\). \\
\textbf{4. Return the scaled LP value.} Solve by [50] with error at most \(2^{-k}/4\) after \(\tau\) scaling. The four error payments sum to at most \(2^{-k}\). \\
\bottomrule
\end{tabularx}
\endgroup\smallskip\par

\textbf{Verified first-order implementation.} For \(m=1\), fixed rational \(0<\gamma<1\) and \(\sigma^2,\tau>0\), one verified binary execution constructs the coefficients and LP and realizes (52). Appendix I specifies its charged arithmetic, control, heap, input/\allowbreak{}output and uniform bounded-LP guarantee.

\subsection{From finite certificates to a physical deadline}

The square-root law becomes a numerical finite-budget guarantee when its remainder constants are supplied. The general bridge (H.11b) uses \(Q=v_0(R+\mathsf L)^2\); it follows from a second-moment inequality, without a normal approximation. For \(m=1,0<a_1=a<B\), a finite dual and the sharper variance sum (56) refine these bounds. Fix an integer \(M\ge1\). Use the common rational source and actual-DG reward approximation of Section 6.3, with errors \(\eta_p,\eta_g\). Let \(h(x,d)\) be a rational table on \([-M,M]\times\{-1,0,1\}\). Write \(\ell_h=\max_{x,d}|h(x+1,d)-h(x,d)|\) and \(S_h=\max_j[\max_xh(x,j)-\min_xh(x,j)]\); the first maximum uses adjacent window integers and the second holds the next memory fixed.

\textbf{Proposition 12 (direct finite-deadline certificate).} Suppose \(u,v\ge0\), \(h\ge0\), \(h(0,0)=0\), and every window state, including its endpoints, and every action satisfy
\[
\widetilde g(x,d,j)+\widetilde K_xh(\cdot,j)
\le u|j|+v+h(x,d).
\tag{54}
\]

For \(e_M\ge(2{V_\infty}/M)e^{-M^2/(2{V_\infty})}\), the original full-transcript resource value obeys
\[
A_{C,T}\le A_{\rm dir}:=
uC+T(v+\eta_g+S_h\eta_p)
 +(T-1)_+(R+\gamma\ell_h)e_M.
\tag{55}
\]

Extend by \(h(c_M(k),d)\). Coupling bounds the actual-row residual by \(\eta_g+S_h\eta_p+(R+\gamma\ell_h)(|k|-M)_+\). The autonomous tail bound pays its expectation; telescoping from zero to a nonnegative terminal potential proves (55).

To convert (55) into privacy, put \(L_n=R(1-\gamma^n)/(1-\gamma)\) and, for \(T\ge1\),
\[
V_T=\sigma^2(R/2+L_{T-1})^2
 +\sum_{n=0}^{T-2}(\sigma^2+1/4)(R+L_n)^2.
\tag{56}
\]

The true hidden-mixture log likelihood is \(R\)-Lipschitz in the output. Combining it with the continuation value on that same output bounds the conditional variance by \((\sigma^2+1/4)(R+L_n)^2\). At the original zero start, the sharper first term applies once. The resulting Bellman-compensated martingale has second moment at most \(V_T\); arbitrary legal policies have an additional nonpositive drift. Cantelli's inequality and reversal of the original policy pair therefore give
\[
\varepsilon_{C,T}^*(\delta)
\le A_{\rm dir}+\sqrt{(1-\delta)V_T/\delta}.
\tag{57}
\]

The zero-budget or zero-deadline value is zero. Appendix H.4 proves the hidden-mixture derivative bound, compensation and source transfer. Appendix H.7 gives a sufficient innovation-source class for the first-order resource law.

\section{Evaluation}

The certificates below evaluate exact compilation, calibration and resources, then compare noise designs under explicit static and online workload contracts.

\subsection{First-order computation and finite-deadline attainability}

This subsection specializes to \(m=1,0<a_1=a<B\). It tests the sharp C2 reduction, the recovered fixed-budget compiler and physical preparation.

Figure 3a compares the reduced compiler with exhaustive \(B^2\) roots in one engine. Complete requests include probability construction, both evaluations and DG transfer, with compile/\allowbreak{}setup charged to both. Both use \(\sigma=3,\varepsilon=2,\delta=10^{-4}\), interval width \(2.5\times10^{-6}\), \(L=16,S=10^{24}\), and threshold denominator \(10^{30}\).

\begin{figure}[!htbp]
\centering
\begingroup
\def\nfM#1#2{\pgfpathmoveto{\pgfpoint{#1bp}{#2bp}}}
\def\nfL#1#2{\pgfpathlineto{\pgfpoint{#1bp}{#2bp}}}
\def\nfQ#1#2#3#4{\pgfpathquadraticcurveto{\pgfpoint{#1bp}{#2bp}}{\pgfpoint{#3bp}{#4bp}}}
\def\nfC#1#2#3#4#5#6{\pgfpathcurveto{\pgfpoint{#1bp}{#2bp}}{\pgfpoint{#3bp}{#4bp}}{\pgfpoint{#5bp}{#6bp}}}
\def\nfT#1#2#3#4#5#6{\pgftransformcm{#1}{#2}{#3}{#4}{\pgfpoint{#5bp}{#6bp}}}
\def\nfB{\begin{pgfscope}}
\def\nfE{\end{pgfscope}}
\def\nfZ{\pgfpathclose}
\def\nfFC{\pgfsetfillcolor}
\def\nfFO{\pgfsetfillopacity}
\def\nfSC{\pgfsetstrokecolor}
\def\nfSO{\pgfsetstrokeopacity}
\def\nfW{\pgfsetlinewidth}
\def\nfU{\pgfusepath}
\def\nfNZ{\pgfsetnonzerorule}
\def\nfEO{\pgfseteorule}
\def\nfBC{\pgfsetbuttcap}
\def\nfRC{\pgfsetroundcap}
\def\nfXC{\pgfsetrectcap}
\def\nfRJ{\pgfsetroundjoin}
\def\nfMJ{\pgfsetmiterjoin}
\def\nfBJ{\pgfsetbeveljoin}
\def\nfDA{\pgfsetdash}
\definecolor{nf007f82}{RGB}{0,127,130}
\definecolor{nf202b33}{RGB}{32,43,51}
\definecolor{nf667782}{RGB}{102,119,130}
\definecolor{nfdce3e6}{RGB}{220,227,230}
\definecolor{nfffffff}{RGB}{255,255,255}
\expandafter\def\csname nfpath0\endcsname{\nfM{0}{0}
\nfL{0}{2.6}}
\expandafter\def\csname nfpath1\endcsname{\nfM{269}{1209}
\nfL{831}{1284}
\nfQ{928}{806}{1161}{595}
\nfQ{1394}{384}{1728}{384}
\nfQ{2125}{384}{2398}{659}
\nfQ{2672}{934}{2672}{1341}
\nfQ{2672}{1728}{2419}{1979}
\nfQ{2166}{2231}{1775}{2231}
\nfQ{1616}{2231}{1378}{2169}
\nfL{1441}{2663}
\nfQ{1497}{2656}{1531}{2656}
\nfQ{1891}{2656}{2178}{2843}
\nfQ{2466}{3031}{2466}{3422}
\nfQ{2466}{3731}{2256}{3934}
\nfQ{2047}{4138}{1716}{4138}
\nfQ{1388}{4138}{1169}{3931}
\nfQ{950}{3725}{888}{3313}
\nfL{325}{3413}
\nfQ{428}{3978}{793}{4289}
\nfQ{1159}{4600}{1703}{4600}
\nfQ{2078}{4600}{2393}{4439}
\nfQ{2709}{4278}{2876}{4000}
\nfQ{3044}{3722}{3044}{3409}
\nfQ{3044}{3113}{2884}{2869}
\nfQ{2725}{2625}{2413}{2481}
\nfQ{2819}{2388}{3044}{2092}
\nfQ{3269}{1797}{3269}{1353}
\nfQ{3269}{753}{2831}{336}
\nfQ{2394}{-81}{1725}{-81}
\nfQ{1122}{-81}{723}{278}
\nfQ{325}{638}{269}{1209}
\nfZ}
\expandafter\def\csname nfpath2\endcsname{\nfM{2384}{0}
\nfL{1822}{0}
\nfL{1822}{3584}
\nfQ{1619}{3391}{1289}{3197}
\nfQ{959}{3003}{697}{2906}
\nfL{697}{3450}
\nfQ{1169}{3672}{1522}{3987}
\nfQ{1875}{4303}{2022}{4600}
\nfL{2384}{4600}
\nfL{2384}{0}
\nfZ}
\expandafter\def\csname nfpath3\endcsname{\nfM{350}{1059}
\nfL{891}{1109}
\nfQ{959}{728}{1153}{556}
\nfQ{1347}{384}{1650}{384}
\nfQ{1909}{384}{2104}{503}
\nfQ{2300}{622}{2425}{820}
\nfQ{2550}{1019}{2634}{1356}
\nfQ{2719}{1694}{2719}{2044}
\nfQ{2719}{2081}{2716}{2156}
\nfQ{2547}{1888}{2255}{1720}
\nfQ{1963}{1553}{1622}{1553}
\nfQ{1053}{1553}{659}{1965}
\nfQ{266}{2378}{266}{3053}
\nfQ{266}{3750}{677}{4175}
\nfQ{1088}{4600}{1706}{4600}
\nfQ{2153}{4600}{2523}{4359}
\nfQ{2894}{4119}{3086}{3673}
\nfQ{3278}{3228}{3278}{2384}
\nfQ{3278}{1506}{3087}{986}
\nfQ{2897}{466}{2520}{194}
\nfQ{2144}{-78}{1638}{-78}
\nfQ{1100}{-78}{759}{220}
\nfQ{419}{519}{350}{1059}
\nfZ
\nfM{2653}{3081}
\nfQ{2653}{3566}{2395}{3850}
\nfQ{2138}{4134}{1775}{4134}
\nfQ{1400}{4134}{1122}{3828}
\nfQ{844}{3522}{844}{3034}
\nfQ{844}{2597}{1108}{2323}
\nfQ{1372}{2050}{1759}{2050}
\nfQ{2150}{2050}{2401}{2323}
\nfQ{2653}{2597}{2653}{3081}
\nfZ}
\expandafter\def\csname nfpath4\endcsname{\nfM{303}{3981}
\nfL{303}{4522}
\nfL{3269}{4522}
\nfL{3269}{4084}
\nfQ{2831}{3619}{2401}{2847}
\nfQ{1972}{2075}{1738}{1259}
\nfQ{1569}{684}{1522}{0}
\nfL{944}{0}
\nfQ{953}{541}{1156}{1306}
\nfQ{1359}{2072}{1739}{2783}
\nfQ{2119}{3494}{2547}{3981}
\nfL{303}{3981}
\nfZ}
\expandafter\def\csname nfpath5\endcsname{\nfM{3222}{541}
\nfL{3222}{0}
\nfL{194}{0}
\nfQ{188}{203}{259}{391}
\nfQ{375}{700}{629}{1000}
\nfQ{884}{1300}{1366}{1694}
\nfQ{2113}{2306}{2375}{2664}
\nfQ{2638}{3022}{2638}{3341}
\nfQ{2638}{3675}{2398}{3904}
\nfQ{2159}{4134}{1775}{4134}
\nfQ{1369}{4134}{1125}{3890}
\nfQ{881}{3647}{878}{3216}
\nfL{300}{3275}
\nfQ{359}{3922}{746}{4261}
\nfQ{1134}{4600}{1788}{4600}
\nfQ{2447}{4600}{2831}{4234}
\nfQ{3216}{3869}{3216}{3328}
\nfQ{3216}{3053}{3103}{2787}
\nfQ{2991}{2522}{2730}{2228}
\nfQ{2469}{1934}{1863}{1422}
\nfQ{1356}{997}{1212}{845}
\nfQ{1069}{694}{975}{541}
\nfL{3222}{541}
\nfZ}
\expandafter\def\csname nfpath6\endcsname{\nfM{266}{1200}
\nfL{856}{1250}
\nfQ{922}{819}{1161}{601}
\nfQ{1400}{384}{1738}{384}
\nfQ{2144}{384}{2425}{690}
\nfQ{2706}{997}{2706}{1503}
\nfQ{2706}{1984}{2436}{2262}
\nfQ{2166}{2541}{1728}{2541}
\nfQ{1456}{2541}{1237}{2417}
\nfQ{1019}{2294}{894}{2097}
\nfL{366}{2166}
\nfL{809}{4519}
\nfL{3088}{4519}
\nfL{3088}{3981}
\nfL{1259}{3981}
\nfL{1013}{2750}
\nfQ{1425}{3038}{1878}{3038}
\nfQ{2478}{3038}{2890}{2622}
\nfQ{3303}{2206}{3303}{1553}
\nfQ{3303}{931}{2941}{478}
\nfQ{2500}{-78}{1738}{-78}
\nfQ{1113}{-78}{717}{272}
\nfQ{322}{622}{266}{1200}
\nfZ}
\expandafter\def\csname nfpath7\endcsname{\nfM{266}{2259}
\nfQ{266}{3072}{433}{3567}
\nfQ{600}{4063}{929}{4331}
\nfQ{1259}{4600}{1759}{4600}
\nfQ{2128}{4600}{2406}{4451}
\nfQ{2684}{4303}{2865}{4023}
\nfQ{3047}{3744}{3150}{3342}
\nfQ{3253}{2941}{3253}{2259}
\nfQ{3253}{1453}{3087}{958}
\nfQ{2922}{463}{2592}{192}
\nfQ{2263}{-78}{1759}{-78}
\nfQ{1097}{-78}{719}{397}
\nfQ{266}{969}{266}{2259}
\nfZ
\nfM{844}{2259}
\nfQ{844}{1131}{1108}{757}
\nfQ{1372}{384}{1759}{384}
\nfQ{2147}{384}{2411}{759}
\nfQ{2675}{1134}{2675}{2259}
\nfQ{2675}{3391}{2411}{3762}
\nfQ{2147}{4134}{1753}{4134}
\nfQ{1366}{4134}{1134}{3806}
\nfQ{844}{3388}{844}{2259}
\nfZ}
\expandafter\def\csname nfpath8\endcsname{\nfM{494}{0}
\nfL{494}{4581}
\nfL{2072}{4581}
\nfQ{2606}{4581}{2888}{4516}
\nfQ{3281}{4425}{3559}{4188}
\nfQ{3922}{3881}{4101}{3404}
\nfQ{4281}{2928}{4281}{2316}
\nfQ{4281}{1794}{4159}{1391}
\nfQ{4038}{988}{3847}{723}
\nfQ{3656}{459}{3429}{307}
\nfQ{3203}{156}{2883}{78}
\nfQ{2563}{0}{2147}{0}
\nfL{494}{0}
\nfZ
\nfM{1100}{541}
\nfL{2078}{541}
\nfQ{2531}{541}{2789}{625}
\nfQ{3047}{709}{3200}{863}
\nfQ{3416}{1078}{3536}{1442}
\nfQ{3656}{1806}{3656}{2325}
\nfQ{3656}{3044}{3420}{3430}
\nfQ{3184}{3816}{2847}{3947}
\nfQ{2603}{4041}{2063}{4041}
\nfL{1100}{4041}
\nfL{1100}{541}
\nfZ}
\expandafter\def\csname nfpath9\endcsname{\nfM{2694}{1069}
\nfL{3275}{997}
\nfQ{3138}{488}{2766}{206}
\nfQ{2394}{-75}{1816}{-75}
\nfQ{1088}{-75}{661}{373}
\nfQ{234}{822}{234}{1631}
\nfQ{234}{2469}{665}{2931}
\nfQ{1097}{3394}{1784}{3394}
\nfQ{2450}{3394}{2872}{2941}
\nfQ{3294}{2488}{3294}{1666}
\nfQ{3294}{1616}{3291}{1516}
\nfL{816}{1516}
\nfQ{847}{969}{1125}{678}
\nfQ{1403}{388}{1819}{388}
\nfQ{2128}{388}{2347}{550}
\nfQ{2566}{713}{2694}{1069}
\nfZ
\nfM{847}{1978}
\nfL{2700}{1978}
\nfQ{2663}{2397}{2488}{2606}
\nfQ{2219}{2931}{1791}{2931}
\nfQ{1403}{2931}{1139}{2672}
\nfQ{875}{2413}{847}{1978}
\nfZ}
\expandafter\def\csname nfpath10\endcsname{\nfM{422}{0}
\nfL{422}{3319}
\nfL{928}{3319}
\nfL{928}{2847}
\nfQ{1294}{3394}{1984}{3394}
\nfQ{2284}{3394}{2536}{3286}
\nfQ{2788}{3178}{2913}{3003}
\nfQ{3038}{2828}{3088}{2588}
\nfQ{3119}{2431}{3119}{2041}
\nfL{3119}{0}
\nfL{2556}{0}
\nfL{2556}{2019}
\nfQ{2556}{2363}{2490}{2533}
\nfQ{2425}{2703}{2258}{2804}
\nfQ{2091}{2906}{1866}{2906}
\nfQ{1506}{2906}{1245}{2678}
\nfQ{984}{2450}{984}{1813}
\nfL{984}{0}
\nfL{422}{0}
\nfZ}
\expandafter\def\csname nfpath11\endcsname{\nfM{213}{1659}
\nfQ{213}{2581}{725}{3025}
\nfQ{1153}{3394}{1769}{3394}
\nfQ{2453}{3394}{2887}{2945}
\nfQ{3322}{2497}{3322}{1706}
\nfQ{3322}{1066}{3130}{698}
\nfQ{2938}{331}{2570}{128}
\nfQ{2203}{-75}{1769}{-75}
\nfQ{1072}{-75}{642}{372}
\nfQ{213}{819}{213}{1659}
\nfZ
\nfM{791}{1659}
\nfQ{791}{1022}{1069}{705}
\nfQ{1347}{388}{1769}{388}
\nfQ{2188}{388}{2466}{706}
\nfQ{2744}{1025}{2744}{1678}
\nfQ{2744}{2294}{2464}{2611}
\nfQ{2184}{2928}{1769}{2928}
\nfQ{1347}{2928}{1069}{2612}
\nfQ{791}{2297}{791}{1659}
\nfZ}
\expandafter\def\csname nfpath12\endcsname{\nfM{422}{0}
\nfL{422}{3319}
\nfL{925}{3319}
\nfL{925}{2853}
\nfQ{1081}{3097}{1340}{3245}
\nfQ{1600}{3394}{1931}{3394}
\nfQ{2300}{3394}{2536}{3241}
\nfQ{2772}{3088}{2869}{2813}
\nfQ{3263}{3394}{3894}{3394}
\nfQ{4388}{3394}{4653}{3120}
\nfQ{4919}{2847}{4919}{2278}
\nfL{4919}{0}
\nfL{4359}{0}
\nfL{4359}{2091}
\nfQ{4359}{2428}{4304}{2576}
\nfQ{4250}{2725}{4106}{2815}
\nfQ{3963}{2906}{3769}{2906}
\nfQ{3419}{2906}{3187}{2673}
\nfQ{2956}{2441}{2956}{1928}
\nfL{2956}{0}
\nfL{2394}{0}
\nfL{2394}{2156}
\nfQ{2394}{2531}{2256}{2718}
\nfQ{2119}{2906}{1806}{2906}
\nfQ{1569}{2906}{1367}{2781}
\nfQ{1166}{2656}{1075}{2415}
\nfQ{984}{2175}{984}{1722}
\nfL{984}{0}
\nfL{422}{0}
\nfZ}
\expandafter\def\csname nfpath13\endcsname{\nfM{425}{3934}
\nfL{425}{4581}
\nfL{988}{4581}
\nfL{988}{3934}
\nfL{425}{3934}
\nfZ
\nfM{425}{0}
\nfL{425}{3319}
\nfL{988}{3319}
\nfL{988}{0}
\nfL{425}{0}
\nfZ}
\expandafter\def\csname nfpath14\endcsname{\nfM{2588}{409}
\nfQ{2275}{144}{1986}{34}
\nfQ{1697}{-75}{1366}{-75}
\nfQ{819}{-75}{525}{192}
\nfQ{231}{459}{231}{875}
\nfQ{231}{1119}{342}{1320}
\nfQ{453}{1522}{633}{1644}
\nfQ{813}{1766}{1038}{1828}
\nfQ{1203}{1872}{1538}{1913}
\nfQ{2219}{1994}{2541}{2106}
\nfQ{2544}{2222}{2544}{2253}
\nfQ{2544}{2597}{2384}{2738}
\nfQ{2169}{2928}{1744}{2928}
\nfQ{1347}{2928}{1158}{2789}
\nfQ{969}{2650}{878}{2297}
\nfL{328}{2372}
\nfQ{403}{2725}{575}{2942}
\nfQ{747}{3159}{1072}{3276}
\nfQ{1397}{3394}{1825}{3394}
\nfQ{2250}{3394}{2515}{3294}
\nfQ{2781}{3194}{2906}{3042}
\nfQ{3031}{2891}{3081}{2659}
\nfQ{3109}{2516}{3109}{2141}
\nfL{3109}{1391}
\nfQ{3109}{606}{3145}{398}
\nfQ{3181}{191}{3288}{0}
\nfL{2700}{0}
\nfQ{2613}{175}{2588}{409}
\nfZ
\nfM{2541}{1666}
\nfQ{2234}{1541}{1622}{1453}
\nfQ{1275}{1403}{1131}{1340}
\nfQ{988}{1278}{909}{1158}
\nfQ{831}{1038}{831}{891}
\nfQ{831}{666}{1001}{516}
\nfQ{1172}{366}{1500}{366}
\nfQ{1825}{366}{2078}{508}
\nfQ{2331}{650}{2450}{897}
\nfQ{2541}{1088}{2541}{1459}
\nfL{2541}{1666}
\nfZ}
\expandafter\def\csname nfpath15\endcsname{\nfM{1650}{503}
\nfL{1731}{6}
\nfQ{1494}{-44}{1306}{-44}
\nfQ{1000}{-44}{831}{53}
\nfQ{663}{150}{594}{308}
\nfQ{525}{466}{525}{972}
\nfL{525}{2881}
\nfL{113}{2881}
\nfL{113}{3319}
\nfL{525}{3319}
\nfL{525}{4141}
\nfL{1084}{4478}
\nfL{1084}{3319}
\nfL{1650}{3319}
\nfL{1650}{2881}
\nfL{1084}{2881}
\nfL{1084}{941}
\nfQ{1084}{700}{1114}{631}
\nfQ{1144}{563}{1211}{522}
\nfQ{1278}{481}{1403}{481}
\nfQ{1497}{481}{1650}{503}
\nfZ}
\expandafter\def\csname nfpath16\endcsname{\nfM{416}{0}
\nfL{416}{3319}
\nfL{922}{3319}
\nfL{922}{2816}
\nfQ{1116}{3169}{1280}{3281}
\nfQ{1444}{3394}{1641}{3394}
\nfQ{1925}{3394}{2219}{3213}
\nfL{2025}{2691}
\nfQ{1819}{2813}{1613}{2813}
\nfQ{1428}{2813}{1281}{2702}
\nfQ{1134}{2591}{1072}{2394}
\nfQ{978}{2094}{978}{1738}
\nfL{978}{0}
\nfL{416}{0}
\nfZ}
\expandafter\def\csname nfpath17\endcsname{}
\expandafter\def\csname nfpath18\endcsname{\nfM{1081}{4666}
\nfL{2694}{4666}
\nfQ{3350}{4666}{3675}{4422}
\nfQ{4000}{4178}{4000}{3688}
\nfQ{4000}{3238}{3720}{2911}
\nfQ{3441}{2584}{2988}{2516}
\nfQ{3375}{2428}{3569}{2181}
\nfQ{3763}{1934}{3763}{1522}
\nfQ{3763}{819}{3242}{409}
\nfQ{2722}{0}{1819}{0}
\nfL{172}{0}
\nfL{1081}{4666}
\nfZ
\nfM{1234}{2228}
\nfL{903}{519}
\nfL{1919}{519}
\nfQ{2491}{519}{2800}{781}
\nfQ{3109}{1044}{3109}{1522}
\nfQ{3109}{1891}{2904}{2059}
\nfQ{2700}{2228}{2247}{2228}
\nfL{1234}{2228}
\nfZ
\nfM{1606}{4147}
\nfL{1331}{2741}
\nfL{2272}{2741}
\nfQ{2775}{2741}{3058}{2959}
\nfQ{3341}{3178}{3341}{3566}
\nfQ{3341}{3869}{3150}{4008}
\nfQ{2959}{4147}{2541}{4147}
\nfL{1606}{4147}
\nfZ}
\expandafter\def\csname nfpath19\endcsname{\nfM{0}{0}
\nfL{-2.6}{0}}
\expandafter\def\csname nfpath20\endcsname{\nfM{3763}{1606}
\nfL{4369}{1453}
\nfQ{4178}{706}{3683}{314}
\nfQ{3188}{-78}{2472}{-78}
\nfQ{1731}{-78}{1267}{223}
\nfQ{803}{525}{561}{1097}
\nfQ{319}{1669}{319}{2325}
\nfQ{319}{3041}{592}{3573}
\nfQ{866}{4106}{1370}{4382}
\nfQ{1875}{4659}{2481}{4659}
\nfQ{3169}{4659}{3637}{4309}
\nfQ{4106}{3959}{4291}{3325}
\nfL{3694}{3184}
\nfQ{3534}{3684}{3231}{3912}
\nfQ{2928}{4141}{2469}{4141}
\nfQ{1941}{4141}{1586}{3887}
\nfQ{1231}{3634}{1087}{3207}
\nfQ{944}{2781}{944}{2328}
\nfQ{944}{1744}{1114}{1308}
\nfQ{1284}{872}{1643}{656}
\nfQ{2003}{441}{2422}{441}
\nfQ{2931}{441}{3284}{734}
\nfQ{3638}{1028}{3763}{1606}
\nfZ}
\expandafter\def\csname nfpath21\endcsname{\nfM{422}{-1272}
\nfL{422}{3319}
\nfL{934}{3319}
\nfL{934}{2888}
\nfQ{1116}{3141}{1344}{3267}
\nfQ{1572}{3394}{1897}{3394}
\nfQ{2322}{3394}{2647}{3175}
\nfQ{2972}{2956}{3137}{2557}
\nfQ{3303}{2159}{3303}{1684}
\nfQ{3303}{1175}{3120}{767}
\nfQ{2938}{359}{2589}{142}
\nfQ{2241}{-75}{1856}{-75}
\nfQ{1575}{-75}{1351}{44}
\nfQ{1128}{163}{984}{344}
\nfL{984}{-1272}
\nfL{422}{-1272}
\nfZ
\nfM{931}{1641}
\nfQ{931}{1000}{1190}{694}
\nfQ{1450}{388}{1819}{388}
\nfQ{2194}{388}{2461}{705}
\nfQ{2728}{1022}{2728}{1688}
\nfQ{2728}{2322}{2467}{2637}
\nfQ{2206}{2953}{1844}{2953}
\nfQ{1484}{2953}{1207}{2617}
\nfQ{931}{2281}{931}{1641}
\nfZ}
\expandafter\def\csname nfpath22\endcsname{\nfM{409}{0}
\nfL{409}{4581}
\nfL{972}{4581}
\nfL{972}{0}
\nfL{409}{0}
\nfZ}
\expandafter\def\csname nfpath23\endcsname{\nfM{2588}{1216}
\nfL{3141}{1144}
\nfQ{3050}{572}{2676}{248}
\nfQ{2303}{-75}{1759}{-75}
\nfQ{1078}{-75}{664}{370}
\nfQ{250}{816}{250}{1647}
\nfQ{250}{2184}{428}{2587}
\nfQ{606}{2991}{970}{3192}
\nfQ{1334}{3394}{1763}{3394}
\nfQ{2303}{3394}{2647}{3120}
\nfQ{2991}{2847}{3088}{2344}
\nfL{2541}{2259}
\nfQ{2463}{2594}{2264}{2762}
\nfQ{2066}{2931}{1784}{2931}
\nfQ{1359}{2931}{1093}{2626}
\nfQ{828}{2322}{828}{1663}
\nfQ{828}{994}{1084}{691}
\nfQ{1341}{388}{1753}{388}
\nfQ{2084}{388}{2306}{591}
\nfQ{2528}{794}{2588}{1216}
\nfZ}
\expandafter\def\csname nfpath24\endcsname{\nfM{197}{991}
\nfL{753}{1078}
\nfQ{800}{744}{1014}{566}
\nfQ{1228}{388}{1613}{388}
\nfQ{2000}{388}{2187}{545}
\nfQ{2375}{703}{2375}{916}
\nfQ{2375}{1106}{2209}{1216}
\nfQ{2094}{1291}{1634}{1406}
\nfQ{1016}{1563}{777}{1677}
\nfQ{538}{1791}{414}{1992}
\nfQ{291}{2194}{291}{2438}
\nfQ{291}{2659}{392}{2848}
\nfQ{494}{3038}{669}{3163}
\nfQ{800}{3259}{1026}{3326}
\nfQ{1253}{3394}{1513}{3394}
\nfQ{1903}{3394}{2198}{3281}
\nfQ{2494}{3169}{2634}{2976}
\nfQ{2775}{2784}{2828}{2463}
\nfL{2278}{2388}
\nfQ{2241}{2644}{2061}{2787}
\nfQ{1881}{2931}{1553}{2931}
\nfQ{1166}{2931}{1000}{2803}
\nfQ{834}{2675}{834}{2503}
\nfQ{834}{2394}{903}{2306}
\nfQ{972}{2216}{1119}{2156}
\nfQ{1203}{2125}{1616}{2013}
\nfQ{2213}{1853}{2448}{1751}
\nfQ{2684}{1650}{2818}{1456}
\nfQ{2953}{1263}{2953}{975}
\nfQ{2953}{694}{2789}{445}
\nfQ{2625}{197}{2315}{61}
\nfQ{2006}{-75}{1616}{-75}
\nfQ{969}{-75}{630}{194}
\nfQ{291}{463}{197}{991}
\nfZ}
\expandafter\def\csname nfpath25\endcsname{\nfM{1497}{-1347}
\nfQ{1031}{-759}{709}{28}
\nfQ{388}{816}{388}{1659}
\nfQ{388}{2403}{628}{3084}
\nfQ{909}{3875}{1497}{4659}
\nfL{1900}{4659}
\nfQ{1522}{4009}{1400}{3731}
\nfQ{1209}{3300}{1100}{2831}
\nfQ{966}{2247}{966}{1656}
\nfQ{966}{153}{1900}{-1347}
\nfL{1497}{-1347}
\nfZ}
\expandafter\def\csname nfpath26\endcsname{\nfM{791}{-1347}
\nfL{388}{-1347}
\nfQ{1322}{153}{1322}{1656}
\nfQ{1322}{2244}{1188}{2822}
\nfQ{1081}{3291}{891}{3722}
\nfQ{769}{4003}{388}{4659}
\nfL{791}{4659}
\nfQ{1378}{3875}{1659}{3084}
\nfQ{1900}{2403}{1900}{1659}
\nfQ{1900}{816}{1576}{28}
\nfQ{1253}{-759}{791}{-1347}
\nfZ}
\expandafter\def\csname nfpath27\endcsname{\nfM{0}{1.581139}
\nfC{0.419323}{1.581139}{0.821528}{1.41454}{1.118034}{1.118034}
\nfC{1.41454}{0.821528}{1.581139}{0.419323}{1.581139}{0}
\nfC{1.581139}{-0.419323}{1.41454}{-0.821528}{1.118034}{-1.118034}
\nfC{0.821528}{-1.41454}{0.419323}{-1.581139}{0}{-1.581139}
\nfC{-0.419323}{-1.581139}{-0.821528}{-1.41454}{-1.118034}{-1.118034}
\nfC{-1.41454}{-0.821528}{-1.581139}{-0.419323}{-1.581139}{0}
\nfC{-1.581139}{0.419323}{-1.41454}{0.821528}{-1.118034}{1.118034}
\nfC{-0.821528}{1.41454}{-0.419323}{1.581139}{0}{1.581139}
\nfZ}
\expandafter\def\csname nfpath28\endcsname{\nfM{0}{1.581139}
\nfC{0.419323}{1.581139}{0.821528}{1.41454}{1.118034}{1.118034}
\nfC{1.41454}{0.821528}{1.581139}{0.419323}{1.581139}{0}
\nfC{1.581139}{-0.419323}{1.41454}{-0.821528}{1.118034}{-1.118034}
\nfC{0.821528}{-1.41454}{0.419323}{-1.581139}{0}{-1.581139}
\nfC{-0.419323}{-1.581139}{-0.821528}{-1.41454}{-1.118034}{-1.118034}
\nfC{-1.41454}{-0.821528}{-1.581139}{-0.419323}{-1.581139}{0}
\nfC{-1.581139}{0.419323}{-1.41454}{0.821528}{-1.118034}{1.118034}
\nfC{-0.821528}{1.41454}{-0.419323}{1.581139}{0}{1.581139}
\nfZ}
\expandafter\def\csname nfpath29\endcsname{\nfM{0}{-2.738613}
\nfL{-2.738613}{2.738613}
\nfL{2.738613}{2.738613}
\nfZ}
\expandafter\def\csname nfpath30\endcsname{\nfM{525}{0}
\nfL{525}{4581}
\nfL{3616}{4581}
\nfL{3616}{4041}
\nfL{1131}{4041}
\nfL{1131}{2622}
\nfL{3281}{2622}
\nfL{3281}{2081}
\nfL{1131}{2081}
\nfL{1131}{0}
\nfL{525}{0}
\nfZ}
\expandafter\def\csname nfpath31\endcsname{\nfM{2597}{0}
\nfL{2597}{488}
\nfQ{2209}{-75}{1544}{-75}
\nfQ{1250}{-75}{995}{37}
\nfQ{741}{150}{617}{320}
\nfQ{494}{491}{444}{738}
\nfQ{409}{903}{409}{1263}
\nfL{409}{3319}
\nfL{972}{3319}
\nfL{972}{1478}
\nfQ{972}{1038}{1006}{884}
\nfQ{1059}{663}{1231}{536}
\nfQ{1403}{409}{1656}{409}
\nfQ{1909}{409}{2131}{539}
\nfQ{2353}{669}{2445}{892}
\nfQ{2538}{1116}{2538}{1541}
\nfL{2538}{3319}
\nfL{3100}{3319}
\nfL{3100}{0}
\nfL{2597}{0}
\nfZ}
\expandafter\def\csname nfpath32\endcsname{\nfM{2575}{0}
\nfL{2575}{419}
\nfQ{2259}{-75}{1647}{-75}
\nfQ{1250}{-75}{917}{144}
\nfQ{584}{363}{401}{755}
\nfQ{219}{1147}{219}{1656}
\nfQ{219}{2153}{384}{2558}
\nfQ{550}{2963}{881}{3178}
\nfQ{1213}{3394}{1622}{3394}
\nfQ{1922}{3394}{2156}{3267}
\nfQ{2391}{3141}{2538}{2938}
\nfL{2538}{4581}
\nfL{3097}{4581}
\nfL{3097}{0}
\nfL{2575}{0}
\nfZ
\nfM{797}{1656}
\nfQ{797}{1019}{1065}{703}
\nfQ{1334}{388}{1700}{388}
\nfQ{2069}{388}{2326}{689}
\nfQ{2584}{991}{2584}{1609}
\nfQ{2584}{2291}{2321}{2609}
\nfQ{2059}{2928}{1675}{2928}
\nfQ{1300}{2928}{1048}{2622}
\nfQ{797}{2316}{797}{1656}
\nfZ}
\expandafter\def\csname nfpath33\endcsname{\nfM{288}{1472}
\nfL{859}{1522}
\nfQ{900}{1178}{1048}{958}
\nfQ{1197}{738}{1509}{602}
\nfQ{1822}{466}{2213}{466}
\nfQ{2559}{466}{2825}{569}
\nfQ{3091}{672}{3220}{851}
\nfQ{3350}{1031}{3350}{1244}
\nfQ{3350}{1459}{3225}{1620}
\nfQ{3100}{1781}{2813}{1891}
\nfQ{2628}{1963}{1997}{2114}
\nfQ{1366}{2266}{1113}{2400}
\nfQ{784}{2572}{623}{2826}
\nfQ{463}{3081}{463}{3397}
\nfQ{463}{3744}{659}{4045}
\nfQ{856}{4347}{1234}{4503}
\nfQ{1613}{4659}{2075}{4659}
\nfQ{2584}{4659}{2973}{4495}
\nfQ{3363}{4331}{3572}{4012}
\nfQ{3781}{3694}{3797}{3291}
\nfL{3216}{3247}
\nfQ{3169}{3681}{2898}{3903}
\nfQ{2628}{4125}{2100}{4125}
\nfQ{1550}{4125}{1298}{3923}
\nfQ{1047}{3722}{1047}{3438}
\nfQ{1047}{3191}{1225}{3031}
\nfQ{1400}{2872}{2139}{2705}
\nfQ{2878}{2538}{3153}{2413}
\nfQ{3553}{2228}{3743}{1945}
\nfQ{3934}{1663}{3934}{1294}
\nfQ{3934}{928}{3725}{604}
\nfQ{3516}{281}{3123}{101}
\nfQ{2731}{-78}{2241}{-78}
\nfQ{1619}{-78}{1198}{103}
\nfQ{778}{284}{539}{648}
\nfQ{300}{1013}{288}{1472}
\nfZ}
\expandafter\def\csname nfpath34\endcsname{\nfM{2069}{0}
\nfL{2069}{1097}
\nfL{81}{1097}
\nfL{81}{1613}
\nfL{2172}{4581}
\nfL{2631}{4581}
\nfL{2631}{1613}
\nfL{3250}{1613}
\nfL{3250}{1097}
\nfL{2631}{1097}
\nfL{2631}{0}
\nfL{2069}{0}
\nfZ
\nfM{2069}{1613}
\nfL{2069}{3678}
\nfL{634}{1613}
\nfL{2069}{1613}
\nfZ}
\expandafter\def\csname nfpath35\endcsname{\nfM{3184}{3459}
\nfL{2625}{3416}
\nfQ{2550}{3747}{2413}{3897}
\nfQ{2184}{4138}{1850}{4138}
\nfQ{1581}{4138}{1378}{3988}
\nfQ{1113}{3794}{959}{3422}
\nfQ{806}{3050}{800}{2363}
\nfQ{1003}{2672}{1297}{2822}
\nfQ{1591}{2972}{1913}{2972}
\nfQ{2475}{2972}{2870}{2558}
\nfQ{3266}{2144}{3266}{1488}
\nfQ{3266}{1056}{3080}{686}
\nfQ{2894}{316}{2569}{119}
\nfQ{2244}{-78}{1831}{-78}
\nfQ{1128}{-78}{684}{439}
\nfQ{241}{956}{241}{2144}
\nfQ{241}{3472}{731}{4075}
\nfQ{1159}{4600}{1884}{4600}
\nfQ{2425}{4600}{2770}{4297}
\nfQ{3116}{3994}{3184}{3459}
\nfZ
\nfM{888}{1484}
\nfQ{888}{1194}{1011}{928}
\nfQ{1134}{663}{1356}{523}
\nfQ{1578}{384}{1822}{384}
\nfQ{2178}{384}{2434}{671}
\nfQ{2691}{959}{2691}{1453}
\nfQ{2691}{1928}{2437}{2201}
\nfQ{2184}{2475}{1800}{2475}
\nfQ{1419}{2475}{1153}{2201}
\nfQ{888}{1928}{888}{1484}
\nfZ}
\expandafter\def\csname nfpath36\endcsname{\nfM{1131}{2484}
\nfQ{781}{2613}{612}{2850}
\nfQ{444}{3088}{444}{3419}
\nfQ{444}{3919}{803}{4259}
\nfQ{1163}{4600}{1759}{4600}
\nfQ{2359}{4600}{2725}{4251}
\nfQ{3091}{3903}{3091}{3403}
\nfQ{3091}{3084}{2923}{2848}
\nfQ{2756}{2613}{2416}{2484}
\nfQ{2838}{2347}{3058}{2040}
\nfQ{3278}{1734}{3278}{1309}
\nfQ{3278}{722}{2862}{322}
\nfQ{2447}{-78}{1769}{-78}
\nfQ{1091}{-78}{675}{323}
\nfQ{259}{725}{259}{1325}
\nfQ{259}{1772}{486}{2073}
\nfQ{713}{2375}{1131}{2484}
\nfZ
\nfM{1019}{3438}
\nfQ{1019}{3113}{1228}{2906}
\nfQ{1438}{2700}{1772}{2700}
\nfQ{2097}{2700}{2305}{2904}
\nfQ{2513}{3109}{2513}{3406}
\nfQ{2513}{3716}{2298}{3927}
\nfQ{2084}{4138}{1766}{4138}
\nfQ{1444}{4138}{1231}{3931}
\nfQ{1019}{3725}{1019}{3438}
\nfZ
\nfM{838}{1322}
\nfQ{838}{1081}{952}{856}
\nfQ{1066}{631}{1291}{507}
\nfQ{1516}{384}{1775}{384}
\nfQ{2178}{384}{2440}{643}
\nfQ{2703}{903}{2703}{1303}
\nfQ{2703}{1709}{2433}{1975}
\nfQ{2163}{2241}{1756}{2241}
\nfQ{1359}{2241}{1098}{1978}
\nfQ{838}{1716}{838}{1322}
\nfZ}
\expandafter\def\csname nfpath37\endcsname{\nfM{494}{0}
\nfL{494}{4581}
\nfL{2222}{4581}
\nfQ{2678}{4581}{2919}{4538}
\nfQ{3256}{4481}{3484}{4323}
\nfQ{3713}{4166}{3852}{3881}
\nfQ{3991}{3597}{3991}{3256}
\nfQ{3991}{2672}{3619}{2267}
\nfQ{3247}{1863}{2275}{1863}
\nfL{1100}{1863}
\nfL{1100}{0}
\nfL{494}{0}
\nfZ
\nfM{1100}{2403}
\nfL{2284}{2403}
\nfQ{2872}{2403}{3119}{2622}
\nfQ{3366}{2841}{3366}{3238}
\nfQ{3366}{3525}{3220}{3729}
\nfQ{3075}{3934}{2838}{4000}
\nfQ{2684}{4041}{2272}{4041}
\nfL{1100}{4041}
\nfL{1100}{2403}
\nfZ}
\expandafter\def\csname nfpath38\endcsname{\nfM{422}{0}
\nfL{422}{4581}
\nfL{984}{4581}
\nfL{984}{2938}
\nfQ{1378}{3394}{1978}{3394}
\nfQ{2347}{3394}{2619}{3248}
\nfQ{2891}{3103}{3008}{2847}
\nfQ{3125}{2591}{3125}{2103}
\nfL{3125}{0}
\nfL{2563}{0}
\nfL{2563}{2103}
\nfQ{2563}{2525}{2380}{2717}
\nfQ{2197}{2909}{1863}{2909}
\nfQ{1613}{2909}{1392}{2779}
\nfQ{1172}{2650}{1078}{2428}
\nfQ{984}{2206}{984}{1816}
\nfL{984}{0}
\nfL{422}{0}
\nfZ}
\expandafter\def\csname nfpath39\endcsname{\nfM{397}{-1278}
\nfL{334}{-750}
\nfQ{519}{-800}{656}{-800}
\nfQ{844}{-800}{956}{-737}
\nfQ{1069}{-675}{1141}{-563}
\nfQ{1194}{-478}{1313}{-144}
\nfQ{1328}{-97}{1363}{-6}
\nfL{103}{3319}
\nfL{709}{3319}
\nfL{1400}{1397}
\nfQ{1534}{1031}{1641}{628}
\nfQ{1738}{1016}{1872}{1384}
\nfL{2581}{3319}
\nfL{3144}{3319}
\nfL{1881}{-56}
\nfQ{1678}{-603}{1566}{-809}
\nfQ{1416}{-1088}{1222}{-1217}
\nfQ{1028}{-1347}{759}{-1347}
\nfQ{597}{-1347}{397}{-1278}
\nfZ}
\expandafter\def\csname nfpath40\endcsname{\nfM{378}{4666}
\nfL{4325}{4666}
\nfL{4225}{4134}
\nfL{2559}{4134}
\nfL{1759}{0}
\nfL{1125}{0}
\nfL{1925}{4134}
\nfL{275}{4134}
\nfL{378}{4666}
\nfZ}
\expandafter\def\csname nfpath41\endcsname{\nfM{3381}{1997}
\nfL{356}{1997}
\nfL{356}{2522}
\nfL{3381}{2522}
\nfL{3381}{1997}
\nfZ}
\expandafter\def\csname nfpath42\endcsname{\nfM{556}{0}
\nfL{556}{2881}
\nfL{59}{2881}
\nfL{59}{3319}
\nfL{556}{3319}
\nfL{556}{3672}
\nfQ{556}{4006}{616}{4169}
\nfQ{697}{4388}{901}{4523}
\nfQ{1106}{4659}{1475}{4659}
\nfQ{1713}{4659}{2000}{4603}
\nfL{1916}{4113}
\nfQ{1741}{4144}{1584}{4144}
\nfQ{1328}{4144}{1222}{4034}
\nfQ{1116}{3925}{1116}{3625}
\nfL{1116}{3319}
\nfL{1763}{3319}
\nfL{1763}{2881}
\nfL{1116}{2881}
\nfL{1116}{0}
\nfL{556}{0}
\nfZ}
\expandafter\def\csname nfpath43\endcsname{\nfM{794}{531}
\nfL{1825}{531}
\nfL{1825}{4091}
\nfL{703}{3866}
\nfL{703}{4441}
\nfL{1819}{4666}
\nfL{2450}{4666}
\nfL{2450}{531}
\nfL{3481}{531}
\nfL{3481}{0}
\nfL{794}{0}
\nfL{794}{531}
\nfZ}
\expandafter\def\csname nfpath44\endcsname{\nfM{678}{2272}
\nfL{4684}{2272}
\nfL{4684}{1741}
\nfL{678}{1741}
\nfL{678}{2272}
\nfZ}
\expandafter\def\csname nfpath45\endcsname{\nfM{1075}{4666}
\nfL{1709}{4666}
\nfL{909}{525}
\nfL{3181}{525}
\nfL{3078}{0}
\nfL{172}{0}
\nfL{1075}{4666}
\nfZ}
\expandafter\def\csname nfpath46\endcsname{\nfM{1625}{4666}
\nfL{2156}{4666}
\nfL{531}{-594}
\nfL{0}{-594}
\nfL{1625}{4666}
\nfZ}
\expandafter\def\csname nfpath47\endcsname{\nfM{991}{4666}
\nfL{1625}{4666}
\nfL{1075}{1831}
\nfQ{1041}{1641}{1027}{1517}
\nfQ{1013}{1394}{1013}{1300}
\nfQ{1013}{869}{1253}{645}
\nfQ{1494}{422}{1959}{422}
\nfQ{2563}{422}{2898}{753}
\nfQ{3234}{1084}{3378}{1831}
\nfL{3928}{4666}
\nfL{4563}{4666}
\nfL{4000}{1753}
\nfQ{3816}{809}{3300}{359}
\nfQ{2784}{-91}{1888}{-91}
\nfQ{1188}{-91}{780}{261}
\nfQ{372}{613}{372}{1216}
\nfQ{372}{1325}{387}{1461}
\nfQ{403}{1597}{434}{1753}
\nfL{991}{4666}
\nfZ}
\expandafter\def\csname nfpath48\endcsname{\nfM{0}{1.732051}
\nfC{0.459345}{1.732051}{0.899939}{1.549551}{1.224745}{1.224745}
\nfC{1.549551}{0.899939}{1.732051}{0.459345}{1.732051}{0}
\nfC{1.732051}{-0.459345}{1.549551}{-0.899939}{1.224745}{-1.224745}
\nfC{0.899939}{-1.549551}{0.459345}{-1.732051}{0}{-1.732051}
\nfC{-0.459345}{-1.732051}{-0.899939}{-1.549551}{-1.224745}{-1.224745}
\nfC{-1.549551}{-0.899939}{-1.732051}{-0.459345}{-1.732051}{0}
\nfC{-1.732051}{0.459345}{-1.549551}{0.899939}{-1.224745}{1.224745}
\nfC{-0.899939}{1.549551}{-0.459345}{1.732051}{0}{1.732051}
\nfZ}
\expandafter\def\csname nfpath49\endcsname{\nfM{0}{2.872281}
\nfC{0.761738}{2.872281}{1.49238}{2.56964}{2.03101}{2.03101}
\nfC{2.56964}{1.49238}{2.872281}{0.761738}{2.872281}{0}
\nfC{2.872281}{-0.761738}{2.56964}{-1.49238}{2.03101}{-2.03101}
\nfC{1.49238}{-2.56964}{0.761738}{-2.872281}{0}{-2.872281}
\nfC{-0.761738}{-2.872281}{-1.49238}{-2.56964}{-2.03101}{-2.03101}
\nfC{-2.56964}{-1.49238}{-2.872281}{-0.761738}{-2.872281}{0}
\nfC{-2.872281}{0.761738}{-2.56964}{1.49238}{-2.03101}{2.03101}
\nfC{-1.49238}{2.56964}{-0.761738}{2.872281}{0}{2.872281}
\nfZ}
\expandafter\def\csname nfpath50\endcsname{\nfM{1116}{2306}
\nfL{319}{2450}
\nfQ{453}{2931}{781}{3162}
\nfQ{1109}{3394}{1756}{3394}
\nfQ{2344}{3394}{2631}{3255}
\nfQ{2919}{3116}{3036}{2902}
\nfQ{3153}{2688}{3153}{2116}
\nfL{3144}{1091}
\nfQ{3144}{653}{3186}{445}
\nfQ{3228}{238}{3344}{0}
\nfL{2475}{0}
\nfQ{2441}{88}{2391}{259}
\nfQ{2369}{338}{2359}{363}
\nfQ{2134}{144}{1878}{34}
\nfQ{1622}{-75}{1331}{-75}
\nfQ{819}{-75}{523}{203}
\nfQ{228}{481}{228}{906}
\nfQ{228}{1188}{362}{1408}
\nfQ{497}{1628}{739}{1745}
\nfQ{981}{1863}{1438}{1950}
\nfQ{2053}{2066}{2291}{2166}
\nfL{2291}{2253}
\nfQ{2291}{2506}{2166}{2614}
\nfQ{2041}{2722}{1694}{2722}
\nfQ{1459}{2722}{1328}{2630}
\nfQ{1197}{2538}{1116}{2306}
\nfZ
\nfM{2291}{1594}
\nfQ{2122}{1538}{1756}{1459}
\nfQ{1391}{1381}{1278}{1306}
\nfQ{1106}{1184}{1106}{997}
\nfQ{1106}{813}{1243}{678}
\nfQ{1381}{544}{1594}{544}
\nfQ{1831}{544}{2047}{700}
\nfQ{2206}{819}{2256}{991}
\nfQ{2291}{1103}{2291}{1419}
\nfL{2291}{1594}
\nfZ}
\expandafter\def\csname nfpath51\endcsname{\nfM{453}{0}
\nfL{453}{4581}
\nfL{1838}{4581}
\nfL{2669}{1456}
\nfL{3491}{4581}
\nfL{4878}{4581}
\nfL{4878}{0}
\nfL{4019}{0}
\nfL{4019}{3606}
\nfL{3109}{0}
\nfL{2219}{0}
\nfL{1313}{3606}
\nfL{1313}{0}
\nfL{453}{0}
\nfZ}
\expandafter\def\csname nfpath52\endcsname{\nfM{1981}{3319}
\nfL{1981}{2619}
\nfL{1381}{2619}
\nfL{1381}{1281}
\nfQ{1381}{875}{1398}{808}
\nfQ{1416}{741}{1477}{697}
\nfQ{1538}{653}{1625}{653}
\nfQ{1747}{653}{1978}{738}
\nfL{2053}{56}
\nfQ{1747}{-75}{1359}{-75}
\nfQ{1122}{-75}{931}{4}
\nfQ{741}{84}{652}{211}
\nfQ{563}{338}{528}{553}
\nfQ{500}{706}{500}{1172}
\nfL{500}{2619}
\nfL{97}{2619}
\nfL{97}{3319}
\nfL{500}{3319}
\nfL{500}{3978}
\nfL{1381}{4491}
\nfL{1381}{3319}
\nfL{1981}{3319}
\nfZ}
\expandafter\def\csname nfpath53\endcsname{\nfM{3353}{2338}
\nfL{2488}{2181}
\nfQ{2444}{2441}{2289}{2572}
\nfQ{2134}{2703}{1888}{2703}
\nfQ{1559}{2703}{1364}{2476}
\nfQ{1169}{2250}{1169}{1719}
\nfQ{1169}{1128}{1367}{884}
\nfQ{1566}{641}{1900}{641}
\nfQ{2150}{641}{2309}{783}
\nfQ{2469}{925}{2534}{1272}
\nfL{3397}{1125}
\nfQ{3263}{531}{2881}{228}
\nfQ{2500}{-75}{1859}{-75}
\nfQ{1131}{-75}{698}{384}
\nfQ{266}{844}{266}{1656}
\nfQ{266}{2478}{700}{2936}
\nfQ{1134}{3394}{1875}{3394}
\nfQ{2481}{3394}{2839}{3133}
\nfQ{3197}{2872}{3353}{2338}
\nfZ}
\expandafter\def\csname nfpath54\endcsname{\nfM{1334}{4581}
\nfL{1334}{2897}
\nfQ{1759}{3394}{2350}{3394}
\nfQ{2653}{3394}{2897}{3281}
\nfQ{3141}{3169}{3264}{2994}
\nfQ{3388}{2819}{3433}{2606}
\nfQ{3478}{2394}{3478}{1947}
\nfL{3478}{0}
\nfL{2600}{0}
\nfL{2600}{1753}
\nfQ{2600}{2275}{2550}{2415}
\nfQ{2500}{2556}{2373}{2639}
\nfQ{2247}{2722}{2056}{2722}
\nfQ{1838}{2722}{1666}{2615}
\nfQ{1494}{2509}{1414}{2295}
\nfQ{1334}{2081}{1334}{1663}
\nfL{1334}{0}
\nfL{456}{0}
\nfL{456}{4581}
\nfL{1334}{4581}
\nfZ}
\expandafter\def\csname nfpath55\endcsname{\nfM{2381}{1056}
\nfL{3256}{909}
\nfQ{3088}{428}{2723}{176}
\nfQ{2359}{-75}{1813}{-75}
\nfQ{947}{-75}{531}{491}
\nfQ{203}{944}{203}{1634}
\nfQ{203}{2459}{634}{2926}
\nfQ{1066}{3394}{1725}{3394}
\nfQ{2466}{3394}{2894}{2905}
\nfQ{3322}{2416}{3303}{1406}
\nfL{1103}{1406}
\nfQ{1113}{1016}{1316}{798}
\nfQ{1519}{581}{1822}{581}
\nfQ{2028}{581}{2168}{693}
\nfQ{2309}{806}{2381}{1056}
\nfZ
\nfM{2431}{1944}
\nfQ{2422}{2325}{2234}{2523}
\nfQ{2047}{2722}{1778}{2722}
\nfQ{1491}{2722}{1303}{2513}
\nfQ{1116}{2303}{1119}{1944}
\nfL{2431}{1944}
\nfZ}
\expandafter\def\csname nfpath56\endcsname{\nfM{3503}{0}
\nfL{2688}{0}
\nfL{2688}{488}
\nfQ{2484}{203}{2207}{64}
\nfQ{1931}{-75}{1650}{-75}
\nfQ{1078}{-75}{670}{386}
\nfQ{263}{847}{263}{1672}
\nfQ{263}{2516}{659}{2955}
\nfQ{1056}{3394}{1663}{3394}
\nfQ{2219}{3394}{2625}{2931}
\nfL{2625}{4581}
\nfL{3503}{4581}
\nfL{3503}{0}
\nfZ
\nfM{1159}{1731}
\nfQ{1159}{1200}{1306}{963}
\nfQ{1519}{619}{1900}{619}
\nfQ{2203}{619}{2415}{876}
\nfQ{2628}{1134}{2628}{1647}
\nfQ{2628}{2219}{2422}{2470}
\nfQ{2216}{2722}{1894}{2722}
\nfQ{1581}{2722}{1370}{2473}
\nfQ{1159}{2225}{1159}{1731}
\nfZ}
\expandafter\def\csname nfpath57\endcsname{}
\expandafter\def\csname nfpath58\endcsname{\nfM{1300}{0}
\nfL{422}{0}
\nfL{422}{3319}
\nfL{1238}{3319}
\nfL{1238}{2847}
\nfQ{1447}{3181}{1614}{3287}
\nfQ{1781}{3394}{1994}{3394}
\nfQ{2294}{3394}{2572}{3228}
\nfL{2300}{2463}
\nfQ{2078}{2606}{1888}{2606}
\nfQ{1703}{2606}{1575}{2504}
\nfQ{1447}{2403}{1373}{2137}
\nfQ{1300}{1872}{1300}{1025}
\nfL{1300}{0}
\nfZ}
\expandafter\def\csname nfpath59\endcsname{\nfM{459}{3769}
\nfL{459}{4581}
\nfL{1338}{4581}
\nfL{1338}{3769}
\nfL{459}{3769}
\nfZ
\nfM{459}{0}
\nfL{459}{3319}
\nfL{1338}{3319}
\nfL{1338}{0}
\nfL{459}{0}
\nfZ}
\expandafter\def\csname nfpath60\endcsname{\nfM{75}{3319}
\nfL{563}{3319}
\nfL{563}{3569}
\nfQ{563}{3988}{652}{4194}
\nfQ{741}{4400}{980}{4529}
\nfQ{1219}{4659}{1584}{4659}
\nfQ{1959}{4659}{2319}{4547}
\nfL{2200}{3934}
\nfQ{1991}{3984}{1797}{3984}
\nfQ{1606}{3984}{1523}{3895}
\nfQ{1441}{3806}{1441}{3553}
\nfL{1441}{3319}
\nfL{2097}{3319}
\nfL{2097}{2628}
\nfL{1441}{2628}
\nfL{1441}{0}
\nfL{563}{0}
\nfL{563}{2628}
\nfL{75}{2628}
\nfL{75}{3319}
\nfZ}
\expandafter\def\csname nfpath61\endcsname{\nfM{256}{1706}
\nfQ{256}{2144}{472}{2553}
\nfQ{688}{2963}{1083}{3178}
\nfQ{1478}{3394}{1966}{3394}
\nfQ{2719}{3394}{3200}{2905}
\nfQ{3681}{2416}{3681}{1669}
\nfQ{3681}{916}{3195}{420}
\nfQ{2709}{-75}{1972}{-75}
\nfQ{1516}{-75}{1102}{131}
\nfQ{688}{338}{472}{736}
\nfQ{256}{1134}{256}{1706}
\nfZ
\nfM{1156}{1659}
\nfQ{1156}{1166}{1390}{903}
\nfQ{1625}{641}{1969}{641}
\nfQ{2313}{641}{2545}{903}
\nfQ{2778}{1166}{2778}{1666}
\nfQ{2778}{2153}{2545}{2415}
\nfQ{2313}{2678}{1969}{2678}
\nfQ{1625}{2678}{1390}{2415}
\nfQ{1156}{2153}{1156}{1659}
\nfZ}
\expandafter\def\csname nfpath62\endcsname{\nfM{3478}{0}
\nfL{2600}{0}
\nfL{2600}{1694}
\nfQ{2600}{2231}{2544}{2389}
\nfQ{2488}{2547}{2361}{2634}
\nfQ{2234}{2722}{2056}{2722}
\nfQ{1828}{2722}{1647}{2597}
\nfQ{1466}{2472}{1398}{2265}
\nfQ{1331}{2059}{1331}{1503}
\nfL{1331}{0}
\nfL{453}{0}
\nfL{453}{3319}
\nfL{1269}{3319}
\nfL{1269}{2831}
\nfQ{1703}{3394}{2363}{3394}
\nfQ{2653}{3394}{2893}{3289}
\nfQ{3134}{3184}{3257}{3021}
\nfQ{3381}{2859}{3429}{2653}
\nfQ{3478}{2447}{3478}{2063}
\nfL{3478}{0}
\nfZ}
\expandafter\def\csname nfpath63\endcsname{\nfM{150}{947}
\nfL{1031}{1081}
\nfQ{1088}{825}{1259}{692}
\nfQ{1431}{559}{1741}{559}
\nfQ{2081}{559}{2253}{684}
\nfQ{2369}{772}{2369}{919}
\nfQ{2369}{1019}{2306}{1084}
\nfQ{2241}{1147}{2013}{1200}
\nfQ{950}{1434}{666}{1628}
\nfQ{272}{1897}{272}{2375}
\nfQ{272}{2806}{612}{3100}
\nfQ{953}{3394}{1669}{3394}
\nfQ{2350}{3394}{2681}{3172}
\nfQ{3013}{2950}{3138}{2516}
\nfL{2309}{2363}
\nfQ{2256}{2556}{2107}{2659}
\nfQ{1959}{2763}{1684}{2763}
\nfQ{1338}{2763}{1188}{2666}
\nfQ{1088}{2597}{1088}{2488}
\nfQ{1088}{2394}{1175}{2328}
\nfQ{1294}{2241}{1995}{2081}
\nfQ{2697}{1922}{2975}{1691}
\nfQ{3250}{1456}{3250}{1038}
\nfQ{3250}{581}{2869}{253}
\nfQ{2488}{-75}{1741}{-75}
\nfQ{1063}{-75}{667}{200}
\nfQ{272}{475}{150}{947}
\nfZ}
\expandafter\def\csname nfpath64\endcsname{\nfM{422}{0}
\nfL{422}{4581}
\nfL{1300}{4581}
\nfL{1300}{2931}
\nfQ{1706}{3394}{2263}{3394}
\nfQ{2869}{3394}{3266}{2955}
\nfQ{3663}{2516}{3663}{1694}
\nfQ{3663}{844}{3258}{384}
\nfQ{2853}{-75}{2275}{-75}
\nfQ{1991}{-75}{1714}{67}
\nfQ{1438}{209}{1238}{488}
\nfL{1238}{0}
\nfL{422}{0}
\nfZ
\nfM{1294}{1731}
\nfQ{1294}{1216}{1456}{969}
\nfQ{1684}{619}{2063}{619}
\nfQ{2353}{619}{2558}{867}
\nfQ{2763}{1116}{2763}{1650}
\nfQ{2763}{2219}{2556}{2470}
\nfQ{2350}{2722}{2028}{2722}
\nfQ{1713}{2722}{1503}{2476}
\nfQ{1294}{2231}{1294}{1731}
\nfZ}
\expandafter\def\csname nfpath65\endcsname{\nfM{472}{0}
\nfL{472}{4581}
\nfL{3613}{4581}
\nfL{3613}{3806}
\nfL{1397}{3806}
\nfL{1397}{2722}
\nfL{3309}{2722}
\nfL{3309}{1947}
\nfL{1397}{1947}
\nfL{1397}{0}
\nfL{472}{0}
\nfZ}
\expandafter\def\csname nfpath66\endcsname{\nfM{359}{1222}
\nfL{359}{2100}
\nfL{2084}{2100}
\nfL{2084}{1222}
\nfL{359}{1222}
\nfZ}
\expandafter\def\csname nfpath67\endcsname{\nfM{459}{0}
\nfL{459}{4581}
\nfL{1338}{4581}
\nfL{1338}{0}
\nfL{459}{0}
\nfZ}
\expandafter\def\csname nfpath68\endcsname{\nfM{1078}{0}
\nfL{28}{3319}
\nfL{881}{3319}
\nfL{1503}{1144}
\nfL{2075}{3319}
\nfL{2922}{3319}
\nfL{3475}{1144}
\nfL{4109}{3319}
\nfL{4975}{3319}
\nfL{3909}{0}
\nfL{3066}{0}
\nfL{2494}{2134}
\nfL{1931}{0}
\nfL{1078}{0}
\nfZ}
\def\nfGA#1#2#3{\nfB
\nfFC{nf202b33}
\nfFO{1}
\nfSC{nf202b33}
\nfSO{1}
\nfW{0.6bp}
\nfBC
\nfRJ
\nfNZ
\pgftransformcm{1}{0}{0}{-1}{\pgfpoint{#1bp}{#2bp}}
\csname nfpath#3\endcsname
\nfU{fill,stroke}
\nfE}
\def\nfGB#1#2#3{\nfB
\nfFC{nf202b33}
\nfFO{1}
\nfNZ
\pgftransformcm{0.00109375}{0}{0}{0.00109375}{\pgfpoint{#1bp}{#2bp}}
\csname nfpath#3\endcsname
\nfU{fill}
\nfE}
\def\nfGC#1#2#3{\nfB
\nfFC{nf202b33}
\nfFO{1}
\nfNZ
\pgftransformcm{0.00128125}{0}{0}{0.00128125}{\pgfpoint{#1bp}{#2bp}}
\csname nfpath#3\endcsname
\nfU{fill}
\nfE}
\def\nfGD#1#2#3{\nfB
\nfFC{nf202b33}
\nfFO{1}
\nfNZ
\pgftransformcm{0}{0.00128125}{-0.00128125}{0}{\pgfpoint{#1bp}{#2bp}}
\csname nfpath#3\endcsname
\nfU{fill}
\nfE}
\def\nfGE#1#2#3{\nfB
\nfFC{nf667782}
\nfFO{1}
\nfNZ
\pgftransformcm{1}{0}{0}{-1}{\pgfpoint{#1bp}{#2bp}}
\csname nfpath#3\endcsname
\nfU{fill}
\nfE}
\def\nfGF#1#2#3{\nfB
\nfFC{nf007f82}
\nfFO{1}
\nfNZ
\pgftransformcm{1}{0}{0}{-1}{\pgfpoint{#1bp}{#2bp}}
\csname nfpath#3\endcsname
\nfU{fill}
\nfE}
\def\nfGG#1#2#3{\nfB
\nfFC{nfffffff}
\nfFO{1}
\nfSC{nf667782}
\nfSO{1}
\nfW{1bp}
\nfBC
\nfRJ
\nfNZ
\pgftransformcm{1}{0}{0}{-1}{\pgfpoint{#1bp}{#2bp}}
\csname nfpath#3\endcsname
\nfU{fill,stroke}
\nfE}
\def\nfGH#1#2#3{\nfB
\nfFC{nf202b33}
\nfFO{1}
\nfNZ
\pgftransformcm{0.00076562}{0}{0}{0.00076562}{\pgfpoint{#1bp}{#2bp}}
\csname nfpath#3\endcsname
\nfU{fill}
\nfE}
\def\nfGI#1#2#3{\nfB
\nfFC{nf202b33}
\nfFO{1}
\nfNZ
\pgftransformcm{0}{0.00089687}{-0.00089687}{0}{\pgfpoint{#1bp}{#2bp}}
\csname nfpath#3\endcsname
\nfU{fill}
\nfE}
\def\nfGJ#1#2#3{\nfB
\nfFC{nf007f82}
\nfFO{1}
\nfSC{nfffffff}
\nfSO{1}
\nfW{0.3bp}
\nfBC
\nfRJ
\nfNZ
\pgftransformcm{1}{0}{0}{-1}{\pgfpoint{#1bp}{#2bp}}
\csname nfpath#3\endcsname
\nfU{fill,stroke}
\nfE}
\def\nfGK#1#2#3{\nfB
\nfFC{nfffffff}
\nfFO{1}
\nfSC{nf007f82}
\nfSO{1}
\nfW{1bp}
\nfBC
\nfRJ
\nfNZ
\pgftransformcm{1}{0}{0}{-1}{\pgfpoint{#1bp}{#2bp}}
\csname nfpath#3\endcsname
\nfU{fill,stroke}
\nfE}
\def\nfGL#1#2#3{\nfB
\nfFC{nf202b33}
\nfFO{1}
\nfNZ
\pgftransformcm{0.00148437}{0}{0}{0.00148437}{\pgfpoint{#1bp}{#2bp}}
\csname nfpath#3\endcsname
\nfU{fill}
\nfE}
\def\nfGM#1#2#3{\nfB
\nfFC{nf202b33}
\nfFO{1}
\nfNZ
\pgftransformcm{0.0013125}{0}{0}{0.0013125}{\pgfpoint{#1bp}{#2bp}}
\csname nfpath#3\endcsname
\nfU{fill}
\nfE}

\endgroup
\caption{\textbf{Compilation cost and physical near-attainment.} Both panels use $\sigma=3,C=2,\varepsilon=2$. \textbf{a}, All 21 completed requests are shown; circles are repetitions, open symbols medians, and bars observed ranges. The first three denominators have three paired trials each; larger denominators have one attempted pair. The common 4.745287-second setup is charged. The triangle marks the producer's 600-second timeout at $B=1009$. Ratios are 3.46 at $B=97$ and 25.40 at $B=257$ (single pair). \textbf{b}, All 15 saved deadlines at $\gamma=1/3$ are plotted. Public waiting precedes the fixed three-release suffix; timeouts and all physical innovations are charged. The witness reaches at least 99.9048\% of the all-horizon value at $T=44$ and 99.9992\% at $T=100$ (Appendix~D.1).}
\label{fig:cost-deadline}
\end{figure}

Every completed pair returns the same exact value and DG interval. Figure 3a shows the support-reduction gains at larger denominators; at \(B=3,31\), identical control sets give essentially equal cost.

Exact evaluation at \(\gamma=336/1009,L=24\) gives
\[
0.000083992151824465512692
\le\Delta_2^*(2)\le
0.000083992151840245375088.
\tag{58}
\]

Regenerative compilation at \(\gamma=1/3,\sigma=3,C=3\), also with \(L=24\), gives
\[
0.001471784810259119
\le\Delta_3^*(2)\le
0.001471784810282800.
\tag{59}
\]

Both paths agree on each finite answer; source and threshold enclosures determine the residual width. The C3 request uses precomputed \(G_2\) (Section 4.2); Appendix F.5 records its certificate.

At \(\gamma=1/3,\sigma=3\), a legal deadline witness approaches the all-horizon C2 value. Public waiting from zero selects when to run the pair \((0,0,0)\) versus \((1,1,0)\). With every timeout branch and every physical innovation charged,
\[
H_{e^2}(P_{44}\Vert Q_{44})
\ge 0.000084231758498895252560
>0.999\,\Delta_2^*(2).
\tag{60}
\]

Figure 3b shows convergence over 15 deadlines; Appendix D.1 gives the stopping recursion.

Complete local PLD curves also expose the cost of discarded dependence. At \(\gamma=1/2,\sigma=2,T=C=3,\varepsilon=5\), the certified upper decreases from \(1.011037\times10^{-4}\) for three one-step envelopes to \(6.410539\times10^{-5}\) for the best two-step partition and \(3.961461\times10^{-5}\) for the full computation. All use the same hidden source and charge three physical draws; Table S1 gives the exact comparisons.

\subsection{Multi-memory commitments and actual-DG calibration}

For \(a=(1,1),B=3,C=2\), all four finite-law commitment levels in (38) differ at \(\lambda=10\). For \(a=(1,1,1),B=4,C=1\), each adjacent difference in (39) equals \(23/2048\), despite the single participation (Figure 4a).

At \(a=(1,1),B=3,C=1,\sigma^2=1\), the finite source uses \(L=8\) and 128-bit lower weights, with certified TV error below \(2.057\times10^{-18}\). Its depth-one and depth-two proof curves have 4,290 and 4,445 breakpoints, respectively; exact knot comparisons verify their order for every prior. Including source and threshold errors gives (41). At \(\varepsilon=5.6\),
\[
\begin{aligned}
D_1(e^{5.6})&\in[1.2704021315917,1.2704021319266]\times10^{-5},\\
\Delta_1^*(5.6)&\in[7.463699392926,7.463699396275]\times10^{-6}.
\end{aligned}
\tag{61}
\]

The complete commitment certifies risk \(10^{-5}\); the shallow proof function lies above that target. Both calculations analyze the same hidden mechanism. Figure 4b shows the corresponding epsilon brackets.

\begin{figure}[!htbp]
\centering
\includegraphics[width=\linewidth]{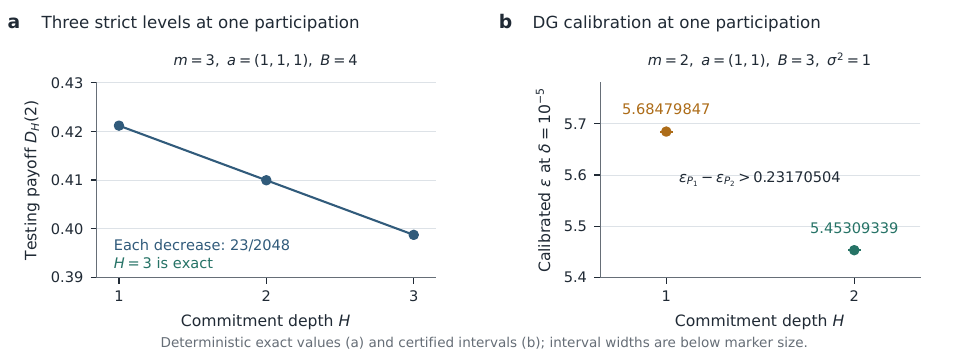}
\caption{\textbf{Memory creates strict commitment and calibration losses.}
\textbf{a}, With $m=3,C=1$ and finite law $(1,2,1)/4$ on $\{-1,0,1\}$, the three testing payoffs decrease by exactly $23/2048$ per level; lines connect the discrete depths.
\textbf{b}, With $m=2,C=1,\sigma^2=1$, the DG epsilon loss at risk $10^{-5}$ exceeds $0.23170504$. Markers show certified interval midpoints; their widths are below marker size. Appendix G.3 gives the complete curves and source transfer.}
\label{fig:calibration-gain}
\end{figure}

These multi-memory examples establish strictness at \(C=2\) for three-point noise and DG calibration at \(C=1\). Runtime and peak memory for DG with \(m\ge2,C\ge2\) remain to be measured; the state counts in Section 5.1 leave rational-curve complexity to be accounted for.

At \(C=8,\gamma=1/3,\sigma=3,\delta=10^{-4}\), depth two exceeds the target at \(\varepsilon=4.91\), while depth three certifies it; at \(4.94\), refining depth two suffices (Table S3). E.5 gives plan costs and larger-budget upper certificates; K.2 gives the fixed-noise comparisons and three-candidate choice used in Section 7.5.

\subsection{Certified multi-memory resource frontiers}

Fix \(a=(1,1),B=3,\sigma^2=1/4\), a different noise scale from the calibration example. A radius-10 window has 3,969 states and 11,907 action rows per request. The certificates use one 160-bit source, 90-bit reward enclosures and rationalized 64-bit candidate witnesses. Exact checking of all actions, flow and dual residuals, source errors and window tails gives

\noindent\begin{minipage}{\linewidth}
\par\smallskip\begingroup
\small
\setlength{\tabcolsep}{3pt}
\renewcommand{\arraystretch}{1.10}
\begin{tabularx}{\linewidth}{@{}>{\hsize=1.000000\hsize\linewidth=\hsize\raggedleft\arraybackslash}X>{\hsize=1.000000\hsize\linewidth=\hsize\raggedleft\arraybackslash}X>{\hsize=1.000000\hsize\linewidth=\hsize\raggedleft\arraybackslash}X@{}}
\toprule
\textbf{\(\tau\)} & \textbf{Lower bound for \(\Phi(\tau)\)} & \textbf{Upper bound} \\
\midrule
1 & 3.828108312556481 & 3.828132912528530 \\
2 & 3.882030842592841 & 3.882058340768777 \\
5 & 3.972779663362445 & 3.972839313104263 \\
\bottomrule
\end{tabularx}
\endgroup\smallskip\par
{\small Table 1. Outward enclosures of the actual hidden-output leading coefficient for two-memory feedback. Physical time is \(T=\lfloor\tau C\rfloor\); the finite-budget law includes its square-root remainder.}
\end{minipage}\par

Hence \(\Phi(5)-\Phi(1)>0.14464675083391\), an increase exceeding 3.7785\%. Ten rational requests supply concavity lower chords and globally valid dual upper lines. Their combined envelope covers every real \(\tau\in[1,5]\), with width below 0.001998 (Figure 5).

\begin{figure}[!htbp]
\centering
\includegraphics[width=\linewidth]{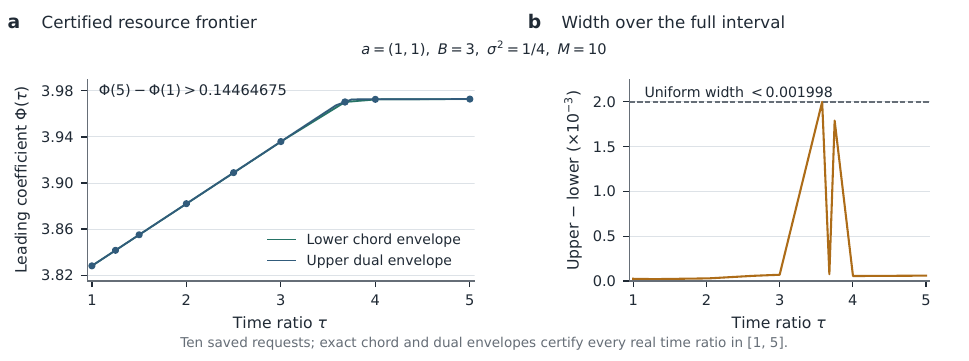}
\caption{\textbf{Physical time raises the leading privacy coefficient.}
\textbf{a}, For $a=(1,1),B=3,\sigma^2=1/4$, lower chords and all-action dual upper lines enclose $\Phi(\tau)$ for every real $\tau\in[1,5]$; dots mark the ten certified requests.
\textbf{b}, Exact checks at 48 combined breakpoints bound the envelope width by $0.001998$. The physical deadline is $T=\lfloor\tau C\rfloor$; Appendix H.6 gives the finite-budget remainder.}
\label{fig:resource-frontier}
\end{figure}

The first-order specialization identifies the same \(\psi,\Phi,a_*\) objects and reproduces the earlier frontier interval. Appendix H.3 gives the numerical cross-check and validation counts; H.5 gives the first-order rate and finite-deadline comparisons.

\subsection{First-order static workload: error, randomness and storage}

Fix \(m=1,0<a_1=a<B\). For deterministic weights \(h_t\), let \(v=\operatorname{Var}(\eta_t)\) and \(r_t=\{\gamma K_{t-1}\}\). The martingale innovations of (2) give the mean squared error (MSE) identity
\[
\mathbb E\!\left[\left(\sum_{t=1}^n h_t(Y_t-x_t)\right)^2\right]
=\sum_{j=1}^n c_j^2\left(v+\mathbb E[r_j(1-r_j)]\right),
\qquad c_j=\sum_{t=j}^n h_t(-\gamma)^{t-j}.
\tag{62}
\]

For a running mean, \(c_j=[1-(-\gamma)^{n-j+1}]/[n(1+\gamma)]\); Appendix G.6 derives (62).

The static comparison protects 100 binary inputs available before 1000 public prefix queries, with \((2,10^{-4})\)-DP under zero-out adjacency. The rank-100 workload realizes cap 56 and minimum gap 4. Table 2 compares \(\gamma=1/2\) feedback with \(A=15I+J\) (implementation in Appendix J.1).

\noindent\begin{minipage}{\linewidth}
\par\smallskip\begingroup
\footnotesize
\setlength{\tabcolsep}{3pt}
\renewcommand{\arraystretch}{1.10}
\begin{tabularx}{\linewidth}{@{}>{\hsize=1.450000\hsize\linewidth=\hsize\raggedright\arraybackslash}X>{\hsize=0.887500\hsize\linewidth=\hsize\raggedleft\arraybackslash}X>{\hsize=0.887500\hsize\linewidth=\hsize\raggedleft\arraybackslash}X>{\hsize=0.887500\hsize\linewidth=\hsize\raggedleft\arraybackslash}X>{\hsize=0.887500\hsize\linewidth=\hsize\raggedleft\arraybackslash}X@{}}
\toprule
\textbf{Finite integer program} & \textbf{Mean MSE} & \textbf{Maximum marginal MSE} & \textbf{Fresh bits} & \textbf{Noise registers (bits)} \\
\midrule
Scalar feedback, scale 16 & \(\approx57087.37\) & \(\approx113947.11\) & 39000 & 9 \\
Structured matrix & \(\le16185.889240\) & \(\le36236.678746\) & 3300 & 1000 \\
\bottomrule
\end{tabularx}
\endgroup\smallskip\par
{\small Table 2. Finite integer implementations under the same complete-vector request. Matrix bounds include rounding. Private noise registers and cumulative distribution function (CDF) tables are counted separately; the latter use 8132 and 15180 bits.}
\end{minipage}\par

The matrix reduces mean and maximum marginal MSE by more than 71.6471\% and 68.1986\% and uses fewer fresh bits; the finite-source feedback program uses less noise-state and source-table storage. The nine-bit register counts that bounded-source implementation. An exact DG register has unbounded support. For data-independent additive unbiased Gaussian designs with analytic calibration [27], a rational primal--dual enclosure of relative width below 0.003410\% bounds optimal mean MSE (Appendix J.2):
\[
3911.022021036<\operatorname{MSE}_{\rm Gaussian}^*
<3911.155354859.
\tag{63}
\]

\subsection{Online prefixes: calibration and finite-memory bounds}

Continue with \(m=1,0<a_1=a<B\). For online prefixes \(S_t=\sum_{i\le t}Y_i\), compare \(T^{-1}\sum_{t\le T}\mathbb E S_t^2\) on zero-query input. Privacy covers predictable signed-unit differences, cap \(C=2\), consecutive participations and every output, at \(\varepsilon=\log(41/4),\delta=10^{-5}\). At \(\gamma=1/3\), the certified \(\sigma^2=9\) design improves DG prefix MSE over depth-one choice \(145/16\) at \(T=64,512,4096\) (Figure 6b; source transfer in Appendix J.3).

\textbf{Proposition 13 (first-order online-prefix separation).} Every member of (2) with \(m=1,0<a_1<B\) satisfying this privacy request, for any \(0<\gamma<1\) and DG scale, has mean prefix MSE strictly greater than \(3(T+1)/16\). At \(T=512\) this exceeds 96.1875, while a buffered linear Toeplitz (BLT) mechanism with two buffers has certified integer-output MSE below 84.104. At \(T=4096\), the corresponding bounds are 768.1875 and 134.863.

\begin{figure}[!htbp]
\centering
\includegraphics[width=\linewidth]{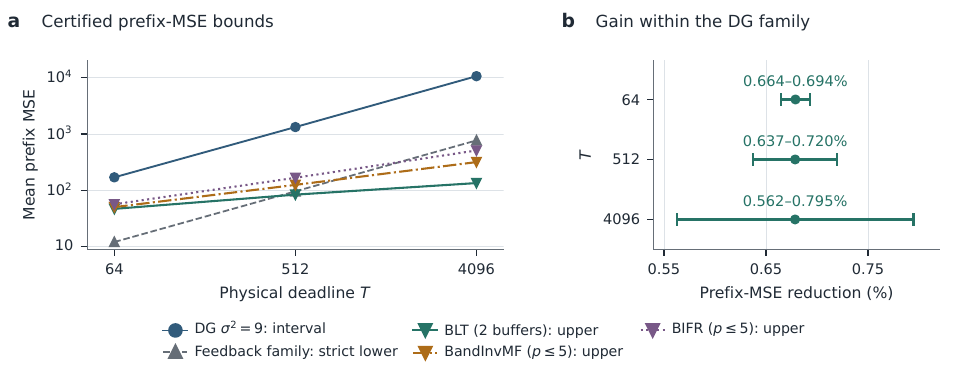}
\caption{\textbf{Calibration gains and a separation from the feedback family.}
All designs use the Section 7.5 privacy and integer-prefix workload.
\textbf{a}, Candidate uppers, strict family lower $3(T+1)/16$, and DG intervals at $\gamma=1/3$; lines guide the eye. BLT uses two buffers throughout; BandInvMF/BIFR use the best retained certificates with bandwidth $p\le5$. At $T=512,4096$, the BLT upper lies below the family lower.
\textbf{b}, MSE reduction relative to fixed $\sigma^2=145/16$. All intervals are deterministic enclosures.}
\label{fig:prefix-utility}
\end{figure}

\textbf{Finite-memory prefix bound.} Appendix J.4 proves \(v(T+1)/[2(1+r)^2]\) for arbitrary finite memory, using the actual innovation variance \(v\). The same request forces \(v>3/2\) at the first release, so every stable finite-memory member obeys the lower bound in Proposition 13. For \(m=2,a=(1,1),B\ge3\), the stronger bound is \(>0.27(T+1)\), hence \(>138.51\) and \(>1106.19\) at these horizons.

The filter explains the growth. With centered martingale innovation \(\xi_t\), (2) becomes \(K_t+\sum_i(a_i/B)K_{t-i}=\xi_t\). Write \(F(z)=1+\sum_i(a_i/B)z^{-i}\); the prefix response is \(1/[(1-z^{-1})F(z)]\). Stability gives \(1-r\le|F(e^{i\omega})|\le1+r\), so \(1/F\) has nonzero gain at zero frequency and the prefix integrator's pole remains. Appendix J.4 turns this observation into the finite-\(T\) bound above using orthogonal innovations; no stationarity assumption is needed.

\section{Discussion}

A shared preparation connects the three results for finite-memory feedback. Complete vector carries preserve its constraints across branches; the full echo sequence identifies when those constraints can be discharged. The commitment hierarchy measures the calibration cost of an earlier release, and the strict examples show that memory length matters even with one participation. Charging actual preparation time yields a second resource parameter and a leading privacy frontier for the same hidden-output process.

The workload comparisons distinguish calibration from architecture. Retaining dependencies can reduce the noise scale needed for a fixed privacy request. The prefix lower bound persists across stable finite memory because the autoregressive response retains nonzero zero-frequency gain. The finite-source comparison also quantifies the storage paid by an alternative design for lower error. Further questions concern tighter support exponents, multi-memory DG computational costs, useful finite-budget constants, source classes for general memory, and partially observed registers.

\clearpage
\typeout{IDEA1_MAIN_PAGES=\number\numexpr\value{page}-1\relax}
\section*{References}
\phantomsection\addcontentsline{toc}{section}{References}
\begingroup\raggedright
\begin{enumerate}[label={[\arabic*]},leftmargin=2.6em,itemsep=0pt,parsep=0pt]

\item Clément L. Canonne, Gautam Kamath, and Thomas Steinke. \emph{The Discrete Gaussian for Differential Privacy}. \href{https://arxiv.org/abs/2004.00010v6}{arXiv:2004.00010v6}, 2024.

\item Ilya Mironov. \emph{Rényi Differential Privacy}. CSF 2017. \href{https://arxiv.org/abs/1702.07476v3}{arXiv:1702.07476v3}.

\item Vitaly Feldman and Tijana Zrnic. \emph{Individual Privacy Accounting via a Rényi Filter}. \href{https://arxiv.org/abs/2008.11193v4}{arXiv:2008.11193v4}, 2022.

\item Guanyan Li, Juanen Li, Zhilei Han, Peixin Wang, Hongfei Fu, and Fei He. \emph{Structural Abstraction and Refinement for Probabilistic Programs}. Proceedings of the ACM on Programming Languages 9(OOPSLA2), Article 337, 2025. \href{https://doi.org/10.1145/3763115}{DOI}.

\item Sílvia Casacuberta, Michael Shoemate, Salil Vadhan, and Connor Wagaman. \emph{Widespread Underestimation of Sensitivity in Differentially Private Libraries and How to Fix It}. CCS 2022. \href{https://arxiv.org/abs/2207.10635v2}{arXiv:2207.10635v2}.

\item Krishnamurthy Dvijotham, H. Brendan McMahan, Krishna Pillutla, Thomas Steinke, and Abhradeep Thakurta. \emph{Efficient and Near-Optimal Noise Generation for Streaming Differential Privacy}. \href{https://arxiv.org/abs/2404.16706v3}{arXiv:2404.16706v3}, 2024.

\item H. Brendan McMahan, Zheng Xu, and Yanxiang Zhang. \emph{A Hassle-free Algorithm for Private Learning in Practice: Don't Use Tree Aggregation, Use BLTs}. \href{https://arxiv.org/abs/2408.08868v3}{arXiv:2408.08868v3}, 2025.

\item Yuxin Wang, Zeyu Ding, Daniel Kifer, and Danfeng Zhang. \emph{CheckDP: An Automated and Integrated Approach for Proving Differential Privacy or Finding Precise Counterexamples}. \href{https://arxiv.org/abs/2008.07485v2}{arXiv:2008.07485v2}, 2020.

\item Gian Pietro Farina, Stephen Chong, and Marco Gaboardi. \emph{Coupled Relational Symbolic Execution for Differential Privacy}. \href{https://arxiv.org/abs/2007.12987v1}{arXiv:2007.12987v1}, 2020.

\item Markus de Medeiros, Muhammad Naveed, Tancrede Lepoint, Temesghen Kahsai, Tristan Ravitch, Stefan Zetzsche, Anjali Joshi, Joseph Tassarotti, Aws Albarghouthi, and Jean-Baptiste Tristan. \emph{Verified Foundations for Differential Privacy}. \href{https://arxiv.org/abs/2412.01671v3}{arXiv:2412.01671v3}, 2025.

\item Nikita P. Kalinin and Rasmus Pagh. \emph{Dithered Gaussian Mechanism for Randomness-Efficient Differential Privacy}. \href{https://arxiv.org/abs/2607.06320v3}{arXiv:2607.06320v3}, 2026.

\item Bishnu Bhusal, Rohit Chadha, A. Prasad Sistla, and Mahesh Viswanathan. \emph{Approximate Algorithms for Verifying Differential Privacy with Gaussian Distributions}. CCS 2025. \href{https://arxiv.org/abs/2509.08804v1}{arXiv:2509.08804v1}.

\item Rohit Chadha, A. Prasad Sistla, Mahesh Viswanathan, and Bishnu Bhusal. \emph{Deciding Differential Privacy of Online Algorithms with Multiple Variables}. \href{https://arxiv.org/abs/2309.06615v1}{arXiv:2309.06615v1}, 2023.

\item Javier Esparza and Andreas Gaiser. \emph{Probabilistic Abstractions with Arbitrary Domains}. SAS 2011 technical report. \href{https://arxiv.org/abs/1106.1364v2}{arXiv:1106.1364v2}.

\item Linus Heck, Tim Quatmann, Jip Spel, Joost-Pieter Katoen, and Sebastian Junges. \emph{Generalized Parameter Lifting: Finer Abstractions for Parametric Markov Chains}. ATVA 2025. \href{https://arxiv.org/abs/2504.05965v2}{arXiv:2504.05965v2}.

\item Tim van Erven and Peter Harremoës. \emph{Rényi Divergence and Kullback--Leibler Divergence}. IEEE Transactions on Information Theory, 2014. \href{https://arxiv.org/abs/1206.2459v2}{arXiv:1206.2459v2}.

\item Anneliese Riess, Juan Felipe Gomez, Flavio Calmon, Julia Schnabel, and Georgios Kaissis. \emph{Optimal Conversion from Rényi Differential Privacy to f-Differential Privacy}. Proceedings of the 43rd International Conference on Machine Learning, PMLR 306:105180--105192, 2026. \href{https://proceedings.mlr.press/v306/riess26a.html}{Published version}.

\item Milos Hauskrecht. \emph{Value-Function Approximations for Partially Observable Markov Decision Processes}. Journal of Artificial Intelligence Research 13, 33--94, 2000. \href{https://arxiv.org/abs/1106.0234v1}{Author manuscript}.

\item Vadym Doroshenko, Badih Ghazi, Pritish Kamath, Ravi Kumar, and Pasin Manurangsi. \emph{Connect the Dots: Tighter Discrete Approximations of Privacy Loss Distributions}. Privacy Enhancing Technologies Symposium, 2022. \href{https://arxiv.org/abs/2207.04380v1}{arXiv:2207.04380v1}.

\item Merlijn Krale, Wietze Koops, Sebastian Junges, Thiago D. Simão, and Nils Jansen. \emph{Tighter Value-Function Approximations for POMDPs}. AAMAS 2025. \href{https://arxiv.org/abs/2502.06523v1}{Author manuscript}.

\item Yuqing Zhu, Jinshuo Dong, and Yu-Xiang Wang. \emph{Optimal Accounting of Differential Privacy via Characteristic Function}. AISTATS 2022, PMLR 151:4782--4817. \href{https://proceedings.mlr.press/v151/zhu22c.html}{Published version}.

\item Günter Rote. \emph{The Convergence Rate of the Sandwich Algorithm for Approximating Convex Functions}. Computing 48:337--361, 1992. \href{https://page.mi.fu-berlin.de/rote/Papers/abstract/The\%2Bconvergence\%2Brate\%2Bof\%2Bthe\%2BSandwich\%2Balgorithm\%2Bfor\%2Bapproximating\%2Bconvex\%2Bfunctions.html}{Author page}.

\item Anthony R. Cassandra, Michael L. Littman, and Nevin Lianwen Zhang. \emph{Incremental Pruning: A Simple, Fast, Exact Method for Partially Observable Markov Decision Processes}. UAI 1997, pp. 54--61. \href{https://arxiv.org/abs/1302.1525v1}{Author manuscript}.

\item Buxin Su, Qiaoshi Yang, Yiding Su, and Chendi Wang. \emph{Unifying Privacy Accounting: Information Equivalence and Information Loss}. \href{https://arxiv.org/abs/2610.02414v1}{arXiv:2610.02414v1}, 2026.

\item Naci Saldi. \emph{Finite-State Approximations to Discounted and Average Cost Constrained Markov Decision Processes}. \href{https://arxiv.org/abs/1807.02994v1}{arXiv:1807.02994v1}, 2018.

\item Ryan McKenna, Gerome Miklau, Michael Hay, and Ashwin Machanavajjhala. \emph{HDMM: Optimizing error of high-dimensional statistical queries under differential privacy}. \href{https://arxiv.org/abs/2106.12118v1}{arXiv:2106.12118v1}, 2021.

\item Borja Balle and Yu-Xiang Wang. \emph{Improving the Gaussian Mechanism for Differential Privacy: Analytical Calibration and Optimal Denoising}. ICML 2018, PMLR 80:394--403. \href{https://proceedings.mlr.press/v80/balle18a.html}{Published version}.

\item Philipp G. Haselwarter, Alejandro Aguirre, Simon Oddershede Gregersen, Kwing Hei Li, Joseph Tassarotti, and Lars Birkedal. \emph{Modular Verification of Differential Privacy in Probabilistic Higher-Order Separation Logic}. Proceedings of the ACM on Programming Languages 10(PLDI), Article 233, 2026. \href{https://doi.org/10.1145/3808311}{DOI}.

\item Justin Whitehouse, Aaditya Ramdas, Ryan Rogers, and Steven Wu. \emph{Fully-Adaptive Composition in Differential Privacy}. ICML 2023, PMLR 202:36990--37007. \href{https://proceedings.mlr.press/v202/whitehouse23a.html}{Published version}.

\item Arie Leizarowitz and Adam Shwartz. \emph{Exact Finite Approximations of Average-Cost Countable Markov Decision Processes}. Automatica 44(6):1480--1487, 2008. \href{https://doi.org/10.1016/j.automatica.2007.09.013}{DOI}.

\item Sirin Nitinawarat and Venugopal V. Veeravalli. \emph{Controlled Sensing for Sequential Multihypothesis Testing with Controlled Markovian Observations and Non-Uniform Control Cost}. Sequential Analysis 34(1):1--24, 2015. \href{https://doi.org/10.1080/07474946.2014.961864}{DOI}.

\item Liyu Chen, Rahul Jain, and Haipeng Luo. \emph{Learning Infinite-horizon Average-reward Markov Decision Process with Constraints}. ICML 2022, PMLR 162:3246--3270. \href{https://proceedings.mlr.press/v162/chen22i.html}{Published version}.

\item Borna Sayedana, Peter E. Caines, and Aditya Mahajan. \emph{Concentration of Cumulative Reward in Markov Decision Processes}. Journal of Machine Learning Research 26(278):1--71, 2025. \href{https://jmlr.org/papers/v26/24-2047.html}{Published version}.

\item Borja Balle, Gilles Barthe, and Marco Gaboardi. \emph{Privacy Amplification by Subsampling: Tight Analyses via Couplings and Divergences}. NeurIPS 2018. \href{https://arxiv.org/abs/1807.01647v2}{arXiv:1807.01647v2}.

\item Christian H. Weiß and Fukang Zhu. \emph{Mean-preserving Rounding Integer-valued ARMA Models}. Journal of Time Series Analysis 46(3):530--551, 2025. \href{https://doi.org/10.1111/jtsa.12774}{DOI}.

\item David B. Brown, James E. Smith, and Peng Sun. \emph{Information Relaxations and Duality in Stochastic Dynamic Programs}. Operations Research 58(4):785--801, 2010. \href{https://doi.org/10.1287/opre.1090.0796}{DOI}.

\item Sung-Wen Lan and I-Hsiang Wang. \emph{Heterogeneous Sequential Hypothesis Testing with Active Source Selection under Budget Constraints}. ISIT 2021, pp. 178--183. \href{https://doi.org/10.1109/ISIT45174.2021.9517747}{DOI}.

\item Antti Koskela, Rachel Emily Redberg, and Yu-Xiang Wang. \emph{Privacy Profiles for Private Selection}. ICML 2024, PMLR 235:25313--25332. \href{https://proceedings.mlr.press/v235/koskela24a.html}{Published version}.

\item Jinho Bok. \emph{Can Rényi Accountants Beat Lossless Accountants for DP-SGD?} arXiv:2610.17120v1, 2026. \href{https://doi.org/10.48550/arXiv.2610.17120}{DOI}.

\item Long Tran, Antti Koskela, Ossi Räisä, and Antti Honkela. \emph{f-Differential Privacy Filters: Validity and Approximate Solutions}. arXiv:2602.06756v2, 2026. \href{https://doi.org/10.48550/arXiv.2602.06756}{DOI}.

\item Matthew Regehr, Bingshan Hu, Ethan Leeman, Pasin Manurangsi, Pierre Tholoniat, and Mathias Lécuyer. \emph{Privacy Filters are Captured by Residues: A Characterization of Free Natural Filters and the Cost of Adaptivity}. arXiv:2602.15815v2, 2026. \href{https://doi.org/10.48550/arXiv.2602.15815}{DOI}.

\item Vitaly Feldman and Moshe Shenfeld. \emph{Efficient Privacy Loss Accounting for Subsampling and Random Allocation}. ICML 2026, PMLR 306:29897--29929. \href{https://proceedings.mlr.press/v306/feldman26a.html}{Published version}.

\item Krishnendu Chatterjee, Ehsan Kafshdar Goharshady, and \DJ{}or\dj{}e \v{Z}ikeli\'{c}. \emph{SuperDP: Differential Privacy Refutation via Supermartingales}. Proceedings of the ACM on Programming Languages 10(PLDI), Article 218, 2026. \href{https://doi.org/10.1145/3808296}{DOI}.

\item Sergey Denisov, H. Brendan McMahan, Keith Rush, Adam Smith, and Abhradeep Guha Thakurta. \emph{Improved Differential Privacy for SGD via Optimal Private Linear Operators on Adaptive Streams}. NeurIPS 2022, volume 35. \href{https://proceedings.neurips.cc/paper_files/paper/2022/hash/271ec4d1a9ff5e6b81a6e21d38b1ba96-Abstract-Conference.html}{Published version}, \href{https://doi.org/10.52202/068431-0428}{DOI}.

\item Nikita P. Kalinin, Ryan McKenna, Rasmus Pagh, and Christoph H. Lampert. \emph{DP-\(\lambda\)CGD: Efficient Noise Correlation for Differentially Private Model Training}. arXiv:2601.22334v3, 2026. \href{https://doi.org/10.48550/arXiv.2601.22334}{DOI}.

\item Nikita P. Kalinin, Ryan McKenna, Jalaj Upadhyay, and Christoph H. Lampert. \emph{Back to Square Roots: An Optimal Bound on the Matrix Factorization Error for Multi-Epoch Differentially Private SGD}. ICLR 2026; arXiv:2505.12128v3. \href{https://openreview.net/forum?id=EEr6cADbZx}{Proceedings}, \href{https://doi.org/10.48550/arXiv.2505.12128}{DOI}.

\item Nikita P. Kalinin, Aki Rehn, Joel Daniel Andersson, Antti Honkela, and Christoph H. Lampert. \emph{Beyond Square Roots: A Memory-Efficient Explicit Factorization for Multi-Epoch Private Learning}. arXiv:2605.18379v2, 2026. \href{https://doi.org/10.48550/arXiv.2605.18379}{DOI}.

\item Eitan Altman and Adam Shwartz. \emph{Markov Decision Problems and State-Action Frequencies}. SIAM Journal on Control and Optimization 29(4):786--809, 1991. \href{https://doi.org/10.1137/0329043}{DOI}.

\item Peter W. Glynn and Sean P. Meyn. \emph{A Liapounov Bound for Solutions of the Poisson Equation}. The Annals of Probability 24(2):916--931, 1996. \href{https://web.stanford.edu/~glynn/papers/1996/GM96.html}{Author copy}.

\item Leonid G. Khachiyan. \emph{A Polynomial Algorithm in Linear Programming}. Doklady Akademii Nauk SSSR 244(5):1093--1096, 1979. \href{https://www.mathnet.ru/eng/dan42319}{Published record}.

\end{enumerate}\endgroup

\clearpage
\typeout{IDEA1_APPENDIX_START=\thepage}
\appendix
\setcounter{figure}{0}\renewcommand{\thefigure}{S\arabic{figure}}\renewcommand{\theHfigure}{S\arabic{figure}}
\section*{Supplementary proofs}
\phantomsection\addcontentsline{toc}{part}{Supplementary proofs}

\section{Preparation, exact support reduction, and error transfer}

Section A.1 proves physical preparation for the finite-memory mechanism (1). Sections A.2--A.6 specialize to \(m=1,0<a_1=a<B\) and prove the sharp C2 result, Theorem 3. In those sections write \(Q_k(n)\) for the transition law in (2). All testing values below are homogeneous correct-testing values: \(V(x,y)=\sup_{\pi,E}\{xP_\pi(E)+yQ_\pi(E^c)\}\), for \(x,y\ge0\). They are convex and positively homogeneous in their weights. The hockey-stick value at threshold \(\lambda\) is \(V(1,\lambda)-\lambda\). For a general innovation law, all output sums run over \(\mathbb Z\); for the finite law they reduce to the complete sets \(I,J\) in (12)--(13).

\subsection{Public preparation and finite deadlines}

Let the innovation law \(\mu\) have full integer support and \(b_0=\mathbb E_\mu|\eta|+1<\infty\). Put \(r=\sum_i|a_i|/B\), \(\alpha_0=1-(1-r)/(2m)\) and \(V(s)=\max_i\alpha_0^{i-1}|s_i|\). Bernoulli's inequality gives \(\alpha_0^m\ge(1+r)/2>r\). The new first coordinate and the shifted old coordinates therefore satisfy
\[
V(S')\le\alpha_0V(s)+|\eta|+1,\qquad
\mathbb E[V(S')\mid s]\le\alpha_0V(s)+b_0.
\tag{A.1}
\]

Choose \(R_0\ge(b_0+1)/(1-\alpha_0)\). Outside the finite set \(F=\{V\le R_0\}\), drift is at most minus one. Summing the stopped drift through \(\tau_F\wedge n\) and then taking limits bounds the expected hitting time of \(F\) by \(V(s)\).

For any target register \(z\), a sequence of \(m\) specified new noise balances reaches \(z\). Write \(T_z\) for its first hitting time. Every such transition has positive probability because the innovation law has full support. From the finite set \(F\), let \(p_z>0\) be a common lower bound on this success probability. An \(m\)-step attempt starting in \(F\) has expected final \(V\) at most \(M_0=R_0+b_0/(1-\alpha_0)\). On failure the unconditional expected return cost is at most \(M_0\). Truncating to \(n\) attempts and summing the geometric bound gives
\[
\mathbb E_s T_z\le V(s)+(m+M_0)/p_z.
\tag{A.1a}
\]

The argument uses conditional success probabilities and does not require independent attempts. In particular every fixed register is reached almost surely.

While \(d=0\) and \(j=0\), the two weights are multiplied by the same transition probability. Every continuation is bounded by the free-preparation value at those weights; homogeneity removes the wait after summing all its outputs. Induction on the physical deadline then bounds every legal experiment by the preparation game, retaining all nonsynchronized rows (21).

Conversely, F.1 supplies finitely many integer representatives for words of length at most \(mC+1\). Realize a near-optimal game policy by public first-hit waits at every synchronized boundary. The register is recoverable from the shared outputs and the fixed common input rule; no additional observation is used. After each hit, a known invertible translation implements the chosen word. Full carries and all echo outputs remain in the transcript.

Let \(R_*\) bound \(V\) on the finite representative set and let \(p\) be the minimum of its positive \(p_z\). There are at most \(C\) preparations and \(D=(m+1)C\) informative innovations. The expected norm before any new preparation is at most \(R_*+Db_0\). Thus one uniform expected completion bound is
\[
L_C=D+C\{R_*+Db_0+(m+M_0)/p\}.
\tag{A.1b}
\]

At deterministic cutoff \(T\), accept only histories that completed the original accepting event; reject unfinished histories and pad completed runs with common inputs. The lost P-mass is at most \(L_C/(T+1)\), while discarding Q-acceptance only increases the payoff. The bound is independent of \(\lambda\ge1\). Taking near-optimal policies proves the all-horizon equality \(\Delta_C^\mu(\log\lambda)=G_C^\mu(1,\lambda)-\lambda\), (30), and then (31) by generalized inversion. Finite-law preparation remains the definition of the comparison game; this physical argument applies to the stated full-support source.

\subsection{Periods, reflection, and pure first rows}

Translation and reflection give
\[
Q_{k+B}(n)=Q_k(n+a),\qquad Q_{-k}(-n)=Q_k(n).
\tag{A.2}
\]

The reflection follows by replacing \((\eta,U)\) with \((-\eta,B-1-U)\). A final echo is therefore \(B\)-periodic in its integer start. Shifting a state with one participation remaining by \(B^2\) shifts its next output by \(-aB\), preserving every subsequent echo phase; an echo followed by synchronized preparation is independent of its new integer start. Thus this continuation is \(B^2\)-periodic. Reflection handles a preceding negative difference without interchanging the testing weights.

At a synchronized positive participation, write \(ak=Bv+R\), \(r=R/B\), and \(f=n+v\). The two current probabilities are
\[
(p_f,q_f)=(1-r)(g(f+1),g(f))+r(g(f+2),g(f+1)).
\tag{A.3}
\]

Hold each successor state \(n=f-v\) fixed and apply convexity of its continuation in \((p_f,q_f)\). Summing bounds this mixed first row by a convex combination of two pure-row sums. Reindexing the second by \(f'=f+1\) gives bases \(-v\) and \(-v-1\). Conversely, pure preparations \(k=Bz\) have base \(-az\). Coprimality makes these bases cover every residue modulo \(B^2\), or modulo \(B\) for one remaining participation. Consequently the synchronized roots equal the pure-root formulas (13) and (16). A pure negative participation gives the same optimum by reflection. Additional observations cannot reduce a testing value, so these formulas also dominate stopping or using less budget. This purification uses convexity after marginalizing the hidden draw.

\subsection{Convexity and the coupled boundary}

For the finite law \(g=\nu\), the complete rows are (12) on \(I=[-L-3,L+1]\cap\mathbb Z\). Summing the raw rounding preimages gives those two affine mixtures. Each row is normalized, including zero masses. The echo \(E\) in (13) is convex in its weights and separately convex in \(r\) on \([0,1-\gamma]\) and \([1-\gamma,1]\). It is continuous at the shared boundary. Reindexing \(f'=f+1\) at the endpoints gives \(E(1;x,y)=E(0;x,y)\).

In the one-participation root (13), all phase breaks occur when \(\{\beta+\gamma f\}\) is zero or \(1-\gamma\). Since \(f\in J=[-L-1,L]\cap\mathbb Z\), they are precisely contained in \(K_0\) of (17). Between consecutive breaks the sum is convex and it is continuous on the circle. Its maximum is therefore attained in \(K_0\), whose points are legal members of \(G_B\).

The map \(k\bmod B^2\mapsto(r(k),s(k))\) in (5) is bijective. Indeed, fix \(R=ak\bmod B\) and its unique representative \(k_0\bmod B\). For \(k=k_0+Bz\),
\[
Bs(k)\equiv-a\lfloor ak_0/B\rfloor-a^2z\pmod B.
\tag{A.4}
\]

Since \(a^2\) is invertible modulo \(B\), the second residue runs through the whole group exactly once. After the current output, \(n=b(k)+f\) has phase \(\{s+\gamma f\}\). An immediate negative participation uses its reflected phase. These facts give exactly the four branches of (14): stopping, echoing into \(G_1\), and either immediate sign. The branch maximum stays outside its output sum.

For fixed \(s,x,y\), every child in (14) is convex in its weights and every current row is affine on the two \(r\)-arcs. Thus \(F\) is convex on those arcs. For fixed \(r,x,y\), it is continuous and periodic in \(s\), and convex between the points
\[
K_1=\{-m\gamma\bmod1:-L-4\le m\le L+2\}.
\tag{A.5}
\]

For the positive child the break indices are \(m=f,f+1\); for the negative child they are \(m=f,f-1\), with \(f\in I\). Their union is contained in (A.5). Maxima and sums preserve convexity on the resulting common intervals. At the first-coordinate boundary, both current rows at \(r=1\) equal their \(r=0\) rows after \(f'=f+1\). The successor phase changes to \(s-\gamma+\gamma f'\). Therefore
\[
F(1,s;x,y)=F(0,\{s-\gamma\};x,y).
\tag{A.6}
\]

The boundary identification translates both coordinates together, carrying the change in \(r\) into \(s\).

\subsection{Reduction of the shared root preparation}

For a root base \(b\), put \(\beta=\{\gamma b\}\) and \(\chi=\{\gamma b(b)\}\). The successor of its pure first row is \(b+f\). Directly expanding its floor gives \(r_f,s_f\) in (16), so a single legal pair \((\beta,\chi)\) is shared by every first output.

Fix \(\chi\). Between consecutive \(K_0\) points, each \(e_f=\lfloor\beta+\gamma f\rfloor\) is constant, \(r_f\) stays on one convex arc, and \(s_f\) is fixed. Each summand and their sum are convex in \(\beta\). At a wrap, \(r_f\) changes from one to zero, \(e_f\) increases by one, and \(s_f\) decreases by \(\gamma\); (A.6) supplies continuity. At the external boundary, the value at \((1,\chi)\) equals that at \((0,\{\chi-\gamma\})\). Maximizing over \(\chi\in G_B\), a group closed under this translation, therefore gives a continuous circular function, convex between the same breaks. An endpoint in \(K_0\) attains its maximum.

Now fix such a legal \(\beta\). The sum is periodic and piecewise convex in \(\chi\), with breakpoints
\[
K_2(\beta)=\{\gamma(e_f-m)\bmod1:f\in J,\ -L-4\le m\le L+2\}.
\tag{A.7}
\]

For \(0\le\beta<1\), \(0<\gamma<1\), and \(f\in J\), one has \(-L-1\le e_f\le L\). Hence \(-2L-3\le e_f-m\le2L+4\), placing every breakpoint in \(K_2\) of (17). All are legal group points. Convexity supplies the upper inequality for the reduced maximum; legality supplies the reverse inequality. Together with the bijection in A.3 this proves Theorem 3 for every \(x,y\ge0\).

The candidate sets are valid simultaneously for every \(x,y\ge0\); the maximizing pair may depend on those weights. The order of decisions matters. Under \(g(-1),g(0),g(1)=(1,2,1)/4\), \(\gamma=1/3\), and \(\lambda=1/2\), the exact finite hockey-stick value is \(5/6\). Allowing the second root coordinate to be reselected after each first output gives \(107/128\). The strict difference \(1/384\) illustrates the extra control excluded by (16).

\subsection{Exact computation and independent reconstruction}

A finite experiment has \(V(x,y)=\sum_f\max(xp_f,yq_f)\). Sorting likelihood ratios yields cumulative supporting pairs \((P(E),Q(E^c))\); zero-probability axes are included directly. Conditional observation sums combine the probability-scaled pairs by Minkowski sums, and pre-output alternatives take the upper envelope of their union [18]. These operations use exact integer coefficients, comparisons, and crossings.

The producer constructs \(G_1\) by summing echo curves. The auditor instead constructs each complete two-output joint experiment and then its testing curve. For integer \(z\), the hidden uniform count is the length of \([0,B)\cap[hB-z,(h+1)B-z)\), as in (A.8); at most two consecutive \(h\) contribute. Applying these counts separately to \(ak\) and \(a(k+1)\) reconstructs both original probability rows without enumerating \(U\). Modular inversion of (A.4) recovers integer representatives and their actual successors. The two paths use separate row, hull, transition, and curve-summation implementations, with the producer using \(K_2\) and the auditor using (A.7).
\[
\#\{0\le U<B:\lfloor(z+U)/B\rfloor=h\}
=\max\{0,\min(B,(h+1)B-z)-\max(0,hB-z)\}.
\tag{A.8}
\]

Writing \(\beta=-m\gamma\) and \(\chi=n\gamma\) modulo one, the reached phases have indices \(f-m\in[-2L-2,2L+1]\) and \(n-e_f\in[-3L-3,3L+5]\). Thus \(O((L+1)^2)\) reached phase pairs and \(O(L+1)\) echo phases suffice. The auditor's second-set candidates also collapse to an interval of \(O(L+1)\) integer indices. Each algorithm constructs \(O((L+1)^3)\) lines before pruning and streams its root queries. Appendix C accounts for all operations and their integer bit lengths.

\subsection{Transfer to the discrete Gaussian}

Suppose \(\operatorname{TV}(g,\nu)\le\tau\). Apply a common strategy to both informative graphs, coupling each innovation and using the same hidden rounding draw until mismatch. There are at most four informative transitions; hence each directional transcript TV is at most \(r_\tau=1-(1-\tau)^4\). For every event and threshold, its payoff differs by at most \((1+\lambda)r_\tau\). Taking suprema over the common policy and event class preserves this bound:
\[
|\Delta_g(\lambda)-\Delta_{\nu}(\lambda)|
\le(1+\lambda)\{1-(1-\tau)^4\}.
\tag{A.9}
\]

This is a coupling of informative graphs. Public waiting has already been eliminated by A.1 for the DG process; no finite-law preparation assumption or coupling over unbounded waits is required. Finally, hockey-stick divergence and its policy supremum are decreasing and 1-Lipschitz in \(\lambda\). With \(v_*=G_2(1,\lambda_-)-\lambda_-\) and \(\lambda_-\le e^\varepsilon\le\lambda_+\), (A.9) gives (28), clipped to \([0,1]\). This proves both endpoints for the actual autonomous DG supremum.

At budget \(C\), at most \(C\) participations and \(C\) echoes give exponent \(2C\) by the same coupling argument. Physical waiting is charged separately across all \(T\) releases in (D.3).

\section{Information and implementation separations}

This appendix concerns the first-order subfamily \(m=1,0<a_1=a<B\), with \(\gamma=a/B\).

\subsection{The finite transcript comparator}

For Theorem 1 fix \(\gamma=1/3\), \(\sigma=3\), and let \((P_3,Q_3)\) be the zero-initial three-release laws for inputs \((0,0,0)\) and \((1,1,0)\). Let \(P_b(1)=23/10000\), \(Q_b(1)=27/100000\). We show, in both directions,
\[
D_\alpha(P_b\Vert Q_b)<D_\alpha(P_3\Vert Q_3)\qquad(\alpha>0).
\tag{B.1}
\]

The map \((\eta,U)\mapsto(-\eta,2-U)\) preserves the noise law and negates the state transcript. The involution \(y\mapsto(1,1,0)-y\) therefore swaps \(P_3,Q_3\), so their directed Rényi divergences coincide.

The first output is a discrete Gaussian and its unit shift. With \(Z(t)=\sum_{z\in\mathbb Z}e^{-(z-t)^2/18}\), completing the square gives
\[
D_\alpha(g\Vert g(\cdot-1))=
\frac{\alpha}{18}+\frac{\log[Z(1-\alpha)/Z(0)]}{\alpha-1},
\qquad D_1(g\Vert g(\cdot-1))=1/18.
\tag{B.2}
\]

Poisson summation expresses \(Z(t)\) as a positive constant times \(1+2\sum_{n\ge1}e^{-18\pi^2n^2}\cos(2\pi nt)\); absolute convergence gives \(Z(t)\le Z(0)\). Thus for \(0<\alpha<1\), (B.2) is at least \(\alpha/18\). Data processing yields \(D_{1/2}(P_3\Vert Q_3)\ge1/36\) and \(D_1(P_3\Vert Q_3)\ge1/18\) [1,16].

\subsection{A finite certificate covering every order above one}

The marginalized kernel is \(Q_k(n)=\frac13\sum_{u=0}^2g(n+\lfloor(k+u)/3\rfloor)\). At conditional states \((k+d,k)\) and next difference \(j\), the same-output probabilities are \(p=Q_{k+d}(n+j)\), \(q=Q_k(n)\). Use \(f=n+\lceil k/3\rceil\). The three layers have \((d,j)=(0,1),(1,1),(1,0)\), with reference states \(k_1=f_1\), \(k_2=f_2-\lceil k_1/3\rceil\). Restrict \(f_1,f_2,f_3\) to \([-24,24]\). This triangular change of coordinates identifies \(49^3\) complete observable transcript atoms, with hidden rounding marginalized.

Directed probability bounds have \(p_{\rm lo}/S\le p\) and \(q\le q_{\rm hi}/S\), with \(S=10^{18}\). At each rational anchor \(a>1\), certified integer weights at scale \(W=10^{60}\) satisfy
\[
0\le w_a/W\le(p_{\rm lo}/S)^a(q_{\rm hi}/S)^{1-a}\le p^a q^{1-a}.
\tag{B.3}
\]

Zero P lower endpoints receive weight zero. Summing products along the stated tree gives \(M_a^{\rm low}=W^{-3}\sum_{f_1,f_2,f_3=-24}^{24}\prod_{t=1}^3w_a(r_{t-1},d_t,j_t,f_t)\), where \(r_{t-1}=k_{t-1}\bmod3\). Every omitted atom has nonnegative moment mass, so a directed lower bound on \(\log M_a^{\rm low}/(a-1)\) is a valid lower bound \(L_a\) on the full divergence. No law is renormalized. The independent check recomputes the raw-U rows, weight inequalities and integer-state tree sums.

For the binary pair, directed evaluation uses
\[
D_a(P_b\Vert Q_b)=\frac{\log[p^a q^{1-a}+(1-p)^a(1-q)^{1-a}]}{a-1}.
\tag{B.4}
\]

At each left anchor \(l\), use a downward bound \(L_l\); at the right anchor \(r\), use an upward bound \(U_r\) for the larger binary direction. At \(l=1\), take \(L_1=1/18\).

The 29 adjoining bands cover \([1,16]\). Their smallest downward-minus-upward displayed margin is \(0.00007398\), attained on \([27/4,55/8]\); hence all comparisons are strict. The accompanying exact certificate records every anchor, row enclosure and integer moment sum.

For \(\alpha\in[l,r]\), monotonicity gives \(D_\alpha(P_b\Vert Q_b)\le U_r<L_l\le D_\alpha(P_3\Vert Q_3)\); the same upper bounds cover the reverse binary direction. The 29 bands cover \((1,16]\) without gaps. Finally,
\[
\begin{aligned}
\max\{D_\infty(P_b\Vert Q_b),D_\infty(Q_b\Vert P_b)\}&<2.142242442919,\\
D_{16}(P_3\Vert Q_3)&>3.158140151588.
\end{aligned}
\tag{B.5}
\]

Monotonicity extends the strict inequality to all \(\alpha\ge16\). This proves (B.1) above one using finite sums and a separate infinite-order bound.

\subsection{Orders at most one and the testing separation}

Directed evaluation gives \(D_1(P_b\Vert Q_b)<0.002899220021595\) and \(D_1(Q_b\Vert P_b)<0.001453658341122\). Both are below \(1/36\). Monotonicity and (B.2) therefore prove (B.1) for \(1/2\le\alpha\le1\). For smaller orders, the identity
\[
D_\alpha(P\Vert Q)=\frac{\alpha}{1-\alpha}D_{1-\alpha}(Q\Vert P)
\tag{B.6}
\]

has a positive multiplier and reduces each direction to the already proved opposite direction [16].

The exact support computation and the DG transfer in A.6, at \(L=24\), \(S=10^{24}\), give the outward enclosure
\[
\Delta_2^*(2)\in
[0.000084311988514885295528,\;0.000084311988530665157923].
\tag{B.7}
\]

This is the same compiler certificate used in the evaluation, with its producer value independently reconstructed from the original rounding law. The binary value is explicit:
\[
H_{e^2}(P_b\Vert Q_b)=\frac{23}{10000}-e^2\frac{27}{100000}
>0.000304954853288724>10^{-4}.
\tag{B.8}
\]

For each positive order, (B.1) places both newly adjoined experiments below a member already in \(\mathcal F\). Thus adjoining them leaves every envelope supremum unchanged, whereas (B.7)--(B.8) increase the testing supremum. An envelope-only converter must accommodate both families and hence the binary value. This proves Theorem 1 by constructing two families with equal divergence envelopes and different privacy profiles.

\subsection{Predictable offsets and an explicit two-release event}

For a common history, the reference state \(k=Y_{t-1}-x^Q_{t-1}(Y_{<t-1})\) is a known conditional coordinate. Choosing \(s=(R-ak)\bmod B\) puts any desired residue \(R\) into the next rounding argument before its fresh randomness. For a positive previous difference, the endpoint residues \(R=0,B-a\) yield
\[
\begin{array}{c|cc}
 &p(f)&q(f)\\\hline
R=0&(1-\gamma)g(f+j)+\gamma g(f+j+1)&g(f)\\
R=B-a&g(f+j+1)&\gamma g(f)+(1-\gamma)g(f+1).
\end{array}
\tag{B.9}
\]

For a negative difference, reflection uses \(s=(-R-ak)\bmod B\) and the opposite physical sign. Synchronized residue zero gives a pure first row immediately. In the graph with freely chosen current phase, each child value depends only on its testing weights and remaining difference and budget. Convexity on the two phase arcs reduces every row maximum to (B.9). A mixed synchronized first row is bounded by its pure components as in A.2. Backward induction gives the upper bound for the offset class. Conversely the residue rule implements every selected endpoint, using only common history, and at most two participations and two echoes suffice. Hence the phase-controlled graph is exact for this offset class at each fixed threshold.

For Proposition 2, use the input vectors and offsets (11). Put \(f_1=Y_1-1\) and \(f_2=Y_2-1+\lceil(Y_1-1)/3\rceil\). This is an invertible public coordinate change. The joint masses become
\[
\begin{aligned}
p(f_1,f_2)&=g(f_1+1)\{(2/3)g(f_2+1)+(1/3)g(f_2+2)\},\\
q(f_1,f_2)&=g(f_1)g(f_2).
\end{aligned}
\tag{B.10}
\]

Accept precisely the pairs in \([-20,20]^2\cap\mathbb Z^2\) for which \(p(f_1,f_2)>e^2q(f_1,f_2)\), and reject all other transcripts. This specifies the event without revealing either rounding draw. It contains 330 atoms. Independent directed summation of the raw-U probabilities gives \(P(A)-e^2Q(A)\ge0.000118891298210654\). The fixed input pair and common offset rule apply on every history. Together with (B.7), this proves the strict contract separation.

In the unrounded model \(K_t=\eta_t-\gamma K_{t-1}-\beta_t(Y_{<t})\), define \(\ell_0=0\), \(\ell_t=-\gamma\ell_{t-1}-\beta_t(Y_{<t})\). Then \(K_t-\ell_t\) follows the offset-free recurrence. The causal map \(Y_t\mapsto Y_t-\ell_t\) is invertible by induction. Transporting both input rules preserves their differences and budget, proving equality of the optimized testing curves in that ideal model.

\subsection{The one-digit comparison}

For budget two, retaining only \(k\bmod B\) fixes \(r(k)\) but leaves \(s(k)\) arbitrary in \(G_B\), by (A.4). Its relaxed remaining-participation value is \(\overline F(r;x,y)=\max_{s\in G_B}F(r,s;x,y)\). The synchronized one-digit root is
\[
G_2^{(1)}(x,y)=\max_{\beta\in G_B}\sum_{f\in J}
\overline F(\{\beta+\gamma f\};xg(f+1),yg(f)).
\tag{B.11}
\]

For the DG graph the sum in (B.11) runs over all integers; \(J\) is its finite-law counterpart. Here the unknown high digit may be chosen after the first output, but its choice precedes the entire next output sum. Every autonomous preparation is represented, so (B.11) is an upper relaxation. Restoring both digits gives exactly (16). Discarding the remaining phase consistency gives the larger offset graph of B.4. These inclusions establish \(\Delta_2^* =\Delta_2^{(2)}\le\Delta_2^{(1)}\le\Delta_2^{(0)}\), where values are evaluated as \(A(1,e^\varepsilon)-e^\varepsilon\).

At the reference point, the existing 1,025-point prior grid and innovation radius 40 certify
\[
\begin{aligned}
\Delta_2^{(0)}(2)&\in[0.000156515262735100,\;0.000157204453248391],\\
\Delta_2^{(1)}(2)&\in[0.000084985942152780,\;0.000085481094193635],\\
\Delta_2^{(2)}(2)&\in[0.000084221293218405,\;0.000084579508162554].
\end{aligned}
\tag{B.12}
\]

Upper endpoints are valid autonomous bounds; lower endpoints at depth less than two concern their relaxation. The one-digit lower endpoint exceeds the autonomous upper endpoint (B.7) by more than \(6.7395\times10^{-7}\), with \(\delta_{\rm sep}=423/5000000\) strictly between them. Hence the compiler certifies this target, whereas any upper bound on the one-digit relaxation exceeds it. Both methods certify \(10^{-4}\). Section 7.1 reports the matched task costs under its stated source and precision settings.

\section{Bit complexity and final-error input selection}

This appendix concerns the first-order subfamily \(m=1,0<a_1=a<B\), with \(\gamma=a/B\).

\subsection{Input, output, and computational model}

Use the explicit symmetric integer weights \(w_{-L},\ldots,w_L\), total \(S\), reduced slope \(a/B\), and threshold \(u/v\) of Theorem 3. Put \(Q_{\rm row}=BS,\ell=L+1,W_{\lambda}=u+v+1\), and
\[
\kappa=1+\lceil\log_2(B+1)\rceil+\lceil\log_2(S+1)\rceil
+\lceil\log_2(W_{\lambda})\rceil+\lceil\log_2(L+2)\rceil.
\tag{C.1}
\]

Both algorithms return \(N/(vQ_{\rm row}^4)\). In exact integer arithmetic, with schoolbook multiplication and division, every probability, curve and threshold intermediate has magnitude at most
\[
W_{\lambda}Q_{\rm row}^4.
\tag{C.2}
\]

Coordinates and counters use \(O(\log B+\log(L+2))\) bits. Consequently each path costs
\[
O\!\left((\ell^4\log(L+2)+\log B)\kappa^2\right)
\quad\text{bit operations},\qquad
O(\ell^3\kappa)\quad\text{storage bits}.
\tag{C.3}
\]

This includes input validation, curve construction, preparation and output. The explicit vector length bounds \(L\); \(L=0\) is allowed. Constructing the finite law and transcendental threshold enclosures is charged separately in C.5.

\subsection{Probability rows and testing curves}

All row probabilities have denominator \(Q_{\rm row}\), with nonnegative numerators satisfying
\[
\sum_f p_f=\sum_f q_f=Q_{\rm row}.
\tag{C.4}
\]

The producer's two mixture terms and the auditor's floor-preimage counts partition this mass. Thus all atoms and partial row sums are at most \(Q_{\rm row}\). Sorting likelihood ratios, including zero-probability axes, produces echo supports in \([0,Q_{\rm row}]^2\).

One-participation joint experiments have mass \(Q_{\rm row}^2\). The auditor accumulates their atoms directly. In the producer, echo edges have increasing first and decreasing second coordinates, so each coordinate's increments telescope to at most \(Q_{\rm row}\). Scaling by the first row gives
\[
\sum_f p_fQ_{\rm row}=\sum_f q_fQ_{\rm row}=Q_{\rm row}^2.
\tag{C.5}
\]

Every Minkowski vertex, equal-slope group and partial accumulator is therefore bounded by \(Q_{\rm row}^2\), without a factor for the number of outputs. Envelopes select supports from the union of alternatives and never add masses across preparations. Hull comparisons multiply coordinate differences, giving bound \(Q_{\rm row}^4\). Auditor crossings are stored as pairs of such differences; comparing two crossings also costs products of size at most \(Q_{\rm row}^4\), with no accumulated rational denominators.

\subsection{Threshold-weighted queries and accumulators}

Root-output weights are \(x=vp_f,y=uq_f\). After the next row they are at most \(vQ_{\rm row}^2,uQ_{\rm row}^2\); evaluating a one-participation support uses products at most \(vQ_{\rm row}^4,uQ_{\rm row}^4\). Signed comparisons obey the same sum bound.

On the three-transition integer scale, the remaining value satisfies
\[
0\le\widehat F(x,y)\le(x+y)Q_{\rm row}^3,\qquad
\sum_f\widehat F(vp_f,uq_f)\le(u+v)Q_{\rm row}^4.
\tag{C.6}
\]

For a one-participation branch, the first inequality is \(\sum_j\widehat G_1(xp_j,yq_j)\le Q_{\rm row}^2\sum_j(xp_j+yq_j)\). Immediate participation uses the echo scale and a final factor \(Q_{\rm row}\); stopping uses \(\max(x,y)Q_{\rm row}^3\). Nonnegative partial sums inherit the bound. Root maximization adds no scale. Subtracting the included stopping term \(uQ_{\rm row}^4\) leaves \(0\le N\le vQ_{\rm row}^4\), proving (C.2) for temporaries as well as the output.

\subsection{State arithmetic, operation counts, and storage}

Each path has \(O(\ell)\) first controls, \(O(\ell)\) second controls, \(O(\ell^2)\) reached phase pairs and \(O(\ell)\) echo phases. The auditor's second candidates reduce to the interval of integers \(e_f-m\) in A.4, so they also number \(O(\ell)\) per first control. Echo curves have \(O(\ell)\) lines; one-participation construction has \(O(\ell^3)\) lines before pruning and \(O(\ell^3\log(L+2))\) operations. Streaming \(O(\ell^2)\) roots, each with \(O(\ell)\) outputs and \(O(\ell)\) binary curve queries, costs \(O(\ell^4\log(L+2))\).

Producer phase expressions are \(O(B(L+1))\). The auditor's representatives lie in \([0,B^2)\), and its floor products are bounded by \(O(B^3+B(L+4)+L+4)\). These quantities have \(O(\kappa)\) bits; Euclidean inversion adds \(O(\log B)\) arithmetic steps. Charging \(O(\kappa^2)\) per operation and storing at most \(O(\ell^3)\) integer cells proves (C.3), including input/\allowbreak{}output and optional final fraction reduction. These mathematical bounds use arbitrary-precision coordinates; executable range restrictions are recorded with the implementation.

\subsection{Connecting the finite input to a final DG error}

Let \(\varepsilon\ge0\), and let rational endpoints satisfy \(\lambda_-\le e^\varepsilon\le\lambda_+\). Equation (28) gives interval width at most
\[
2(1+\lambda_+)\{1-(1-\tau)^4\}+\lambda_+-\lambda_-
\le8(1+\lambda_+)\tau+\lambda_+-\lambda_-.               \tag{C.7}
\]

For a rational target \(0<\zeta\le1\), choose \(\lambda_+-\lambda_-\le\zeta/2\) and \(0<\tau\le\zeta/[16(1+\lambda_+)]\). Assume \(\sigma\ge1\). Choose
\[
L\ge\left\lceil\sigma\sqrt{2\log(4\sigma/\tau)}\right\rceil,
\qquad S\ge\left\lceil8L/\tau\right\rceil.
\tag{C.8}
\]

The normalizer is at least one and the density decreases beyond zero, so
\[
\sum_{|n|>L}g_\sigma(n)
\le2\int_L^\infty e^{-x^2/(2\sigma^2)}\,dx
\le\frac{2\sigma^2}{L}e^{-L^2/(2\sigma^2)}.             \tag{C.9}
\]

Taking \(L\ge\lceil\sigma\sqrt{2\log(4\sigma/\tau)}\rceil\) makes \(L\ge\sigma\) and the last expression at most \(\tau/2\). For \(n=1,\ldots,L\), obtain a certified rational (for example dyadic) interval \([l_n,h_n]\) containing \(g_\sigma(n)\), of width at most \(1/S\), and use \(l_n^+=\max(0,l_n)\). Set
\[
w_n=w_{-n}=\lfloor S l_n^+\rfloor,\qquad
w_0=S-2\sum_{n=1}^Lw_n.                                \tag{C.10}
\]

These weights are nonnegative and normalized because \(w_n/S\le g_\sigma(n)\). Each nonzero retained atom loses less than \(2/S\). All removed mass moves to zero, so its total is exactly the total-variation distance. Consequently
\[
\operatorname{TV}(\nu,g_\sigma)
\le\sum_{|n|>L}g_\sigma(n)+4L/S\le\tau
\quad\text{if }S\ge\lceil8L/\tau\rceil.                 \tag{C.11}
\]

The lower-enclosure construction remains effective when \(Sg_\sigma(n)\) is arbitrarily close to an integer. The evaluated generator instead certifies a unique floor at its chosen precision, reporting unresolved cases explicitly.

Choosing dyadic threshold endpoints of gap at most \(\zeta/2\) permits denominator bit length \(O(1+\log(1/\zeta))\) and numerator bit length \(O(1+\varepsilon+\log(1/\zeta))\). For \(\lambda_+\le e^\varepsilon+1\), choose \(\tau=\zeta/[16(1+\lambda_+)]\) and take \(L,S\) to be the ceilings of the sufficient bounds above (or constant-factor enlargements). Then
\[
\begin{aligned}
L&=O\!\left(\sigma\sqrt{\log(\sigma+1)+\varepsilon+\log(1/\zeta)}\right),\\
\log S&=O(\log(L+1)+\varepsilon+\log(1/\zeta)).
\end{aligned}           \tag{C.12}
\]

The total query cost is the sum of Gaussian and exponential enclosure construction, the finite computation in (C.3), error transfer, and the chosen serialization. The enclosure cost depends on the representation of \(\sigma\), and \(L\) depends on its numerical value. The polynomial guarantee in Theorem 3 applies to the explicit finite rational input.

\section{Finite deadlines and numerical controls}

This appendix concerns the first-order subfamily \(m=1,0<a_1=a<B\), with \(\gamma=a/B\).

\subsection{A complete physical-horizon witness}

Fix the reference parameters and the finite law \(q\) of radius 24 and scale \(10^{24}\), with \(\operatorname{TV}(q,g_3)\le\tau\). Its common-zero-input kernel \(Q_k^q(n)\) leaves \(\mathcal K=[-36,36]\cap\mathbb Z\) invariant: for \(k\in\mathcal K\), the rounded feedback lies in \([-12,12]\) and the innovation in \([-24,24]\). All three original U values are summed.

Let \(P_k^q,Q_k^q\) be the complete three-release laws starting synchronized at \(k\), for inputs \((0,0,0)\) and \((1,1,0)\). At threshold \(\lambda_-\), define
\[
\begin{aligned}
h(k)&=H_{\lambda_-}(P_k^q\Vert Q_k^q),\qquad V_0(k)=h(k),\\
V_s(k)&=\max\{h(k),\sum_nQ_k^q(n)V_{s-1}(n)\}.
\end{aligned}
\tag{D.1}
\]

The two alternatives start the fixed suffix now or observe one more common release. The prefix laws agree and expose the synchronized state, so this is a public stopping rule. At the waiting deadline it runs the suffix from the actual state, retaining every timeout branch. The terminal event is the likelihood test for the complete suffix conditional on that start. Common-input padding makes the horizon exactly \(T=s+3\); the event ignores the padding. The waiting recursion uses actual integer states, even though the suffix laws have a finite translation period.

Round \(h(k)\) and each expectation down to multiples of \(1/G\), with \(G=10^{70}\), producing \(\underline V_s\). Normalization of each row and the fact that a maximum is nonexpansive give inductively
\[
0\le V_s(k)-\underline V_s(k)\le(s+1)/G.
\tag{D.2}
\]

The stored maximizing choices themselves implement a policy with payoff at least \(\underline V_s(k)\): a stopping choice uses the exact terminal test, and a waiting choice averages child policies whose values dominate their stored lower values. The constructed policy therefore attains the certified lower bound.

Outside \(\mathcal K\), extend the policy by starting the suffix immediately; define its event on all integer transcripts by the same likelihood comparison, including zero masses. Coupling innovations across all \(T\) physical releases, with common U draws until mismatch, bounds each world's transcript TV by \(T\tau\). Therefore the explicit event obeys
\[
H_{e^2}(P_T\Vert Q_T)\ge
\max\{0,\underline V_{T-3}(0)-(\lambda_+-\lambda_-)
-(1+\lambda_+)T\tau\}.
\tag{D.3}
\]

The threshold term uses \(Q(A)\le1\). Here the coupling covers all physical releases, including waiting; A.6 charges the informative graph's participating and echo releases.

\par\smallskip\begingroup
\small
\setlength{\tabcolsep}{3pt}
\renewcommand{\arraystretch}{1.10}
\begin{tabularx}{\linewidth}{@{}>{\hsize=1.000000\hsize\linewidth=\hsize\raggedleft\arraybackslash}X>{\hsize=1.000000\hsize\linewidth=\hsize\raggedleft\arraybackslash}X>{\hsize=1.000000\hsize\linewidth=\hsize\raggedleft\arraybackslash}X@{}}
\toprule
\textbf{Total releases \(T\)} & \textbf{Certified DG event lower bound} & \textbf{Ratio to the upper endpoint (B.7)} \\
\midrule
3 & 0.000082848959100245135254 & 0.982647432993 \\
44 & 0.000084231758498895252560 & 0.999048414903 \\
100 & 0.000084311323694656676833 & 0.999992114573 \\
\bottomrule
\end{tabularx}
\endgroup\smallskip\par

Both lower bounds and ratios are rounded downward. At \(T=44\), the correction in (D.3) is at most \(8.6789243169\times10^{-14}\). Thus 44 releases suffice for more than 99.9 percent of the all-horizon supremum. This is a concrete witness, without a minimal-horizon or globally optimal finite-policy claim. The artifact retains the complete stopping table and an independent reconstruction of all suffix atoms and stopping decisions.

\subsection{Complete-transcript accounting controls}

For \(\gamma=1/2,\sigma=2,T=C=3,\varepsilon=5\), the first complete block starts at zero. Each later length-\(b\) block uses a common testing envelope over \(k\bmod2^b\), \(d\in\{-1,0,1\}\), and all legal adaptive continuations. Its dominating pair bounds every conditional history at every nonnegative threshold, as required by [21, Theorem 10]. All partitions share one finite source and one transfer charge for the three physical draws.

\noindent\begin{minipage}{\linewidth}
\par\smallskip\begingroup
\small
\setlength{\tabcolsep}{3pt}
\renewcommand{\arraystretch}{1.10}
\begin{tabularx}{\linewidth}{@{}>{\hsize=1.000000\hsize\linewidth=\hsize\raggedright\arraybackslash}X>{\hsize=1.000000\hsize\linewidth=\hsize\raggedleft\arraybackslash}X@{}}
\toprule
\textbf{Retained physical block lengths} & \textbf{Certified DG upper} \\
\midrule
\(1+1+1\) & \(0.000101103642523998\) \\
\(1+2\) & \(0.000065716695011075\) \\
\(2+1\) & \(0.000064105386413249\) \\
Full three-step optimum & \(0.000039614602845959\) \\
\bottomrule
\end{tabularx}
\endgroup\smallskip\par
{\small Table S1. Certified DG upper bounds under the same physical contract. Either two-step partition certifies \(10^{-4}\); only the reported full computation certifies every target in \([4\times10^{-5},6\times10^{-5}]\). An independent complete-transcript consumer checks the full strategy optimum. Physical block length and retained preparation depth are different quantities.}
\end{minipage}\par

\subsection{Why two coordinates do not suffice for C3}

For \(b(k)=-\lfloor ak/B\rfloor\), shifting \(k\) by \(B^2\) preserves both stored phases but gives
\[
b(k+B^2)=b(k)-aB,\qquad b(b(k+B^2))=b(b(k))+a^2.
\tag{D.4}
\]

The next phase \(\{ab(b(k))/B\}\) consequently changes by \(a^3/B\) modulo one. This change can affect the testing value. With \(\gamma=1/3\) and \(q(-1),q(0),q(1)=(1,2,1)/4\), literal hidden-U summation gives
\[
\begin{aligned}
F_2^+(0;1,2)&=4723/1728,\qquad F_2^+(9;1,2)=2351/864,\\
F_2^+(0;1,2)-F_2^+(9;1,2)&=7/576.
\end{aligned}
\tag{D.5}
\]

Here \(F_2^+\) is the regenerated correct-testing value with previous difference +1 and two participations left: total budget three including the first participation. The states share \((r,s)=(0,0)\), and the weights occur up to scale \(1/4\) after a first participation at threshold one. The recurrence covers stopping, echo-and-preparation, and either immediate sign, each chosen before its output sum. Richer carry coordinates distinguish the states that this representation identifies.

The hockey-stick continuation is \(H_{2,1}(k;x,y)=F_2^+(k;x,y)-y\). Thus \(H_{2,1}(0;1,2)=1267/1728\) and \(H_{2,1}(9;1,2)=623/864\): both values decrease by two, preserving the gap \(7/576\).

\section{Selective commitment and weighted certificates}

Sections E.1--E.2 and E.4--E.5 give the first-order fixed-cut construction and certificates, with \(m=1,0<a_1=a<B\) and \(\gamma=a/B\). Section E.3 proves Theorem 8 for general finite informative rank. Appendix G treats the analyst minima and convex closure of the recursive hierarchy. The finite innovation law is normalized on \([-L,L]\), with integer scale \(S\) and row denominator \(BS\). Actions precede their outputs; stopping contributes \(\max(x,y)\).

\subsection{Complete preparation words and carry propagation}

Put \(b(k)=-\lfloor ak/B\rfloor\). An integer reference base \(k\) determines the digit word \(\mathbf z=(z_0,\ldots,z_{q-1})\), where \(z_i=ab^i(k)\bmod B\). Such words bijectively encode \(k\bmod B^q\). To see this inductively, adding \(tB^j\) to \(k\) leaves the first \(j\) digits unchanged, while changing digit \(j\) by \(a(-a)^jt\bmod B\). Coprimality makes this a permutation of the final digit. Hence every finite digit word has an integer representative.

A pending state stores \((\mathbf z,e,d,c)\): a committed word, an integer offset, the preceding signed difference, and the remaining participation budget. Its physical reference state is \(k+e\); the other state is \(k+e+d\). For current action \(j\), express the next common reference state as \(b(k)+n\). With \(R_d=z_0+a(e+d)\), write \(R_d=Bv_d+r_d\), \(0\le r_d<B\). The complete row is
\[
\begin{aligned}
q(n)&=(1-r_0/B)\nu(n+v_0)+(r_0/B)\nu(n+v_0+1),\\
p(n)&=(1-r_d/B)\nu(n+j+v_d)+(r_d/B)\nu(n+j+v_d+1).
\end{aligned}
\tag{E.1}
\]

For \(j\ne0\), the next offset is \(n\), the new difference is \(j\), and the first committed digit is consumed. For \(j=0\), the states synchronize and the remaining word is released. These rows follow by counting the original hidden uniform variable, including negative states and zero masses.

The offset can be removed while transforming the entire remaining word. Set \(c_{-1}=-e\) and, successively,
\[
c_i=\left\lfloor\frac{z_i-ac_{i-1}}B\right\rfloor,
\qquad z_i'=z_i-ac_{i-1}-Bc_i\in\{0,\ldots,B-1\}.
\tag{E.2}
\]

The games with pending states \((\mathbf z,e,d,c)\) and \((\mathbf z',0,d,c)\) are related by a common deterministic translation of their observations. Indeed, if the offsets before step \(i\) satisfy \(e_i'=e_i+c_{i-1}\), then in either world
\[
z_i'+a(e_i'+d_i)=z_i+a(e_i+d_i)-Bc_i,
\qquad n_i'=n_i+c_i.
\tag{E.3}
\]

Here \(d_i\) is the preceding difference in the first world and zero in the reference world. The same fresh innovation and raw \(U\) establish the output identity. A nonzero action preserves the offset relation for the next step; an echo returns both games to synchronization. These translations depend on the already committed word and offset, and are bijections of the observation labels. Translating policies and events in both directions proves equality of the two testing games.

At synchronization, convexity in the testing weights purifies the first row exactly as in (A.3). With \(c\) participations available, choose the complete \(c\)-digit word before the pure first participation, whose probabilities are \(\nu(f+j),\nu(f)\). Every such word is legal: a physical synchronized state \(B\ell\) has pure first-row phase and reference successor base \(-a\ell\); multiplication by \(-a\) permutes residues modulo \(B^c\). An uninterrupted run has at most \(c-1\) further participations and a final echo, so the \(c\) committed digits suffice. This defines \(G_C\).

For discrete-Gaussian noise, A.1 supplies each selected synchronized integer representative by public waiting. There are at most \(C\) waits, each finite almost surely. At a deterministic cutoff, rejecting every unfinished history gives convergence of both event probabilities to the informative-game probabilities. Thus the full-commitment DG value equals the actual all-finite-horizon supremum. For a finite innovation law, synchronized preparation remains part of the comparison game's definition.

\subsection{Inclusion, refinement, and sparse graph size}

Consider a full-word policy at a synchronized node. A policy in the cut-set game stores that word in its strategy memory, chooses its prescribed first prefix, and at every permitted renewal exposes the prescribed remaining prefix after the observed carries. The entire digit alphabet is available at each renewal. Equation (E.3) preserves both world probabilities, and no prefix is changed before it is consumed. This simulates the complete policy and event, proving \(G_C\le U_{\mathcal S}\).

If \(\mathcal S_0\subseteq\mathcal S_1\), each block allowed by \(\mathcal S_0\) is a union of blocks allowed by \(\mathcal S_1\), including runs truncated by an echo. The finer policy can retain the coarse block and expose the corresponding subblocks. The same simulation proves \(U_{\mathcal S_0}\le U_{\mathcal S_1}\). With no cuts, no additional renewal occurs before synchronization, giving equality with \(G_C\). With every cut, each block has one digit. This proves (32).

For \(C\ge1\), let \(q=C-1-|\mathcal S|\) count the removed cuts. Normalize all pending offsets by (E.2) before state enumeration. When the removed cuts have no adjacent elements, a pending word has length one or two. There are at most \(C+1\) synchronized keys, \(2|\mathcal S|\) renewal keys, and \(2CB\) one-digit pending keys. A two-digit word can start at level \(r\) only when cut \(r-1\) was removed, so at most \(2qB^2\) two-digit keys are required. This gives
\[
N\le C+1+2|\mathcal S|+2CB+2qB^2,
\qquad
E=O\!\left((L+1)(CB+qB^2)\right)
\tag{E.4}
\]

Summing root prefix choices gives \(2[(C-q)B+qB^2]\), including both signs. Renewal choices have the same order, and each pending state has at most three complete actions. Each normalized row has \(O(L+1)\) outputs. Therefore all action-output branches number \(O((L+1)(CB+qB^2))\). This is a structural count; constructing and evaluating the rows additionally uses the source and integer arithmetic.

Synchronized nodes have rank \(2c\), and pending or renewal nodes have rank \(2c+1\). A renewal combines the new prefix with its current action in one complete row, so every probabilistic edge descends in rank. There are at most \(2C\) informative innovations. Coupling two innovation laws of total-variation distance at most \(\tau_\nu\) therefore changes each world's event probability by at most
\[
d_C=1-(1-\tau_\nu)^{2C}.
\tag{E.5}
\]

For a fixed policy and event, the hockey-stick payoff changes by at most \((1+\lambda)d_C\). Taking the corresponding suprema preserves the bound. Whole-policy inclusion, monotonicity in \(\lambda\), and the directed threshold enclosure give the fixed-cut analogue of (28). For a finite-game lower \(\underline\delta_{\nu,\mathcal S}(\lambda_+)\), the same argument yields the comparison-game lower
\[
\mathscr D_{\mathcal S}(\varepsilon)\ge
\max\{0,\underline\delta_{\nu,\mathcal S}(\lambda_+)
-(1+\lambda_+)d_C\}.
\tag{E.6}
\]

The uniform two-digit benchmark starts a new length-two partition at every synchronization. The order in (32) is defined by nested global cut sets; its proof applies to those sets directly.

\subsection{Weighted error, refinement depth, and sample count}

Fix a complete-row comparison graph of maximum informative rank \(D\), with rank decreasing on every probability edge. For the finite-memory cut games, (23) gives \(D=(m+1)C\); prefix choices are combined with their next complete row. Let \(f_v(t)=V_v(t,1-t)\). These curves are convex, lie between \(\max(t,1-t)\) and one, and equal one at both axes. For each complete normalized row, omit zero-mass branches and write
\[
\sum_o m_ow(t_o)=\sum_o\min(tp_o,(1-t)q_o)
\le\min(t,1-t)=w(t).
\tag{E.7}
\]

Hence an error of at most \(e w\) in every child remains at most \(e w\) after the observation sum. Maximization over complete current actions preserves that bound.

At one node, fix all child lower envelopes. Its exact lower-backup oracle \(F\) is a maximum of lines \((At+B_0(1-t))/(BS)^h\), with integer coefficients between zero and \((BS)^h\). A selected line is the event pair of one feasible policy and globally supports \(F\). Its slope lies in \([-1,1]\), and selected supporting slopes are nondecreasing with the prior.

At depth \(k\), let \(a_k=2^{-k}\). For an interior cell \([ja_k,(j+1)a_k]\), define \(q_j=\min(j,2^k-j-1)\ge1\). Throughout the cell, \(w\ge q_ja_k\). If its endpoint support slopes differ by \(\Delta s\), the chord-to-two-support gap is at most \(a_k\Delta s/4\): the maximum is at the intersection of the endpoint supports, and the bound follows from the product of its two distances to the endpoints. A failed weighted test therefore requires \(\Delta s>4\xi q_j\).

Disjoint failed cells share a total support-slope variation of at most two. Each positive distance index occurs at most twice. If there are \(m_k\) failed interior cells, then
\[
\frac{m_k^2}{4}\le\sum_{j\,\mathrm{failed}}q_j
<\frac1{2\xi},
\qquad m_k<\sqrt{2/\xi}.
\tag{E.8}
\]

At most two more cells touch the axes. Each split adds one midpoint, giving at most \(2+\lceil\sqrt{2/\xi}\rceil\) new samples per depth.

A proper crossing of two distinct integer-coefficient lines is at distance at least \(1/(2(BS)^h)\) from either axis. Thus \(F\) is affine in the corresponding endpoint neighborhoods. On a cell entirely inside either affine region, the two-support gap is zero, including when an axis has a tied support. Once \(a_k\le1/(4(BS)^h)\), any cell with a nonzero gap is at distance at least \(1/(4(BS)^h)\) from its nearest axis; its absolute gap is at most \(a_k/2\). Its weighted gap is consequently at most \(2(BS)^ha_k\). The depth \(K_h\) in (E.9) makes this at most \(\xi/2\). Starting from samples \(0,1/2,1\) and summing the per-depth split bound gives
\[
M_h\le3+K_h\left(2+\left\lceil\sqrt{2/\xi}\right\rceil\right),
\qquad
K_h=\left\lceil\log_2(4(BS)^h/\xi)\right\rceil.
\tag{E.9}
\]

At a sampled prior, round the upper advantage above \(\max(t,1-t)\) upward in units of \(w(t)/\mathsf Q\). This costs at most \(w(t)/\mathsf Q\). On an interval, convexity bounds the true value by its endpoint chord. The local chord-to-lower gap is at most \(\xi w\), and concavity of \(w\) bounds interpolation of the endpoint errors. Thus
\[
e_v\le\max_{u\,\mathrm{child}}e_u+\xi+1/\mathsf Q,
\qquad e_{\mathrm{terminal}}=0.
\tag{E.10}
\]

\[
\begin{aligned}
\overline f(t)-\underline f(t)&\le D(\xi+1/\mathsf Q)w(t),\\
(1+\lambda)\{\overline f(t_\lambda)-\underline f(t_\lambda)\}
&\le D(\xi+1/\mathsf Q),
\qquad t_\lambda=(1+\lambda)^{-1},\quad\lambda\ge1.
\end{aligned}
\tag{E.11}
\]

Each edge propagates weighted error through (E.7) and adds at most one local allowance. Rank induction gives the first inequality for any memory dimension. The conversion uses \((1+\lambda)w((1+\lambda)^{-1})=\min(1,\lambda)\), proving the second part. Axis values are exact; the local weighted tests use the one-sided limits at an axis, the intersection of endpoint supports, and \(t=1/2\). Between these points the ratio of two affine functions is monotone or constant, so these checks cover the whole interval.

The consumer enforces the sample and depth bounds as well as the whole-action and quantization obligations. For a verified graph, numerical work also includes all branch backups, child-envelope queries, input bytes, and integer arithmetic; event coefficients at a rank-\(h\) oracle use \(O(h\log(BS))\) bits. Equation (E.9) bounds samples per fixed oracle within that computation.

\subsection{Feasible minorants and exact testing-value aggregation}

The pointwise lower path stores pairs \((a,b)\) satisfying, for one common policy \(\pi\) and event \(A\),
\[
0\le a/\mathsf Q\le P_\pi(A),
\qquad0\le b/\mathsf Q\le Q_\pi(A^c).
\tag{E.12}
\]

Fix a complete current action and, after each output, choose one whole child pair with its witness. Summing the two coordinates with their respective branch probabilities and rounding each sum downward preserves (E.12). Whole candidate pairs compete by testing score. A zero prior-weighted branch may use an empty or full continuation event to give zero aggregate coordinates. Induction also bounds each coordinate's rounding deficit relative to the chosen witness by \(h/\mathsf Q\) at rank \(h\).

Every saved pair defines a valid line at every later prior. Its hockey-stick query is therefore bounded below by \(\max\{0,\max_{(a,b)}(a/\mathsf Q+\lambda b/\mathsf Q-\lambda)\}\). This uses supporting lines throughout, retaining the same policy in both coordinates.

The graph classes reported in Section 7 use an exact aggregation. Within a complete action, group outputs only when they have the same child class and proportional likelihood vectors: \((p_i,q_i)=g_i(p_*,q_*)\). Homogeneity gives
\[
\sum_i V(xp_i,yq_i)=V\!\left(x\sum_i p_i,y\sum_i q_i\right).
\tag{E.13}
\]

Induction over rank then permits identical sets of complete action signatures to share one class. A lower witness lifts by choosing a matching concrete action and the same child witness on every branch in its likelihood group. This preserves the testing value for every prior. The independent consumer reconstructs the physical rows from raw-U interval counts and checks their complete action sets against the submitted classes before checking numerical tables. Each complete signature includes the likelihood vector, which records the information revealed about the world.

\subsection{The certified decision records}

\noindent\begin{minipage}{\linewidth}
\par\smallskip\begingroup
\footnotesize
\setlength{\tabcolsep}{3pt}
\renewcommand{\arraystretch}{1.10}
\begin{tabularx}{\linewidth}{@{}>{\hsize=0.380000\hsize\linewidth=\hsize\raggedleft\arraybackslash}X>{\hsize=1.800000\hsize\linewidth=\hsize\raggedright\arraybackslash}X>{\hsize=0.550000\hsize\linewidth=\hsize\raggedleft\arraybackslash}X>{\hsize=0.450000\hsize\linewidth=\hsize\raggedleft\arraybackslash}X>{\hsize=1.650000\hsize\linewidth=\hsize\raggedleft\arraybackslash}X>{\hsize=1.170000\hsize\linewidth=\hsize\raggedleft\arraybackslash}X@{}}
\toprule
\textbf{\(C\)} & \textbf{Comparison} & \textbf{Classes} & \textbf{\(\varepsilon\)} & \textbf{DG bound \((\times10^{-5})\)} & \textbf{Cold cost (s)} \\
\midrule
8 & All cuts & 40 & 5 & \(\ge10.78232\) & --- \\
8 & Remove cut 1 & 45 & 5 & \(\le9.96722\) & 21.526 \\
8 & Remove cuts 1, 6 & 50 & 5 & \(\le9.30794\) & 5.240 \\
8 & Uniform two digits & 99 & 5 & \(\le8.42898\) & 8.621 \\
16 & All cuts & 80 & 8 & \(\le6.21318\) & --- \\
32 & Selected cuts & 200 & 12 & \(\le9.98828\) & --- \\
64 & Selected cuts & 475 & 19 & \(\le9.15396\) & --- \\
\bottomrule
\end{tabularx}
\endgroup\smallskip\par
{\small Table S2. Outward bounds at \(\gamma=1/3,\sigma=3,\delta=10^{-4}\). C8 cold medians cover three serial repetitions of fixed plans, including failed refinements. Final grids use 2947 prior points for the one-cut model and 739 for the others. Rows C16--64 give upper certificates with the source and precision settings in Appendix E.5.}
\end{minipage}\par

The C8 certificates bind the public parameters, cut rule, rational innovation law, threshold interval, graph classes and numerical payload. Define \(\mathcal S_{\rm all}=\{0,\ldots,6\}\), \(\mathcal S_* =\mathcal S_{\rm all}\setminus\{1\}\) and \(\mathscr D_{\mathcal S}(\varepsilon)=U_{\mathcal S}^{\rm DG}(1,e^\varepsilon)-e^\varepsilon\). Independent source reconstruction and separate upper-table and lower-pair consumers give
\[
\mathscr D_{\mathcal S_*}(5)\le0.000099672197505791<10^{-4},\qquad
\mathscr D_{\mathcal S_{\rm all}}(5)\ge0.000107823264112084>10^{-4}.
\tag{E.13a}
\]

The one-cut comparison also has lower bound \(0.000099474142144503\), rounded downward. Whole-policy inclusion and informative-source transfer establish these statements for the DG mechanism.

The unique all-cut model has comparison capacity strictly above \(10^{-4}\); deleting cut 1 produces capacity strictly below it. These two strict inequalities prove the minimum-deletion statement for the specified global-cut family. Table S2 uses the same query to compare certification costs. All failed mesh attempts are included, and the selected configurations are held fixed during the three rotated repetitions.

C8 cold cost includes source generation, graph construction, serialization, solving, independent checks, query evaluation and the common 2.483-second compile charge. Nine decisions use 15 mesh requests; the one-cut model first fails on two coarser meshes. Larger-budget selected/\allowbreak{}uniform-two-digit class counts are 80/\allowbreak{}211, 200/\allowbreak{}435 and 475/\allowbreak{}883. At C32, removed cuts are one modulo four; at C64 they are odd. Neither set has adjacent cuts.

The full directed upper endpoints behind Table S2 are
\[
\begin{array}{c|c|l}
C&\varepsilon&\text{DG upper endpoint}\\\hline
16&8&0.000062131798629194\\
32&12&0.000099882780447748\\
64&19&0.000091539571900235
\end{array}
\tag{E.14}
\]

Each endpoint is below \(10^{-4}\). The delivered package retains the complete C8 certificates and the larger-budget generation records; its final extracted replay includes regeneration of the C64 result. Probability sources and numerical meshes are recorded per request. The numerical endpoints are sufficient upper certificates for the original mechanism through whole-policy inclusion and informative-innovation coupling.

\section{Finite-memory exact compilation and source transfer}

We prove Theorem 4 under (1), for a normalized rational law \(\nu(z)=w_z/S\) on \([-L,L]\). The law may be asymmetric and may have zero atoms. Preparations are free only at synchronized nodes of the comparison game. Appendix A.1 connects this game to full-support physical sources.

\subsection{Complete-word realization and vector carry}

Let \(\mathbf b\) and \(M\) be defined as in Section 4.2. Mathematical floors commute with integer translations, giving \(\mathbf b(s+B^qv)=\mathbf b(s)+B^{q-1}Mv\). Changing the initial register by \(B^tu e_1\) preserves the first \(t\) digits. Modulo \(B\), only the first row of \(M\) can be nonzero, so its next phase changes by \((-1)^t a_1^{t+1}u\). This coefficient is invertible. Starting at zero, choose one unique \(u_t\in\{0,\ldots,B-1\}\) at each digit and add \(B^t u_t e_1\). Induction constructs every desired length-\(H\) word, with representative \((k,0,\ldots,0)\), \(0\le k<B^H\). This proves the word bijection and its bit bound.

For integer displacement \(u_i\), (22) implies
\[
\phi_i+a^\top(w_i+u_i)=\theta_i+a^\top w_i-Bk_i.
\tag{F.1}
\]

With the same innovation and rounding coin, the new first offsets differ by \(k_i\); the remaining offsets shift. The relation is simultaneous in both worlds. It yields a common deterministic bijection on complete histories, along which causal actions and terminal events transport in both directions. Taking \(u_0=-w\) normalizes the current offset; taking \(u_0=0\) normalizes closed phase endpoints. Given a retained prefix and an output, the map on unread digits is triangular translation modulo \(B\), so every compatible completion is preserved.

At synchronization, fix a whole future word and a causal policy/\allowbreak{}event. Its score is affine in the first phase on \([0,B]\); optimizing policies makes it convex. An endpoint is optimal. The endpoint \(B\) carries to phase zero together with a transformed complete suffix, which the word construction realizes. Hence a pure first row is exact, with its sign and all subsequent digits selected before that output.

For completeness, a separate sufficient condition is \(\gcd(a_m,B)=1\). Suppose a register \(r\) realizes a length-\(H\) suffix. Replacing it by \(r'=r+B^Ht e_1\) keeps that suffix. A predecessor \(s=(r'_2,\ldots,r'_m,s_m)\) with first phase \(\theta\) solves
\[
a_m s_m=\theta-Br'_1-\sum_{i<m}a_i r'_{i+1}.
\tag{F.2}
\]

Choose \(t\) from the congruence modulo \(|a_m|\); \(B^{H+1}\) is invertible there. For \(|a_m|=1\), use \(t=0\). The resulting floor gives \(\mathbf b(s)=r'\). Induction realizes the full word, and \(S_{H+1}\le2BS_H+2B^{H+2}\) bounds its coordinate bit length by \(O_m(H\log B)\). The main statements and executed certificate contracts use the first-tap condition in (1).

\subsection{Exact rows, synchronization and finite support cover}

Counting \(U\) in its two rounding intervals gives exactly (21), including negative floors, zero masses and nonsymmetric laws. The difference update is (4). The number \(h(d)\) is the number of zero actions needed to flush the most recent nonzero difference. At a nonsynchronized node a zero action decreases it by one; a nonzero action spends a participation and sets it to \(m\). Both ranks in (23) therefore decrease on each such edge. The synchronized pure first output reduces \(\mathcal R\) by one and leaves \(\mathcal K=m(c-1)+m=mc\), so at most \(mc\) digits and \((m+1)c-1\) informative outputs remain.

Let \(F(c,d,\vartheta,w;x,y)\) be the pending value for a precommitted word. For each allowed action \(j\), sum complete rows, then maximize over actions and stopping. The child is \(G_{c-|j|}\) when \(d_j=0\), and otherwise the pending value with the unchanged tail, new offset and difference. The synchronized root is
\[
G_c(x,y)=\max\left\{\max(x,y),\
\max_{j=\pm1,\vartheta}\sum_f
F(c-1,je_1,\vartheta,fe_1;x\nu(f+j),y\nu(f))\right\}.
\tag{F.3}
\]

The word maximum is outside the sum. Rank induction proves that every legal policy/\allowbreak{}event is bounded by this recursion. Selecting complete maximizing actions and child events gives the reverse direction. At synchronization each output branch has equal registers and may start a new preparation, which proves (24). Before synchronization the whole unused word remains shared.

After a pure first output, an own-world offset begins at \((\eta,0,\ldots,0)\). On the continuous phase extension \([0,B]\), an offset box of radius \(R_t\) has successor radius at most \(L+rR_t+2\le L+R_t+2\). Over at most \(mC+1\) steps this is \(O_{m,C}(\ell)\), where \(\ell=L+3\). At position \(t\), use the integer breakpoint set
\[
T_t=\{0,B\}\cup\{(-a^\top w)\bmod B:w\text{ in the own-offset box}\},
\qquad |T_t|=O_{m,C}(\ell^m).
\tag{F.4}
\]

A fixed complete policy/\allowbreak{}event, indexed by common offset histories, has a multiaffine score on each product cell: every path uses each digit once, and each row is affine between its floor boundaries. Subsequent choices after synchronization belong to the fixed policy and no longer depend on the initial word. Taking the policy supremum gives separate convexity. Moving each coordinate to a cell endpoint cannot lower the optimum. Complete carry normalizes those vertices to legal words. Legality gives the reverse inequality, so the cover is exact.

At each position one may use all \(B\) digits or the smaller critical set, giving at most
\[
\prod_{t=1}^{mc}\min\{B,|T_t|\}
\le\min\{B,O_{m,C}(\ell^m)\}^{mc}
\tag{F.5}
\]

preparations. A branch with only one world of positive mass is already perfectly distinguishable and returns \(\max(x,y)\).

\subsection{Complete flush curves, prefix counts and bit complexity}

At zero remaining budget, \(h(d)\le m\) common outputs remain. For their complete joint row \((P_{\mathbf f},Q_{\mathbf f})\), the exact continuation is
\[
E(x,y)=\sum_{\mathbf f}\max(xP_{\mathbf f},yQ_{\mathbf f}).
\tag{F.6}
\]

Stopping earlier is represented by a cylinder event, so full flushing dominates it. Each curve has \(O(\ell^m)\) leaves. Its raw suffix has at most \(m\) positions with \(O(\ell^m)\) possibilities each, its offset has \(O(\ell^m)\) values, and its difference has fixed dimension. There are therefore \(O(\ell^A)\) complete normalized keys, \(A=m(m+1)\). Sorting rational likelihood breakpoints constructs them in \(O(\ell^{A+m}\log(\ell+1))\) operations and \(O(\ell^{A+m})\) stored lines. New weights are queried in logarithmic time. The single-budget root curve has \(O(\ell^{A+1})\) lines.

In a budget-\(c\) root query, the final participation occurs by informative time \(m(c-1)+1\); the rest uses a stored flush curve. Enumerating a complete preparation and its local tree costs \(O_{m,C}(\ell^{Ac-m+1}\log(\ell+1))\). If the first full synchronization leaves \(q\ge2\) participations, at most \(m(c-q)\) digits and that many outputs have been consumed. The digits contribute exponent \(m^2(c-q)\), and outputs contribute \(m(c-q)\). The quadratic term in \(m\) comes from allowing \(O(\ell^m)\) critical choices independently at each consumed digit. This covering argument supplies an upper bound; it does not prove that all offset combinations are necessary. A complete echo key therefore has \(O_{m,C}(\ell^{A(c-q)})\) possibilities. Each uncached key calls \(G_q\) for \(O(\ell)\) full outputs; \(q=1\) uses its stored curve. Thus
\[
\begin{aligned}
T_2&=O_{m,C}(\ell^{2A-m+1}\log(\ell+1)),\\
T_c&\le O_{m,C}(\ell^{Ac-m+1}\log(\ell+1))
+\sum_{q=2}^{c-1}O_{m,C}(\ell^{A(c-q)+1})T_q.
\end{aligned}
\tag{F.7}
\]

For \(E_c=(A+1)c-m-1\), the local exponent is at most \(E_c\), and \(A(c-q)+1+E_q=Ac+q-m\le E_c\). Also \(A+m\le E_2\). Induction proves the time exponent in (27). The largest local dictionary corresponds to \(q=2\), with exponent \(A(C-2)\); dictionaries are released on return and stack depth is at most \(C\). Combining their storage with flush curves proves the space exponent.

Every row has denominator \(BS\); padded complete paths have common denominator \((BS)^D\). Event-support coordinates are probabilities in \([0,1]\), so their scaled integer numerators are bounded by \((x+y)(BS)^D\). Preparation representatives use \(O(mC\log B)\) bits and offsets \(O_{m,C}(\log\ell)\). Curve intersections and comparisons use a fixed multiple of these bit lengths. Schoolbook arithmetic costs \(O_{m,C}(\kappa_{m,C}^2)\) per operation; modular inversion, normalization and binary input/\allowbreak{}output are included. This proves Theorem 4 for the specified organization. Literal rounding enumeration and optional extra global caches have their own implementation costs.

\subsection{DG transfer, moments and terminating calibration}

For two sources at TV distance \(\tau\), couple innovations maximally under one common word, policy, event and sequence of rounding coins. Histories and carries agree until the first mismatch. At most \(D\) informative innovations are used, so each directional transcript TV is at most \(1-(1-\tau)^D\). Testing scores differ by at most \((1+\lambda)\) times this quantity. Taking suprema and then using A.1 for the full-support source proves (28).

For the DG, lower dyadic weights \(b_z/2^b\le e^{-z^2/(2\sigma^2)}\) on \([-L,L]\), their sum \(A_b\), and an upper normalizer \(Z_\sigma^+\) give the normalized finite source \(\nu(z)=b_z/\sum b_z\) with
\[
\operatorname{TV}(g_\sigma,\nu)\le1-A_b/Z_\sigma^+,
\quad
2\sum_{z>L}e^{-z^2/(2\sigma^2)}
\le\frac{2e^{-(L+1)^2/(2\sigma^2)}}{1-e^{-(2L+3)/(2\sigma^2)}}.
\tag{F.8}
\]

Range-reduced rational series bound exponentials and logarithms, with outward remainder estimates. The normalized source is shared across all rows.

For a current common rounding coin, the innovation shift is \(j+\lfloor(a^\top s+a^\top d+U)/B\rfloor-\lfloor(a^\top s+U)/B\rfloor\), of magnitude at most two. Completing the square at an integer shift \(k\) gives
\[
\sum_z\frac{g_\sigma(z-k)^2}{g_\sigma(z)}=e^{k^2/\sigma^2}\le e^{4/\sigma^2}.
\tag{F.9}
\]

Hidden-coin marginalization and decreasing-rank induction give (7). The event inequalities there yield (8). For \(D_C(\lambda)=\Delta_C^*(\log\lambda)\), a near-optimal event at \(y>x\ge1\) and \(P(E)^2\le M_CQ(E)\) prove (29); also \(D_C(y)\le M_C/(4y)\).

Start an inverse bracket at \(\lambda_-=1\), doubling a rational upper endpoint until certified feasible. At the two interior thirds \(p<q\), refine source and threshold enclosures until the lower risk at \(p\) rejects or the upper risk at \(q\) accepts. If both true risks equaled the positive target, (29) would be violated. Thus a decision occurs and reduces the width by at least a factor \(2/3\). Keeping the lower endpoint at one includes a zero inverse without testing real equality.

For bracket width \(W\), either \(D_C(q)<\delta/2\), or \(D_C(p)-D_C(q)\ge W\delta^2/(12M_C)\). Accuracy proportional to this positive gap suffices. Reaching epsilon width \(2^{-k}\) requires \(O(k)\) rounds with \(O(k)\)-bit source and transcendental precision at fixed parameters. One can take \(L=O(\sqrt{k+1})\) and \(b=O(k+\log(k+2))\). Theorem 4 then gives polynomial bit cost for exact-profile calibration at fixed \(m,C,a,B,\sigma^2,\delta\).

\subsection{First-order exact queries and residual DG uncertainty}

The following records specialize to \(m=1,0<a_1=a<B\).

Apply the transfer bound (28) to the exact finite answer, using the source and threshold enclosures below.

In the released source construction, every retained nonzero atom is a lower bound on its DG mass; all removed mass is assigned to zero. Hence \(\operatorname{TV}(\nu,g_\sigma)=\nu(0)-g_\sigma(0)\), bounded above using a certified upper normalizer. Separate rational exponential constructions check the atom and threshold enclosures. Both exact native paths then agree on the finite rational optimum. The recorded C3 queries, both using the \(G_2\) base-curve variant, are

\par\smallskip\begingroup
\small
\setlength{\tabcolsep}{3pt}
\renewcommand{\arraystretch}{1.10}
\begin{tabularx}{\linewidth}{@{}>{\hsize=1.000000\hsize\linewidth=\hsize\raggedright\arraybackslash}X>{\hsize=1.000000\hsize\linewidth=\hsize\raggedleft\arraybackslash}X>{\hsize=1.000000\hsize\linewidth=\hsize\raggedleft\arraybackslash}X@{}}
\toprule
\textbf{Query} & \textbf{DG lower display} & \textbf{DG upper display} \\
\midrule
\(\gamma=1/3\), \(\sigma=3\), \(\varepsilon=2\), \(L=24\) & 0.001471784810259119 & 0.001471784810282800 \\
\(\gamma=336/1009\), \(\sigma=1\), \(\varepsilon=4\), \(L=6\) & 0.110138846336863912 & 0.110138858532574647 \\
\bottomrule
\end{tabularx}
\endgroup\smallskip\par

Their exact rational widths are below \(2.368\times10^{-14}\) and \(1.220\times10^{-8}\). The second row uses \(\sigma=1\) and is distinct from the earlier \(\sigma=3\) large-denominator task. Both are intervals for the original hidden mechanism's all-deterministic-finite-horizon supremum, and both lower endpoints exceed \(10^{-4}\). The release retains the source, request, threshold, two finite-value bindings and completion status of each attempt. These records establish exact finite evaluation and the stated interval transfer.

The finite controls compare 174 scalar queries across ten mechanisms with a separately reconstructed full-period reference, covering asymmetric and sparse laws and budgets through five. The tested C3 implementation enforces \(B\le10^6,L\le64,C\le12\). Its completed DG requests use the precomputed \(G_2\) path. These records establish exact-value agreement and DG enclosures. A systematic runtime and memory comparison between the streaming and precomputed-\(G_2\) configurations is left to future work. The C2 support-reduction timings in Section 7.1 answer a separate algorithmic question.

\section{The commitment hierarchy, calibration, and workload consequences}

\subsection{Complete proof groups and exact endpoints}

At a synchronized root \(G_c\), the pure first row precedes a shared future word of length \(mc\). A nonsynchronized free node \(R_{c,d}\) has no recorded prefix but its concrete target chooses a full compatible suffix before the current output. A pending node records a prefix of length at most \(mc+h(d)\), with all its compatible completions. Let \(\mathfrak F_s(t)\) be the supremum over these concrete policies, completions and events. The family is prior-independent, so \(\mathfrak F_s\) is convex, has endpoints one and lies between the stopping lines and one.

At a free node choose a proof-group length \(1\le\ell\le\min\{H,mc+h(d)\}\); at a synchronized root use \(1\le\ell\le\min\{H,mc\}\). Inside a group maximize over the entire prefix and complete current action before its output. For each branch, carry the whole unread prefix with initial displacement \(-fe_1\), update \(c,d\), and continue pending until the prefix ends or the registers synchronize. A shortening caused by an action removes only digits that no continuation can use.

Each concrete full completion occurs in every allowed group's prefix enumeration. The carry bijection preserves its child fiber. Hence dominating children imply
\[
\mathfrak F_s(t)\le\mathcal B_{s,g}(P_{\rm child})(t)
\quad\text{for every group }g.
\tag{G.1}
\]

For convex \(f\le u\), Jensen gives \(f(t)\le\sum_i\alpha_i f(t_i)\le\sum_i\alpha_i u(t_i)\) at every mixture with barycenter \(t\). Taking the infimum proves \(f\le\operatorname{Cvx}u\). Apply this to (G.1) and induct on \(\mathcal R(c,d)\). Flatten the zero-time prefix choice into its first probability edge; every such edge decreases rank. This proves soundness.

Increasing \(H\) adds groups, and complete sum, maximum, minimum and convex minorant are monotone. Induction gives \(P_{H+1}\le P_H\). When \(H\ge mC\), a root can retain its full \(mc\)-digit word. After the pure first row, \(\mathcal K(c,d)\) decreases on each edge, so that word never renews before true synchronization. Subsequent roots have smaller budget and are exact by induction, so this group's value is \(\mathcal G_c\). Soundness proves equality.

A finer fixed-cut schedule can simulate a coarser schedule by remembering its whole word, exposing only the demanded prefix at each boundary. The transported completions and both probabilities agree, proving Theorem 5. Each next block of a fixed schedule of length at most \(H\) is an allowed group. The same induction proves \(P_H\le U_{\mathcal S}\), and convexity proves the portfolio statement in (35).

There are \(C+1\) synchronized nodes and at most \(C(3^m-1)\) free nodes, with prefixes counted by (36). Rational finite-source curves remain rational piecewise linear. A child knot \(s\) under row \((p,q)\) pulls back to
\[
t=\frac{sq}{p(1-s)+sq},\qquad
v=(tp+(1-t)q)F(s).
\tag{G.2}
\]

Zero-weight branches use endpoint lines. Sums use the common knot refinement; maxima use the upper envelope of supports; the convex minorant of group minima is the lower hull of all group-curve vertices. A convex function below those vertices is below every segment. This gives an exact finite full-curve algorithm. Curve intersections and convexification generate their own rational denominators.

\subsection{Source stability and calibration at every depth}

For convex \(V\le F\le1\), \(F(0)=F(1)=1\), an interior support slope satisfies \((F(t)-1)/t\le s\le(1-F(t))/(1-t)\). Its coordinates \(A=F(t)+(1-t)s\), \(B=F(t)-ts\) therefore lie in \([0,1]\). Interior supports and the two stopping lines suffice to represent \(F\).

For fixed child functions a complete backup selects one such support per output, giving \(\sum_f[tp_fA_f+(1-t)q_fB_f]\). A source TV change of \(\tau\) changes each world's current row by at most \(\tau\), hence the backup by at most \(t\tau+(1-t)\tau=\tau\). A child error \(e\) adds at most \(e\) because complete row mass is one. Maximum and minimum are nonexpansive, and so is the convex minorant by monotonicity and translation equivariance. Rank induction gives
\[
\|P_H^\mu-P_H^\nu\|_\infty\le D\tau,
\qquad D=(m+1)C.
\tag{G.3}
\]

Bounded support coordinates justify the countable output sums, including DG. Homogeneous extension proves (37). The same argument applies to fixed schedules and root-convexified portfolios on their common complete semantics.

Depth one is a digit-renewal game with one group, so the row moment and informative-rank bound yield \(0\le D_H(\lambda)\le D_1(\lambda)\le M_C/(4\lambda)\). Supporting coordinates make \(D_H\) convex, nonincreasing and 1-Lipschitz. For \(1\le x<y\), put \(d=D_H(y)>0\) and \(z=M_C/(2d)\). Then \(z\ge2y\), \(D_H(z)\le d/2\), and convex secants give
\[
D_H(x)-D_H(y)\ge(y-x)\frac{d-D_H(z)}{z-y}
\ge(y-x)d^2/M_C.
\tag{G.4}
\]

The two-interior-point inverse algorithm of F.4 consequently terminates for every depth using (37). Its full-curve arithmetic is separate from the sharp point-query organization of Theorem 4. Monotonicity and the exact \(mC\) endpoint prove existence and the upper bound \(mC\) for (40).

\subsection{Strict memory examples, DG calibration and scalar recovery}

The identities (38) and (39) are obtained by exact rational row generation, complete digit lifts, piecewise-linear sums, whole-action maxima and lower hulls of complete proof groups. All root curves are compared at their joint knots. For the two-memory example at normalized prior \(1/11\), the smallest adjacent gap is \(277/912384\). Since \(D=6\), source replacement can shrink a gap by at most \(12\tau\). The mixture \((1-10^{-5})\nu_0+10^{-5}g_1\) has TV distance at most \(10^{-5}<277/21897216\), preserving at least half the gap. It has full support and finite first moment, so A.1 supplies physical preparation. The three-memory example has adjacent homogeneous gaps \(23/2048\) and an exact third level with \(C=1\).

For the pure-DG two-memory example, use \(L=8\), 128-bit lower weights and \(\tau<2.057\times10^{-18}\). A threshold bracket \(\lambda^-\le e^\varepsilon\le\lambda^+\) yields lower risk \(D_H^\nu(\lambda^+)-(1+\lambda^+)D\tau\) and upper risk \(D_H^\nu(\lambda^-)+(1+\lambda^+)D\tau\). Exact integer comparisons verify the two endpoints in (41), the risks in (61), and
\[
\varepsilon_{P_1}(1.1\times10^{-5})\in(5.65320445,5.65320448].
\tag{G.5}
\]

By monotonicity in risk, throughout \([10^{-5},1.1\times10^{-5}]\) the depth-one loss exceeds \(5.65320445-5.45309341=0.20011104\). Depth two is exact, proving (41). Literal-state hidden-rounding enumeration and the phase/\allowbreak{}carry recursion agree on the finite source.

The specialization \(m=1,0<a_1=a<B\) recovers the original nodes, complete carries, group minimization and convex closure. Its full-depth endpoint is \(H\ge C\). A cut labeled by remaining budget \(b\) in a segment starting at budget \(c\) corresponds to position \(c-1-b\). For \(a=1,B=2,C=4\) and uniform \(\nu_0\) on \(\{-1,0,1\}\), exact reconstruction gives
\[
(\mathcal G_4,P_3,P_2,Q_2,P_1)(1/2)
=\frac{(3737,3740,3746,3748,3753)}{3888}.
\tag{G.5a}
\]

The portfolio \(Q_2\) fixes one of two lengths at \(G_2,G_3,G_4,R_{1,\pm1},R_{2,\pm1}\), giving 128 selectors, then convexifies their root minimum. Its value is \(937/972\). The minimum consecutive gap in (G.5a) is \(1/1944\). Each endpoint changes by at most \(8\tau\), proving strictness for \(\tau<1/31104\) and a gap at least \(1/3888\) for \(\tau\le1/62208\). A.1 applies to any chosen full-support mixture at a fixed positive weight. Appendix K.2 records the first-order actual-DG calibration witnesses.

A local identity explains the common-preparation loss. Fix a causal prefix whose branch probabilities \(p_z,q_z\) are independent of an unused origin suffix \(\theta\). Put \(m_z=tp_z+(1-t)q_z\), \(t_z=tp_z/m_z\), let \(\kappa_z\) be the complete carry and \(r_z=\max_\theta F_{z,\kappa_z(\theta)}(t_z)\). Releasing that cut changes its value by
\[
\min_\theta\sum_{z:m_z>0}m_z
\left[r_z-F_{z,\kappa_z(\theta)}(t_z)\right].
\tag{G.6}
\]

This is the sum of branch maxima minus the maximum under one common suffix. For finitely many suffixes it vanishes exactly when one origin suffix optimizes every positive-mass branch. Complete carries distinguish origin compatibility from equal canonical labels. Subsequent optimization over prefixes, groups or convexification is accounted for by the full root curves above.

\subsection{Numerical sandwiches and structural diagnosis}

Suppose each child has a verified convex lower envelope represented by affine lines. Fix one complete current action and select one whole child line per output. Summing its two coordinates with that branch's P and Q probabilities gives one affine lower line for that action's backup. Flooring both coordinates weakens it on all priors. For each group \(g\), let \(L_g\) be the maximum of these lines and both stopping lines. Then \(L_g\le\mathcal B_{s,g}(P_{\rm child})\). Any affine line \(\ell\) below every \(L_g\) satisfies
\[
\ell\le\operatorname{Cvx}\!\left(\min_g L_g\right)
\le\operatorname{Cvx}\!\left(\min_g\mathcal B_{s,g}(P_{\rm child})\right)=P_s.
\tag{G.7}
\]

An independent consumer can check this without recreating the producer's cross-group convex hull. If \(L_g=\max_i\ell_i\), a violation is a prior where all strict affine inequalities \(\ell(t)>\ell_i(t)\) hold. Exact intersection of their intervals with \([0,1]\) decides whether such a prior exists. Checking every line against every authenticated group and then inducting on decreasing rank proves the lower envelope. Each check retains the current action and child support as whole probability pairs.

Upper certificates check complete backups at supplied prior points, using child upper chords and exact outward arithmetic. Convexity of the target makes interpolation between dominating sample values safe. The upper and lower functions together provide an observed precision contract. On the common refinement of their knots, including \(1/2\), their difference is affine, so its largest absolute width is attained at a knot. Dividing by \(w(t)=\min(t,1-t)\) gives a ratio of affine functions on each piece; its extrema are at piece ends or one-sided axis limits. At an axis, nonzero width gives infinite weighted width; zero width is evaluated through the slope limit.

These checks certify \(P_H\) directly, including new rational coordinates introduced by convex closure. Coordinate flooring has error at most \(1/\mathsf Q\); a weighted \(w/\mathsf Q\) error budget therefore requires weighted checking. Measured joint widths supply this precision contract, and (42)--(43) separate numerical and structural loss.

For authenticated C8 rows at \(\gamma=1/3,\sigma=3\), Table S3 distinguishes structural failure at \(\varepsilon=4.91\) from insufficient numerical resolution at \(4.94\). A depth-three certificate resolves the former; refining depth two resolves the latter, after DG transfer.

\noindent\begin{minipage}{\linewidth}
\par\smallskip\begingroup
\small
\setlength{\tabcolsep}{3pt}
\renewcommand{\arraystretch}{1.10}
\begin{tabularx}{\linewidth}{@{}>{\hsize=1.000000\hsize\linewidth=\hsize\raggedright\arraybackslash}X>{\hsize=1.000000\hsize\linewidth=\hsize\raggedright\arraybackslash}X>{\hsize=1.000000\hsize\linewidth=\hsize\raggedleft\arraybackslash}X>{\hsize=1.000000\hsize\linewidth=\hsize\raggedleft\arraybackslash}X@{}}
\toprule
\textbf{Query} & \textbf{Proof and mesh} & \textbf{Certified finite-source lower} & \textbf{Certified finite-source upper} \\
\midrule
\(\varepsilon=4.91\) & \(H=2\), r4 & \(1.06568375\times10^{-4}\) & \(1.09961070\times10^{-4}\) \\
\(\varepsilon=4.91\) & \(H=3\), r4 & \(9.61060134\times10^{-5}\) & \(9.92731400\times10^{-5}\) \\
\(\varepsilon=4.94\) & \(H=2\), r4 & \(9.74830305\times10^{-5}\) & \(1.00662301\times10^{-4}\) \\
\(\varepsilon=4.94\) & \(H=2\), r5 & \(9.81775632\times10^{-5}\) & \(9.89361532\times10^{-5}\) \\
\bottomrule
\end{tabularx}
\endgroup\smallskip\par
{\small Table S3. Proof-function diagnosis at a fixed source contract. Values are outward decimal displays of deterministic finite-source brackets. The accepted decisions use exact endpoints plus DG transfer; rows two and four certify the actual target. The dyadic r5 mesh refines r4. The lower endpoints enclose the proof function; G.5 constructs feasible-event lower bounds for the mechanism.}
\end{minipage}\par

At C4 and prior \(1/2\), accepted depth-one through depth-four brackets at tolerance \(1/20000\) give a full-depth upper \(3737/3888\). Subtraction via (43) gives positive structural lower gaps \(1/243,1/432,1/1296\) at depths one, two and three, respectively.

\subsection{First-order physical witnesses and inverse calibration}

The lawful lower class chooses a public integer target from a finite set at synchronization, waits with identical inputs until it is reached, and then applies a predictable unit sign. Subsequent informative transitions retain literal integer states and the actual hidden-rounding rows; phases and suffixes follow those integer transitions. In the DG chain, the target is hit almost surely by the drift and full-support argument. The finite comparison policy is extended to all other outputs by a fixed stopping decision, then embedded into the DG process. Deterministic horizon truncation keeps all failure branches, as in A.1. Thus a lower bound on this authenticated strategy class is a lower bound on the physical all-finite-horizon value.

Let \((A_E/\mathsf Q,B_E/\mathsf Q)\) be simultaneous minorants of \((P(E),Q(E^c))\) for one lawful finite strategy/\allowbreak{}event, and let \(r=1-(1-\tau)^{2C}\) be its per-world coupling loss. Its transferred payoff at threshold \(\lambda\) is at least
\[
\frac{A_E}{\mathsf Q}-r
-\lambda\left(1-\frac{B_E}{\mathsf Q}+r\right).
\tag{G.8}
\]

When the denominator is positive, this exceeds \(\delta\) below the rational threshold \(\lambda_{\rm crit}=(A_E/\mathsf Q-r-\delta)/(1-B_E/\mathsf Q+r)\). Maximizing that ratio over accepted whole pairs never combines coordinates from different events. Each pair is authenticated against one source, strategy and event, giving a physical lower bound.

For an upper-root chord \(f(t)=ta_s+(1-t)b_s\), its untruncated hockey-stick expression is \(a_s+(b_s-1)\lambda\) on the chord's own threshold domain. Use the upper source's coupling loss, which can differ from the lower source's. For \(\delta<1\), safety including zero/\allowbreak{}one clipping and the source correction is equivalent on that domain to
\[
r(1+\lambda)\le\delta,
\qquad a_s+r+(b_s-1+r)\lambda\le\delta.
\tag{G.9}
\]

Exact intersection with every chord's valid domain and \(\lambda\ge1\) finds a sufficient rational threshold without assuming that a rounded numerical upper is globally monotone. Directed rational exponential bounds verify \(e^{\varepsilon_l}<\lambda_{\rm crit}\) on the lower side and \(e^{\varepsilon_u}\ge\lambda_{\rm safe}\) on the upper. The actual profile is nonincreasing and one-Lipschitz in \(\lambda\), uniformly over strategies and horizons. A strict lower violation therefore persists in a neighborhood, giving the strict lower bound on \(\varepsilon_C^*\); the certified upper gives its inclusive upper bound.
\[
4.833477<\varepsilon_8^*(10^{-4})\le4.875282,
\qquad
7.486700<\varepsilon_{16}^*(10^{-4})\le7.582102.
\tag{G.10}
\]

The records underlying these bounds bind \(a/B=1/3\), \(\sigma=3\), \(\delta=1/10000\), and budgets 8 and 16. Lower sources and strategies use target set \(\{0\}\) and K28/\allowbreak{}r5; the upper models use K24/\allowbreak{}r4 with depths 4 and 3, respectively. Both numerical scales are \(2^{60}\). The full recorded audit reconstructs source probabilities and complete original-state semantics, consumes the numerical payload, extracts its authenticated root data, and checks the rational inverse.  If \(l<\varepsilon_C^*\le u\) with \(l>0\), then \(u/\varepsilon_C^*-1<(u-l)/l\), giving relative margins below 0.864906\% and 1.274287\%, respectively.

\subsection{A workload identity for the first-order generator}

Let \(\xi_t=\lfloor(aK_{t-1}+U_t)/B\rfloor-\gamma K_{t-1}\). Conditional on the past, direct counting of the hidden uniforms gives \(\mathbb E[\xi_t\mid\mathcal F_{t-1}]=0\) and \(\mathbb E[\xi_t^2\mid\mathcal F_{t-1}]=r_t(1-r_t)\), where \(r_t=\{\gamma K_{t-1}\}\). With centered innovations independent of the past and of \(U_t\), \(M_t=\eta_t-\xi_t\) is a square-integrable martingale difference of variance \(v_t=v+\mathbb E[r_t(1-r_t)]\in[v,v+1/4]\).

Iterating \(K_t=-\gamma K_{t-1}+M_t\) from zero gives \(K_t=\sum_{j=1}^t(-\gamma)^{t-j}M_j\). For fixed deterministic weights, interchange the two finite sums to obtain \(\sum_t h_tK_t=\sum_j c_jM_j\), with \(c_j\) as in (62). Distinct martingale differences are orthogonal, proving
\[
v\sum_jc_j^2\le
\mathbb E\!\left[\left(\sum_t h_tK_t\right)^2\right]
=\sum_jc_j^2v_j
\le(v+1/4)\sum_jc_j^2.
\tag{G.11}
\]

For \(h_t=1/n\), summing a geometric series gives the coefficient in Section 7.4 and \(n\sum_jc_j^2\to(1+\gamma)^{-2}\). In contrast, an individual release obeys \(\mathbb E[K_t^2]=\gamma^2\mathbb E[K_{t-1}^2]+v_t\). The martingale calculation applies to the evolving chain with its dependent rounding residuals. It uses deterministic workload weights; predictable inputs cancel in \(Y_t-x_t\). The formulas evaluate a fixed innovation law. Section 7.4 calibrates the designs to a common privacy request before comparing utility.

\section{Physical resource laws and finite certificates}

\subsection{Autonomous noise, regeneration and the testing inverse}

Subtracting the predictable P-input from each release is a triangular bijection on complete transcripts. Its P-law is the autonomous register chain (2); the Q-reference register is \(s-d\), and its current noise output is \(n-j\). Reversing the worlds gives another legal signed policy with the same physical deadline and participation cap. Conversely, with \(x^P=0\), the P-register is recoverable from public outputs, and a causal rule on the register and action history defines an allowed input pair.

Temporarily retaining the fresh rounding coin pairs integer DG translates whose shifts have magnitude at most two, because \(|a^\top d|/B<1\) and \(|j|\le1\). Marginalization gives \(0\le g\le R=2/\sigma^2\); a synchronized idle row has zero loss. Each participation affects at most its own and the next \(m\) rows.

\textbf{Weighted contraction and tails.} Under a common innovation and rounding coin, the rounded difference has constant sign and expected magnitude \(|a^\top(s-\widetilde s)|/B\). Let \(A\) be the signed companion matrix with first row \(-a^\top/B\), and let \(|A|\) be its entrywise absolute value. For the weights in (48), \((w^\top|A|)_i=w_i-\chi\le\alpha w_i\), since \(w_i\le1\). Therefore
\[
\mathbb E\|S'-\widetilde S'\|_w\le\alpha\|s-\widetilde s\|_w,
\qquad \mathbb E S_{t,i}^2\le{V_\infty},
\qquad {V_\infty}=\frac{v_0}{1-\alpha^2}.
\tag{H.1}
\]

For the moment claim, write \(S_t=AS_{t-1}+e_1\xi_t\). The innovation plus centered rounding error \(\xi_t\) has conditional sub-Gaussian proxy \(v_0=\sigma^2+1/4\). The coefficient \(|e_1^\top A^ke_1|\le\alpha^k\), so iterated conditional moment-generating functions give proxy \({V_\infty}\) for the first coordinate; the other coordinates are its time shifts. Thus every coordinate has tail \(2e^{-x^2/(2{V_\infty})}\), uniformly at deterministic physical times. These bounds hold under every action rule because the register chain is autonomous.

All \(m\)-step register transitions are positive: specifying the next \(m\) actual noise balances reaches any target register, with positive DG probability. The tail bound makes zero-start Cesàro laws tight. A total-variation convergent subsequence and the Cesàro telescoping identity give an invariant probability \(\mu\); irreducibility makes it unique and fully supported. Its coordinates inherit the same moments. Appendix A.1 gives finite expected public hitting times; H.6 strengthens these to geometric bounds for the selected controllers.

The first-order return estimate used by the binary refinement has explicit rational constants. With \(q=\sigma^2\), choose positive integers \(S,K\) satisfying \(S^2\ge q\), \((K+1)^2\ge4v_0/(1-\gamma^2)\), and set
\[
\begin{aligned}
M_r&=\lceil\gamma K\rceil+1,& n_r&=\lceil M_r^2/q\rceil,\\
p_0&=2^{-n_r}/(1+3S),& e_0&=\min\{p_0/2,(1-\gamma^2)/2\},\\
q_0&=1-e_0,&c_0&=v_0/e_0.
\end{aligned}
\tag{H.2}
\]

On \(|k|\le K\), each uniform component reaches zero with probability at least \(p_0\): the normalizer is below \(1+3S\) and \(2^{-n_r}\le e^{-M_r^2/(2q)}\). The kernel killed at zero contracts the function \(1+k^2/c_0\) off zero by \(q_0\). Its excess is at most \(2e_0-p_0\) inside the window, and the quadratic drift pays at least \(2e_0\) outside. Hence \(P_k(\tau_0>J)\le(1+k^2/c_0)q_0^J\).

\textbf{Budget values.} Public first-hit waiting, deterministic cutoffs and bounded informative rewards prove \(V_c(s,0)=A_c\) in both directions. From a nonsynchronized register, at most \(m\) common outputs flush the difference, giving \(V_c(s,d)\ge A_c\). Copying the original policy's virtual actions from \((s,0)\) leaves its P-register path unchanged and changes only the first \(m\) rewards. Thus \(V_c(s,d)\le A_c+mR\). Deleting one participation with its virtual history retained similarly gives \(0\le A_c-A_{c-1}\le(m+1)R\).

Concatenate a near-optimal budget-\(b\) block, a full echo, a public return to zero and a near-optimal budget-\(c\) block. Truncating the return proves the lower inequality in (45). For the upper inequality, stop just after the \(b\)-th participation: the prefix is a postprocessing of a deterministically padded budget-\(b\) policy, while its pending continuation has value at most \(A_c+mR\). Superadditivity gives \(a_*=\sup_c A_c/c\); iterating the upper defect at a fixed block size gives \(0\le ca_*-A_c\le mR\).

\textbf{Likelihood concentration.} For a hidden conditional row loss \(X\), DG moments and Hölder give
\[
\begin{aligned}
\mathbb E e^{tX}&\le e^{Rt(t+1)} &&(t\ge0),\\
\mathbb E e^{-tX}&\le1 &&(0\le t\le1),\\
\mathbb E e^{-tX}&\le e^{Rt(t-1)} &&(t\ge1).
\end{aligned}
\tag{H.3}
\]

If \(X'\) is a conditionally independent copy, Jensen bounds the centered moment-generating function by that of \(X-X'\). For \(b_0=(1+R)^{-1}\), the nonnegative even series gives
\[
\mathbb E\cosh(t(X-X'))-1
\le(t/b_0)^2\{\mathbb E\cosh(b_0(X-X'))-1\},
\qquad |t|\le b_0.
\tag{H.4}
\]

The bracket is at most \(e-1\) by (H.3). Thus the centered row has sub-Gaussian proxy \(v_*=2(e-1)(1+R)^2\) for \(|t|\le b_0\). For larger \(|t|\), use (H.3), \(0\le\mathbb EX\le R\), and \(R|t|\le R(1+R)t^2\); the same proxy suffices.

Compensate accumulated likelihood by \(V_{c_t}(S_t,d_t)\), as in reward-plus-potential concentration [33]. Conditional on the chosen action, the next continuation has range width at most \(mR\). Hölder combines its Hoeffding proxy with that of the row loss without assuming independence. Synchronized idle increments vanish, and at most \((m+1)C\) informative increments remain. The centered martingale \(N_T\) therefore satisfies
\[
\mathcal L_T\le A_C+N_T,\qquad
\mathbb E e^{tN_T}\le e^{\kappa Ct^2/2},\qquad
\kappa=(m+1)(\sqrt{v_*}+mR/2)^2.
\tag{H.5}
\]

The likelihood upper tail proves \(\varepsilon_C^*\le A_C+O_\delta(\sqrt C)\). Choose a finite policy with KL at least \(A_C-\eta\). Its nonnegative gap \(A_C+N_T-\mathcal L_T\) has mean at most \(\eta\). With \(\beta=(1-\delta)/2\), Markov and the martingale lower tail give an event of probability at least \(1-\beta-\sqrt\eta\) on which
\[
\mathcal L_T\ge A_C-\sqrt{2\kappa C\log(1/\beta)}-\sqrt\eta.
\tag{H.6}
\]

A threshold smaller by \(-\log\beta\) has testing payoff tending to \((1-\beta)^2>\delta\). Let \(\eta\downarrow0\) and clip the inverse at zero. This proves Theorem 9. The same estimates give its leading rate for risks with \(\log(1/\delta_C)=o(C)\) and \(\log(1/(1-\delta_C))=o(C)\). The superlinear-deadline conclusion follows from the frontier limit proved in H.6.

\subsection{Price values, compact occupations and the testing bridge}

We specialize the occupation and average-reward tools in [25,30,32,48] to the original hidden-output kernel. At participation price \(u\ge0\),
\[
V_0^u=0,\qquad
V_t^u(s,d)=\max_{j\in\{-1,0,1\}}
\{g(s,d,j)-u|j|+\mathsf K_sV_{t-1}^u(\cdot,d_j)\}.
\tag{H.7}
\]

Here the kernel also shifts the old register coordinates, and \(J_t(u)=V_t^u(0,0)\). Distinct integer registers have weighted distance at least \(w_{\min}\), so the bounded reward is \(R/w_{\min}\)-Lipschitz. Comparing the same action and using (H.1) yields spatial constant at most \(\mathsf L\). To compare different initial difference memories, copy the virtual action history: it agrees after \(m\) steps, the participation costs agree throughout, and only the first \(m\) rewards can differ. Hence
\[
|V_t^u(s,d)-J_t(u)|\le\mathsf L\|s\|_w+mR,
\qquad |J_{t+n}(u)-J_t(u)-J_n(u)|\le{b_\psi}.
\tag{H.8}
\]

The second inequality splits at a deterministic time and uses \(\mathbb E\|S_t\|_w\le W\sqrt{{V_\infty}}\). Applying the sub- and superadditive bounds to \(J_t\pm{b_\psi}\) gives \(\psi(u)\) and \(|J_t-t\psi|\le{b_\psi}\). The limit is nonnegative, convex, nonincreasing and 1-Lipschitz. A path with \(N\) participations has reward at most \(U_0N\), where \(U_0=(m+1)R\), so \(J_t(u)=\psi(u)=0\) for \(u\ge U_0\).

Define \(\mathcal Z\) by nonnegativity, total mass one and the original flow equations
\[
\sum_jz(s',d',j)
=\sum_{s,d,j}z(s,d,j)K(s,s')\mathbf1_{\{d'=d_j\}},
\qquad \sum_{d,j}z(s,d,j)=\mu(s).
\tag{H.9}
\]

The marginal identity follows by summing flow over the memories and using uniqueness of \(\mu\). It makes \(\mathcal Z\) totally bounded in \(\ell^1\), since each register has only \(3^{m+1}\) memory/\allowbreak{}action entries. The flow operator is continuous in that norm, so \(\mathcal Z\) is compact. Bounded reward and cost therefore attain their constrained maxima.

For a discounted price value, subtract its value at \((0,0)\). The spatial and memory bounds in (H.8) hold uniformly in the discount. A diagonal subsequence as the discount tends to one, with rowwise dominated convergence, gives
\[
\psi(u)+h_u(s,d)=\max_j\{g(s,d,j)-u|j|+\mathsf K_sh_u(\cdot,d_j)\},
\qquad |h_u(s,d)|\le\mathsf L\|s\|_w+mR.
\tag{H.10}
\]

Telescoping from zero identifies the scalar with \(\psi(u)\). Integrating against any occupation cancels the integrable bias. A maximizing stationary rule has tight Cesàro occupations and an invariant limit attaining equality. The compact convex cost/\allowbreak{}reward image contains the idle point, which strictly satisfies every positive participation allowance. Finite-dimensional separation then gives
\[
\psi(u)=\max_{z\in\mathcal Z}\{r(z)-uc(z)\},\qquad
\Phi(\tau)=\min_{0\le u\le U_0}\{u+\tau\psi(u)\}
=\tau\max_{z\in\mathcal Z,\ c(z)\le1/\tau}r(z).
\tag{H.11}
\]

For any hard-budget policy, \(A_{C,T}\le uC+J_T(u)\). At \(T=\lfloor\tau C\rfloor\), the minimizing price and \(\psi\ge0\) imply \(A_{C,T}\le C\Phi(\tau)+{b_\psi}\). Section H.6 constructs the matching hard-budget lower bound from an attaining occupation.

\textbf{From KL to testing at a fixed deadline.} Let \(B_{c,t}\) be the maximum expected KL with \(t\) releases and cap \(c\):
\[
B_{c,0}=0,\qquad
B_{c,t}(s,d)=\max_{|j|\le c}
\{g(s,d,j)+\mathsf K_sB_{c-|j|,t-1}(\cdot,d_j)\},
\qquad j\in\{-1,0,1\}.
\tag{H.11a}
\]

A deterministic policy attains each finite-horizon maximum, and \(A_{C,T}=B_{C,T}(0,0)\). Its spatial Lipschitz constant is at most \(\mathsf L\), uniformly in \(c,t\).

The true hidden-mixture row loss is \(R\)-Lipschitz in the next noise balance. To see this, align the P-row as \((1-p)g_\sigma(f)+pg_\sigma(f+1)\) and the Q-row as \((1-q)g_\sigma(f+\Delta)+qg_\sigma(f+\Delta+1)\), where \(|\Delta|\le2\). With \(W(p,f)=pe^{-(f+1/2)/\sigma^2}/(1-p+pe^{-(f+1/2)/\sigma^2})\), the real extension of its derivative is \([\Delta+W(q,f+\Delta)-W(p,f)]/\sigma^2\). For \(|\Delta|\le1\) its magnitude is at most \(R\). At \(\Delta=2\), the carry forces \(q<p\), making the posterior difference nonpositive; at \(\Delta=-2\) it forces \(q>p\), giving the reverse order. These cover signed coefficients and negative states.

Only the new first coordinate is random, with weight one and conditional variance at most \(v_0\). The row likelihood plus the continuation on that same output is therefore \((R+\mathsf L)\)-Lipschitz. The independent-copy variance identity gives \(Q=v_0(R+\mathsf L)^2\). Centering its increments and telescoping yields \(\mathcal L_T\le A_{C,T}+M_T\), \(\mathbb EM_T=0\), \(\mathbb EM_T^2\le TQ\). The maximizing policy has equality because every Bellman gap vanishes.

For \(\beta=(1-\delta)/2\), Cantelli's upper tail and the maximizing policy's Chebyshev lower tail imply
\[
\max\{0,A_{C,T}-\sqrt{TQ/\beta}+\log\beta\}
\le\varepsilon_{C,T}^*(\delta)
\le A_{C,T}+\sqrt{(1-\delta)TQ/\delta}.
\tag{H.11b}
\]

For the lower bound, the event \(\mathcal L_T\ge A_{C,T}-\sqrt{TQ/\beta}\) has P-probability at least \(1-\beta\). Lowering its testing threshold by \(-\log\beta\) gives payoff at least \((1-\beta)^2>\delta\). Reversing the input pair repeats the argument in that world's autonomous coordinates. The first-order constants in H.4 sharpen this bridge.

\subsection{Window approximation and polynomial precision}

First use the actual DG in the computational window \([-M,M]^m\). Coordinatewise clamping is nonexpansive under the common-noise coupling. Its Bellman values retain spatial constant \(\mathsf L\) and memory span \(mR\). Extend a window value by lookup at \([s]_M\). Comparing a true backup at \(s\) with that at \([s]_M\) costs at most \((R/w_{\min}+\alpha\mathsf L)\|s-[s]_M\|_w=\mathsf L\|s-[s]_M\|_w\). All rewards remain the original full-output KL in (44).

The comparison holds for every true action and for the true policy selected by window lookup. Iteration in both directions along the autonomous true path gives
\[
|J_T(u)-J_T^M(u)|\le T\mathsf L W e_M,
\qquad e_M=\frac{2{V_\infty}}{M}e^{-M^2/(2{V_\infty})}.
\tag{H.12}
\]

Indeed the coordinate tail integral bounds \(\mathbb E(|S_{t,i}|-M)_+\) by \(e_M\), uniformly in deterministic time.

Next replace the source by one common rational law, initially retaining the exact reward. The same coupling preserves the spatial bound. With the next memory and shifted old coordinates fixed, only the first coordinate varies; its value span is at most \(2M\mathsf L\). A source TV error \(\eta_p\) therefore costs at most \(2M\mathsf L\eta_p\) per step. A separate uniform reward perturbation costs \(\eta_g\), with rewards clipped to \([0,R]\) and synchronized-idle reward zero. Passing to rates and minimizing over prices proves (51).

The finite occupation LP equals (53). Normalize finite discounted values at \((0,0)\). Their uniform spatial and memory bounds give an all-state average Bellman equation in the discount limit. Telescoping identifies the same price gain from every initial state; integrating against invariant occupations gives its LP representation. Finite LP duality with the idle occupation then gives the constrained value. This argument also covers common finite sources whose transition graph is reducible.

\textbf{Complete coefficient construction.} For target \(\xi=2^{-k}\), increase the integer window until a rational upper tail bound makes \(\tau\mathsf LWe_M\le\xi/4\); \(M=O(\sqrt{k+1})\) suffices. Take \(\eta_g\le\xi/(4\tau)\) and \(\eta_p\le\xi/(8\tau M\mathsf L)\). Rational exponential enclosures and Gaussian tail bounds produce lower masses on a finite support, with the remainder assigned to zero, giving one normalized source for every row.

Enclose the actual-DG rewards separately. Form the hidden mixtures before taking logarithms, retaining both positive and negative summands. For aligned output \(f\), the derivative argument in H.2 and the mixture ratios at zero give \(|\ell(f)|\le(2|f|+4)/\sigma^2\). Since \(|f|\le|\eta|+1\), the absolute omitted reward outside \(|f|\le F\), with \(K=F-1>0\), is at most
\[
\frac{4K+4\sigma^2/K+12}{\sigma^2}
 e^{-K^2/(2\sigma^2)}.
\tag{H.12a}
\]

This follows by integrating the DG tail to bound \(\mathbb E[|\eta|\mathbf1_{\{|\eta|>K\}}]\) by \(2(K+\sigma^2/K)e^{-K^2/(2\sigma^2)}\). Thus \(F=O(\sqrt{k+1})\) suffices. Rational series with explicit remainders evaluate exponentials and logarithms to \(O(k+1)\) bits. The logarithm of the reciprocal smallest retained mass is also \(O(k+1)\), keeping every coefficient polynomial in precision.

There are \(3^{m+1}(2M+1)^m\) occupation variables. Store normalization, flow, participation and explicit boxes \(0\le z\le1\). The conservative dense input size is \(O((k+1)^{m+1})\) bits at fixed mechanism parameters. The idle invariant occupation proves feasibility and normalization proves boundedness. A polynomial-bit rational LP algorithm [50] computes the objective to \(\xi/4\) after the known \(\tau\) scaling, or recovers the exact rational optimum. The three source/\allowbreak{}window payments and the LP tolerance prove (52). All stopping comparisons use outward rational bounds.

\textbf{First-order binary refinement.} For \(m=1\), the existing implementation has \(9(2M+1)\) variables and \(15(2M+1)+3\) stored rows before redundant boxes, with dense payload \(O((k+1)^2)\). Its coefficient evaluator constructs and caches the source and complete reward table. The LP evaluator uses finite-precision ellipsoid decisions, denominator clearing and rational relaxation, value bisection and bounded continued-fraction recovery of the exact rational optimum. A denominator bound separates candidate values. Its correctness also covers infeasibility of a boxed input.

Both evaluators are finite first-order instruction trees. Their binary-addressed heap charges control, address comparisons, traversal, allocation, input loading and complete output serialization. For input cells \(I\), source execution cost \(T_0\) and output cells \(O\), the proved simulation cost is \(32769(I+T_0+O+6)^4\). Each term is polynomial in \(k\). Multiplying the LP output by \(\tau\) and applying (51) proves the accuracy of this same run; Appendix I records the terminal declarations.

\textbf{Residual-aware certificates.} Let a candidate occupation \(z\) be nonnegative and normalized, with flow residual \(e=\mathrm{out}-\mathrm{in}\). The finite price bias has bound \(K_M=\mathsf LMW+mR\). Integrating its Bellman inequality gives the valid finite-model lower bound
\[
L_M=\tau\{\langle\widetilde g,z\rangle-K_M\|e\|_1\}
-U_0\bigl(\tau\langle|j|,z\rangle-1\bigr)_+.
\tag{H.12b}
\]

For rational \(u\ge0,b,h\), compute the maximum nonnegative all-action residual \(\zeta=\max_{s,d,j}[\widetilde g+\widetilde{\mathcal P}h-h-u|j|-b]_+\). The upper bound is \(U_M=u+\tau(b+\zeta)\). Paying the right side \(E_M\) of (51) on both sides encloses the true frontier. In particular,
\[
\left|\Phi(\tau)-\frac{L_M+U_M}{2}\right|
\le E_M+\frac{U_M-L_M}{2}.
\tag{H.13}
\]

The multi-memory certificates in Section 7.3 use these checks. Ten rational requests give lower chords by concavity, while every dual remains a valid affine upper bound across the whole time-ratio interval. Exact comparisons at their combined breakpoints establish the continuous enclosure. Replaying the ten multi-memory certificates and one scalar regression checks 119,259 action rows. The regression at \(a=1,B=3,\sigma^2=1/4,\tau=2\) gives endpoints \(2.509717617551541\) and \(2.509717617560597\), overlapping the earlier first-order interval in Table S4.

Five earlier first-order certificates retain their finite primal--dual gaps. Lean instances verify their original-source transfer and midpoint inequalities, and independent arithmetic reconstructs 1,113 flow equalities and 3,339 dual inequalities.

\noindent\begin{minipage}{\linewidth}
\par\smallskip\begingroup
\footnotesize
\setlength{\tabcolsep}{3pt}
\renewcommand{\arraystretch}{1.10}
\begin{tabularx}{\linewidth}{@{}>{\hsize=1.150000\hsize\linewidth=\hsize\raggedright\arraybackslash}X>{\hsize=0.350000\hsize\linewidth=\hsize\raggedleft\arraybackslash}X>{\hsize=0.350000\hsize\linewidth=\hsize\raggedleft\arraybackslash}X>{\hsize=1.575000\hsize\linewidth=\hsize\raggedleft\arraybackslash}X>{\hsize=1.575000\hsize\linewidth=\hsize\raggedleft\arraybackslash}X@{}}
\toprule
\textbf{\( (\gamma,\sigma^2,\tau) \)} & \textbf{\(k\)} & \textbf{\(M\)} & \textbf{Frontier lower} & \textbf{Frontier upper} \\
\midrule
\((1/3,1/4,2)\) & 0 & 10 & 1.8231621827021824 & 3.1564955160356631 \\
\((1/3,1/4,2)\) & 8 & 11 & 2.5070801802359043 & 2.5122885135954458 \\
\((1/3,1/4,2)\) & 40 & 12 & 2.5097176175597534 & 2.5097176175611344 \\
\((2/3,1,3)\) & 12 & 34 & 1.2350823192545407 & 1.2354078410420640 \\
\((9/10,1/4,2)\) & 8 & 116 & 5.2591832237619530 & 5.2643916394215952 \\
\bottomrule
\end{tabularx}
\endgroup\smallskip\par
{\small Table S4. First-order original-frontier enclosures. Each exact midpoint has error at most \(2^{-k}\), including source, window and LP-gap payments. The initial \(k=40\) candidate remains in the artifact with a half-gap above the requested error; the table shows the accepted certificate. These instances and the general binary execution theorem validate their respective supplied-data and algorithmic statements.}
\end{minipage}\par

\subsection{First-order finite-deadline testing and source residuals}

This subsection fixes \(m=1,0<a_1=a<B\).

\textbf{The actual hard-budget value and hidden loss.} Let \(B_{c,t}(k,d)\) be the optimal KL of the next \(t\) physical outputs with \(c\) participations left and previous difference \(d\). Thus
\[
B_{c,0}(k,d)=0,\qquad
B_{c,t}(k,d)=\max_{j\in\{-1,0,1\},\,|j|\le c}
\{g(k,d,j)+\mathsf K_kB_{c-|j|,t-1}(\cdot,j)\}.
\tag{H.14}
\]

Finite action maxima give a deterministic predictable optimizer. Induction in \(t\), simultaneously for every budget, gives period \(B^t\) in \(k\): shifting by \(B^t\) shifts the next integer by \(-aB^{t-1}\). A consistent tie rule therefore gives a finite nonstationary residue table. The finite-table representation is defined by real-valued Bellman comparisons. Removing predictable common queries by the integer translation \(y\mapsto y-x_t^P\) identifies this recursion with the full-history value and gives \(A_{C,T}=B_{C,T}(0,0)\).

For the output loss \(\ell(y)\), write \(m=\lfloor\gamma k\rfloor\), \(n=\lfloor\gamma(k-d)\rfloor-j\), and let \(\rho,\rho'\) be the corresponding fractional parts. The two actual rows are \((1-\rho)g_\sigma(y+m)+\rho g_\sigma(y+m+1)\) and the analogous mixture with \(n,\rho'\). Extend these positive expressions to real \(y\) for differentiation only. Set
\[
W(\rho,x)=\frac{\rho e^{-(x+1/2)/\sigma^2}}
 {1-\rho+\rho e^{-(x+1/2)/\sigma^2}},\qquad
\ell'(y)=\frac{n-m+W(\rho',y+n)-W(\rho,y+m)}{\sigma^2}.
\tag{H.15}
\]

The posterior \(W\) increases in \(\rho\), decreases in \(x\), and lies in \([0,1]\). For \(n-m\in\{-1,0,1\}\), the derivative numerator has magnitude at most two. If \(n-m=2\), the floor relation forces \(d=j=-1\) and \(\rho'=\rho+\gamma-1<\rho\), so the posterior difference is nonpositive. If \(n-m=-2\), it forces \(d=j=1\) and \(\rho'=\rho-\gamma+1>\rho\), giving the reverse order. These exhaust all cases, including negative states. Hence \(\ell\) is \(R\)-Lipschitz on the actual integer outputs. At \(k=d=0\), the two rows are single DG translates by \(j\), so \(R/2\) suffices.

The DG variance bound [1, Corollary 17] and independence of the fresh noise and rounding give \(\operatorname{Var}(K'\mid k)\le\sigma^2+1/4\). Comparing the same legal action in (H.14), with the same successor budget and memory, gives spatial Lipschitz constant \(L_n=R(1-\gamma^n)/(1-\gamma)\). Thus the likelihood plus continuation on the same next output has conditional variance at most \((\sigma^2+1/4)(R+L_n)^2\), by the independent-copy bound \(\operatorname{Var}f(X)\le\operatorname{Lip}(f)^2\operatorname{Var}X\). No independence between the two summands is needed.

For a legal policy, let \(D_i=\ell_i+B_{c_i,T-i}(K_i,j_i)-B_{c_{i-1},T-i+1}(K_{i-1},j_{i-1})\), where \(c_i\) is the budget after row \(i\). Put \(\rho_i=\mathbb E[D_i\mid\mathcal F_{i-1}]\le0\) and \(M_T=\sum_i(D_i-\rho_i)\). Orthogonality and the original first row give
\[
\mathcal L_T=A_{C,T}+M_T+\sum_{i=1}^T\rho_i
\le A_{C,T}+M_T,\qquad
\mathbb E M_T=0,\qquad \mathbb E M_T^2\le V_T.
\tag{H.16}
\]

Here \(V_T\) is (56); the original-state improvement is used once, and the sum covers physical slots. For the optimizer every \(\rho_i=0\). For a general policy the variance bound (H.16) applies to the compensating martingale. Reversing the query pair preserves the public transcript, cap and deadline, maps the residual to \(k-d\), and negates the previous and current differences. The same argument then uses the reverse world's own autonomous source. Applying Cantelli in each ordered direction proves (57), with the same risk \(\delta\).

The optimizer also supplies the lower testing bridge. With \(\beta=(1-\delta)/2\), Chebyshev gives probability at least \(1-\beta\) to \(\{\mathcal L_T\ge A_{C,T}-\sqrt{V_T/\beta}\}\). At epsilon smaller by \(-\log\beta\), its hockey-stick payoff is at least \((1-\beta)^2>\delta\). Consequently
\[
\max\{0,A_{C,T}-\sqrt{V_T/\beta}+\log\beta\}
\le\varepsilon_{C,T}^*(\delta)
\le A_{C,T}+\sqrt{(1-\delta)V_T/\delta}.
\tag{H.17}
\]

Finite transcript boxes approximate the event and preserve its strict margin. An alternative estimate uses only the row second moment and continuation variance. For DG it gives \(\mathbb E M_T^2\le TQ_*\), where \(Q_*=2(e^{4/\sigma^2}+1)+2\mathsf L^2v_0\). Replacing \(V_T\) by \(TQ_*\) and using Chebyshev on both sides gives the bridge used in H.6. This finite-moment argument transfers to H.7 with its source-specific constants.

\textbf{Proof of Proposition 12.} Extend \(h\) by \(H_h(k,d)=h(c_M(k),d)\), keeping the actual state and output unbounded. Comparing \(k\) with \(x=c_M(k)\) costs at most \(R|k-x|\) in reward and \(\gamma\ell_h|k-x|\) in expected potential. At \(x\), common-source total variation costs \(S_h\eta_p\), since the next memory \(j\) is fixed. The checked finite row therefore implies, for every integer state and action,
\[
g(k,d,j)+\mathsf K_kH_h(\cdot,j)
\le u|j|+v+H_h(k,d)+\eta_g+S_h\eta_p
 +(R+\gamma\ell_h)(|k|-M)_+.
\tag{H.18}
\]

The source mgf bound [1, Lemma 16] and centered rounding in an interval of length one give \(\mathbb E[e^{sK'}\mid k]\le e^{-s\gamma k}e^{v_0s^2/2}\). Iteration from zero gives \(\mathbb E e^{sK_i}\le e^{{V_\infty}s^2/2}\). Integrating its two-sided tail yields \(\mathbb E(|K_i|-M)_+\le e_M\). At \(i=0\) the excess is zero. These are unconditional bounds at deterministic physical times, which remain valid for adaptive actions because the P-state chain is autonomous.

Sum (H.18) in expectation over \(i=0,\ldots,T-1\). The cap pays at most \(uC\); the potential telescopes from zero to a nonnegative terminal value; the tail costs at most \((T-1)_+(R+\gamma\ell_h)e_M\). The chain rule for all published outputs gives (55). The consumer checks the table's nonnegativity after its origin is fixed. Since every finite action row is checked, additional participation-spacing restrictions only reduce the admissible policies.

For the four selected tables, \(\ell_h\le\mathsf L\) and \(S_h\le2M\mathsf L\). Using \(R+\gamma\mathsf L=\mathsf L\) and \(T\le\tau C\) shows that (55) is no larger than the previous \(C\Phi^++D^+\) certificate. Exact rational comparisons establish strict improvement at all 56 evaluated requests. The table uses upward rounding for both bounds.

\noindent\begin{minipage}{\linewidth}
\par\smallskip\begingroup
\small
\setlength{\tabcolsep}{3pt}
\renewcommand{\arraystretch}{1.10}
\begin{tabularx}{\linewidth}{@{}>{\hsize=1.000000\hsize\linewidth=\hsize\raggedright\arraybackslash}X>{\hsize=1.000000\hsize\linewidth=\hsize\raggedleft\arraybackslash}X>{\hsize=1.000000\hsize\linewidth=\hsize\raggedleft\arraybackslash}X>{\hsize=1.000000\hsize\linewidth=\hsize\raggedleft\arraybackslash}X@{}}
\toprule
\textbf{\((\gamma,\sigma^2,\tau)\)} & \textbf{Window \(M\)} & \textbf{Previous upper} & \textbf{Direct upper} \\
\midrule
\((1/2,1/4,3/2)\) & 4 & 3.367294527 & 3.367104912 \\
\((2/3,1,3)\) & 7 & 1.282059630 & 1.281803467 \\
\((1/2,1/4,5)\) & 4 & 3.507157885 & 3.506871453 \\
\((1/2,1/2,20)\) & 5 & 2.059410315 & 2.059143268 \\
\bottomrule
\end{tabularx}
\endgroup\smallskip\par
{\small Table S5. Certified upper bounds on \(\varepsilon_{C,\lfloor\tau C\rfloor}^*/C\), with \(C=10^6\) and \(\delta=0.1\). Both columns use the same previously certified variance radius. The full set has these four contexts, \(C\in\{10^2,\ldots,10^8\}\), and \(\delta\in\{10^{-1},10^{-4}\}\). Separate 14 requests using a different synthesized source are outside this comparison.}
\end{minipage}\par

The finite coefficients enclose the infinite-DG rewards and source tails; the consumer reconstructs the original hidden-U transitions and verifies all 396 dual rows, the sign and origin conditions, the spans and request bindings. Each finite-deadline bound follows directly from the supplied potential and the original-source residuals.

\subsection{First-order rate certificates and the disclosed-rounding reference}

Fix \(m=1,0<a_1=a<B\). At \(\gamma=1/2,\sigma=1\), rational certificates give \(1.02784055<a_*<1.02784057\).

A nonnegative bounded potential with \(h(k,0)=0\) and
\[
g(k,d,0)\le h(k,d),\qquad
g(k,d,j)+\mathsf K_k h(\cdot,j)-h(k,d)\le r|j|
\quad(j=\pm1)
\tag{H.19}
\]

telescopes to KL at most \(rC\). At these parameters, the upper certificate uses a period-32 rational potential and checks all 64 predecessor residues, three previous actions and both participation signs. For the lower, prepare the actual integer \(\bar k_0=(1-(-2)^C)/3\) and take \(C\) positive participations. Its deterministic center has \(C\) odd phases, and the residual follows \(E_{t+1}=Z_{t+1}-\lfloor(E_t+1+U_{t+1})/2\rfloor\). A rational potential on 25 residual states, extended by zero, supplies a row lower bound. A uniform residual sub-Gaussian tail bounds excursions outside those states. Telescoping yields cumulative KL at least \(Cr-O(1)\); capped public preparation approaches this value in the all-horizon supremum. Directed arithmetic verifies strict margins at the two rate endpoints above.

For the reference that publishes \((U_t,Y_t)\) after choosing the current action, an integer DG translation has exact KL equal to squared shift divided by \(2\sigma^2\). Averaging the carry gives the conditional reward
\[
g_{\rm rev}(d,j)=\frac{|j|+2\gamma dj+\gamma|d|}{2\sigma^2}.
\tag{H.20}
\]

On a realized action word with \(n\) participations, \(S\) nonzero segments, \(O\) opposite-sign adjacencies and terminal indicator \(b_T=\mathbf1_{j_T\ne0}\), its total numerator is \((1+3\gamma)n-2\gamma S-4\gamma O-\gamma b_T\). The maximum under cap \(C\ge1\) is \((1+3\gamma)C-2\gamma\), attained by consecutive same-sign participations and one echo. This identity also bounds adaptive policies pathwise. Conditional centering combines DG sub-Gaussianity with the bounded conditional shift mean and gives total proxy \(O(C)\). The likelihood argument of H.1 then proves the disclosed rate \(\rho=(1+3\gamma)/(2\sigma^2)\). At \(\gamma=1/2,\sigma=1\), \(\rho=1.25\), so disclosing the fresh coin costs \((\rho-a_*)C+O_\delta(\sqrt C)\), with coefficient in \((0.22215943,0.22215945)\).

At \(\gamma=1/2,\sigma^2=1/4\), a separate radius-12 window supplies one all-action Bellman upper and two fixed-policy Poisson lower certificates. Their affine bounds enclose \(\Phi(\tau)\) for every real \(\tau\in[1,10]\), with gap below \(1.7\times10^{-12}\); the leading coefficient increases by at least 6.889\% over that interval. Table S4 retains five further precision requests. Four normalized potentials tighten all 56 corresponding finite-deadline endpoints; for \(T=5C,C=10^7,\delta=0.1\) at this scale, the upper bound on \(\varepsilon_{C,T}^*/C\) decreases from 3.429296962 to 3.429029487. Table S5 gives four representative settings.

\subsection{Closed-class entry, Poisson control and the hard-budget lower bound}

Fix the mechanism, \(\tau>0\) and \(\delta\in(0,1)\). We construct finite constants \(K_*,B_*\), independent of \(C\), for the lower bound in Theorem 10. The construction fixes an optimal occupation before taking the budget limit.

\textbf{1. At most two fixed classes.} Recover a stationary randomized rule from an attaining occupation in (H.11). Its positive-mass support is closed and decomposes into positive recurrent irreducible classes. Each class's normalized invariant distribution has register marginal \(\mu\). Since \(\mu(0)>0\), each class contains a state \((0,d)\); disjointness therefore bounds their number by \(3^m\). This argument allows correlations between the register and difference memory, such as the constraint \(d_1=\mathbf1_{\{s_2=0\}}\) in Section 6.2.

Take the cost/\allowbreak{}reward pair of each of these finitely many classes. The LP over mixture weights has total mass one and one cost constraint. An optimal extreme point uses at most two classes: one if cost is slack, and at most two when it is tight. Any such mixture is an actual occupation, so its optimum equals the original optimum. Fix the selected \(k\le2\) rules, classes and weights before varying \(C\), with
\[
\sum_{i=1}^k\theta_i=1,\qquad
\bar c=\sum_i\theta_i c_i\le1/\tau,\qquad
\bar r=\sum_i\theta_i r_i=\Phi(\tau)/\tau.
\tag{H.21}
\]

\textbf{2. Quadratic drift and fourth moments.} The signed companion matrix \(A\) is stable by (H.1). Let \(P=\sum_{n\ge0}(A^n)^\top A^n\), so \(P\ge I\) and \(A^\top PA=P-I\). Put \(V(s)=1+s^\top Ps\) and \(\lambda=\lambda_{\max}(P)\). Conditional centering of \(\xi_t\) gives
\[
\begin{aligned}
\mathbb E[V(S')\mid s]&\le\rho V(s)+b,
&\rho&=1-1/\lambda,& b&=1/\lambda+v_0P_{11},\\
\sup_t\mathbb E V(S_t)^2&\le M_V:=2+32\lambda^2m^2{V_\infty}^2.
\end{aligned}
\tag{H.22}
\]

If necessary increase \(\rho\) into \((0,1)\), preserving the inequality. The second line follows by tail integration: \(\mathbb E S_{t,i}^4\le16{V_\infty}^2\), hence \(\mathbb E\|S_t\|_2^4\le16m^2{V_\infty}^2\). All expectations are at deterministic physical times, including under the entry and virtual controllers below.

\textbf{3. Hitting inside a selected class.} Fix a class and a target \(o=(0,d_o)\). Set \(B_0=b/(1-\rho)>0\) and let \(F\) be its finite subset with \(V\le4B_0/(1-\rho)\). It is nonempty because its invariant expectation of \(V\) is at most \(B_0\). Irreducibility and finiteness give \(L_0\ge1,p_0>0\) such that every state in \(F\) hits \(o\) by time \(L_0\) with probability at least \(p_0\). Different paths may have different lengths; the statement concerns hitting by a time and includes periodic classes.

Choose
\[
0<\eta\le\min\{p_0/4,(1-\rho)/2\},\qquad
b_h=\eta/B_0,\qquad f=1+b_hV.
\tag{H.23}
\]

For \(\tau_o=\inf\{t\ge0:X_t=o\}\), iterated drift yields
\[
\mathbb E_x[f(X_{L_0})\mathbf1_{\{\tau_o>L_0\}}]
\le1-p_0\mathbf1_F(x)+b_h\{\rho^{L_0}V(s)+B_0\}
\le(1-\eta)f(x).
\tag{H.23a}
\]

For the second inequality, subtract its right side. On \(F\) the difference is at most \(2\eta-p_0+b_h(\rho^{L_0}-1+\eta)V\le0\). Outside \(F\), it is at most \(2\eta-b_h(1-\rho)V/2\le0\). These use \(\rho^{L_0}\le\rho\), the two bounds on \(\eta\), and \(b_hB_0=\eta\). Iterating the killed kernel gives
\[
\Pr_x(\tau_o>nL_0)\le(1-\eta)^n[1+b_hV(s)],\qquad
\mathbb E_x\tau_o\le(L_0/\eta)[1+b_hV(s)].
\tag{H.23b}
\]

Summing this tail against successive powers gives finite moments of every fixed order, bounded by a constant times \(1+V(s)\). From \(o\), take one step and then apply the bound to obtain the same conclusion for the first positive return time.

\textbf{4. The Poisson equation at every class state.} For the fixed rule's bounded state-averaged cost or reward \(f_0\), with invariant mean \(\bar f_0\), define
\[
h(x)=\mathbb E_x\sum_{t<\tau_o}(f_0(X_t)-\bar f_0),\qquad h(o)=0.
\tag{H.24}
\]

Equation (H.23b) gives \(|h(x)|\le\kappa(1+V(s))\), and first-step conditioning proves \(h-P_\pi h=f_0-\bar f_0\) away from \(o\). To establish it at \(o\), divide the expected state-visit counts during a positive return cycle by its finite mean length. Summing transitions and canceling the equal initial and final state shows that this probability is invariant. Uniqueness in the class identifies it with the class invariant law. The centered cycle sum therefore has expectation zero. Decomposing it after its first step gives \(f_0(o)-\bar f_0+P_\pi h(o)=0\), the required target equation. The growth bound and drift justify all expectations. This is the Poisson construction of [49] with the present hitting and moment conditions verified.

\textbf{5. Legal entry from an arbitrary actual state.} Repeat an \(m\)-step deterministic action block that writes the components of \(d_o\) in reverse order. At every block end the actual difference memory equals \(d_o\). Switch to the class rule exactly when the register is also zero; otherwise repeat. Every action, output and elapsed slot is part of the execution.

At block boundaries, \(K^mV\le\rho^mV+b\sum_{i<m}\rho^i\), and \(K^m(s,0)>0\). A finite level set has a positive common success probability. Applying the killed-kernel argument to this block chain proves a geometric tail and \(\mathbb E_s\sigma\le K(1+V(s))\) for the entry time. From a position inside a block, complete at most \(m\) slots first; drift gives the same bound with a larger constant. A common constant covers all finitely many entry positions.

\textbf{6. Potentials during unfinished entry.} Augment the analysis state by the block position while entering, or by the class mode after entry. Before entry define \(h(x)=\mathbb E_x\sum_{t<\sigma}(f_t-\bar f_0)+h(o)\), with center equal to the target class mean; after entry use (H.24). Here \(f_t\) is the controller's state-averaged cost or reward. The entry-time bound gives \(|h(x)|\le\kappa(1+V(s))\) throughout the augmented space. First-step conditioning proves one Poisson equation across all modes. It includes trajectories whose entry has not finished at the end of a deterministic stage.

\textbf{7. Deterministic-stage fluctuations.} Consider a stage starting at deterministic physical time \(a\), of length \(n\), using class \(i\)'s entry controller and then its rule. Let \(c_t=|j_t|\) be the sampled action cost, and take the filtration before the new action coin and innovation. The cost potential gives martingale differences
\[
D_t=c_t-c_i+h(X_{t+1})-h(X_t),\qquad
\mathbb E D_t^2\le3+6H_{c,i}^2,
\tag{H.24a}
\]

where \(H_{c,i}\) bounds \(\|h(X_t)\|_2\) at all deterministic times by (H.22). The actual action-coin fluctuation is included in \(D_t\). Orthogonality gives \(\mathbb E(\sum D_t)^2\le n(3+6H_{c,i}^2)\). Telescoping and the two endpoint potentials yield
\[
\|N_i-nc_i\|_2\le A_i\sqrt n+B_i,
\qquad \mathbb E G_i\ge nr_i-J_i,
\quad
A_i=\sqrt{3+6H_{c,i}^2},\ B_i=2H_{c,i},\ J_i=2H_{g,i}.
\tag{H.25}
\]

The reward estimate uses its corresponding potential bound \(H_{g,i}\). These constants depend on the fixed controllers but are independent of \(a,n,C\).

\textbf{8. Fixed stages and rounding costs.} Let \(T=\lfloor\tau C\rfloor\) and \(n_i=\lfloor\theta_iT\rfloor\). Run the \(k\le2\) controllers in a fixed order for these deterministic lengths, preserving the actual register and memory at boundaries, then use zero actions for the remaining at most \(k-1\) slots. Put \(A_\Sigma=\sum_iA_i\), \(B_\Sigma=\sum_iB_i\), \(J_\Sigma=\sum_iJ_i\). The uncapped virtual policy satisfies
\[
\begin{aligned}
q_C&:=\sum_i n_ic_i\le T\bar c\le C,\\
\mathbb E(N-C)_+&\le\sum_i\|N_i-n_ic_i\|_2
\le A_\Sigma\sqrt T+B_\Sigma,\\
\mathbb E G^{\rm vir}&\ge T\bar r-kR-J_\Sigma
\ge C\Phi(\tau)-(k+1)R-J_\Sigma.
\end{aligned}
\tag{H.25a}
\]

The cost bound uses the \(L^2\) triangle inequality, so it also covers dependent stages. The reward bound pays both phase-length and deadline rounding; zero-action padding has nonnegative KL.

\textbf{9. Retaining the virtual history under the cap.} Run this controller on the same autonomous P-register path, recording its full virtual actions and difference memory. Keep only its first \(C\) nonzero actions in the actual input pair; replace later nonzero actions by zero, while the virtual rule continues from its own history. Induction shows that the actual participation count at every time is the minimum of the virtual count and \(C\). At the deadline the deletion count is \((N-C)_+\).

This is a legal history rule. Choose \(x^P=0\), so the register is the public output history, and compute the virtual memory from earlier virtual actions. For every fixed auxiliary action seed, \(x_t^Q=j_t^{\rm cap}\) is a deterministic function of the public past before fresh mechanism randomness.

\textbf{10. Finite-memory loss per deletion.} Both rewards are evaluated on the same P-register path. A row can change only if the current action or one of the previous \(m\) actions was deleted. If \(D\) is the set of deleted slots, affected rows belong to \(\bigcup_{s\in D}\{s,\ldots,s+m\}\cap\{1,\ldots,T\}\), of size at most \((m+1)|D|\). Since both rewards lie in \([0,R]\),
\[
G^{\rm vir}-G^{\rm cap}\le U_0(N-C)_+,
\qquad U_0=(m+1)R.
\tag{H.25b}
\]

Overlapping affected windows reduce this bound. All counted rows remain inside the original deadline.

\textbf{11. Fixed-seed realization and the KL lower bound.} The auxiliary action seed is independent of mechanism randomness. Each fixed-seed policy is deterministic and obeys the cap on every history. The KL chain rule gives its KL as \(\mathbb E_{\rm noise}G^{\rm cap}\). Therefore \(A_{C,T}\) is at least the supremum over seeds, and hence at least the seed average. The reward bound \(TR\) justifies exchanging expectations. Finite transcript boxes approximate these policies with arbitrarily small KL loss when finite tables are required.

Combining (H.25a)--(H.25b) gives
\[
\begin{aligned}
A_{C,\lfloor\tau C\rfloor}&\ge C\Phi(\tau)-K_*\sqrt C-B_*,\\
K_*&=U_0A_\Sigma\sqrt\tau,\qquad
B_*=(k+1)R+J_\Sigma+U_0B_\Sigma.
\end{aligned}
\tag{H.26}
\]

These constants include actual entry, unfinished entry, stage endpoints, integer rounding and excess-action deletion.

\textbf{12. The two-sided privacy law.} Combine (H.26), the price upper bound and the true-likelihood bridge (H.11b). With \(Q=v_0(R+\mathsf L)^2\), \(\beta=(1-\delta)/2\),
\[
\begin{aligned}
\max\{0,C\Phi(\tau)-(K_*+\sqrt{\tau Q/\beta})\sqrt C-B_*+\log\beta\}
&\le\varepsilon_{C,\lfloor\tau C\rfloor}^*(\delta),\\
\varepsilon_{C,\lfloor\tau C\rfloor}^*(\delta)
&\le C\Phi(\tau)+{b_\psi}+\sqrt{(1-\delta)\tau CQ/\delta}.
\end{aligned}
\tag{H.27}
\]

The lower test has strict margin \((1-\beta)^2-\delta=(1-\delta)^2/4\). This proves Theorem 10 for fixed parameters and risk. The Poisson constants belong to the selected optimal controllers. Supplied finite-deadline certificates in H.4 give separately computable bounds for their stated requests.

\textbf{Frontier limit and saturation.} The price representation makes \(\Phi\) nondecreasing and concave. For \(\tau\le1\), the cost constraint is redundant, giving \(\Phi=\tau\psi(0)\). The inequality \(A_{C,T}\le A_C\) and (H.26) give \(\Phi\le a_*\). A fixed near-optimal budget block, followed by a full echo and public return to zero, has finite mean cycle length. Its expected visit frequencies satisfy the original flow equations by cancellation of the equal cycle endpoints, even when the controller has stage memory. Its KL-to-cost ratio is at least that of the block. Thus \(\sup_{c(z)>0}r(z)/c(z)\ge a_*\).

Conversely, at \(\tau=1/c(z)\), (H.11) and \(\Phi\le a_*\) bound this ratio by \(a_*\). A zero-cost occupation is supported on zero actions and zero difference memory and has zero reward. Mixing a fixed positive-cost occupation with the idle occupation makes it feasible for sufficiently large \(\tau\), proving \(\Phi(\tau)\to a_*\) and \(a_*=\min\{u:\psi(u)=0\}\). For \(a_*>0\), equality in \(\tau r(z)\le\tau a_*c(z)\le a_*\), together with attainment, proves (50). For \(a_*=0\) the frontier is zero. If \(T_C/C\to\infty\), eventually \(T_C\ge\lfloor\tau C\rfloor\) for every fixed \(\tau\); monotonicity, Theorem 9 and then \(\tau\to\infty\) give the superlinear-deadline rate.

For \(m=1,0<a_1=a<B\), the state, hidden reward, occupation equations and physical policy class reduce exactly to the first-order objects. Also \(w_1=1\), \(\alpha=\gamma\) and \({V_\infty}=(\sigma^2+1/4)/(1-\gamma^2)\). This recovers their \(\psi,\Phi,a_*\), with the sharper first-order return and deadline constants retained in H.1 and H.4.

\subsection{A sufficient first-order innovation-source class}

Fix \(m=1,0<a_1=a<B\). Keep the same integer recurrence, hidden uniform rounding, signed actions, one-step action memory and complete observation, and replace only the innovation mass by \(\nu\). Suppose \(\nu(z)>0\) for every integer, \(\mathbb E_\nu Z=0\), \(s_2=\mathbb E_\nu Z^2<\infty\), \(m_4=\mathbb E_\nu Z^4<\infty\), and there is a finite constant \(\Xi\) satisfying
\[
\Xi\ge\max_{|h|\le2}\sum_{z\in\mathbb Z}
\frac{\nu(z)^2}{\nu(z+h)}<\infty.
\tag{H.28}
\]

Marginalizing the hidden rounding cannot increase the order-two row moment. Jensen and \((\log x)^2\le x+x^{-1}\) give \(0\le g\le R_\nu:=\log\Xi\) and \(\mathbb E_P\ell^2\le\Xi+1\), in both orders. The common-rounding coupling is unchanged. Put \(v_\nu=s_2+1/4\), \(V_{\infty,\nu}=v_\nu/(1-\gamma^2)\), and \(\mathsf L_\nu=R_\nu/(1-\gamma)\). Centered rounding gives the same second-moment drift; Minkowski gives the uniform fourth-moment bound \(((m_4^{1/4}+1)/(1-\gamma))^4\).

Define \(J_t^\nu(u)\) by (47) for the same zero-initial policy class with source \(\nu\), and let \(\psi_\nu(u)=\lim_{t\to\infty}J_t^\nu(u)/t\) and \(\Phi_\nu(\tau)=\inf_{u\ge0}\{u+\tau\psi_\nu(u)\}\). These facts supply every source-dependent step of H.6: the common invariant marginal, compact occupations, linearly bounded discounted biases, positive hitting paths within each selected closed class, quadratic Poisson potentials and finite unconditional \(L^2\) bounds. For testing, use the finite-moment compensation with \(Q_\nu=2(\Xi+1)+2\mathsf L_\nu^2v_\nu\). Therefore
\[
\varepsilon^\nu_{C,\lfloor\tau C\rfloor}(\delta)
=C\Phi_\nu(\tau)+O_{\gamma,\nu,\tau,\delta}(\sqrt C)
\tag{H.29}
\]

for every fixed \(\tau>0\) and \(\delta\in(0,1)\). The centered two-sided geometric law \(\nu_r(z)=(1-r)r^{|z|}/(1+r)\), \(0<r<1\), satisfies these hypotheses: all polynomial moments are finite and its shift ratios are bounded. This extends the resource law for the same recurrence.

\section{Proof coverage and formal refinements}

\subsection{Finite-memory proofs and selected Lean lemmas}

Appendices A, E.3, F, G and H prove the finite-memory preparation, compiler, refinement, commitment and resource results; J.4 gives the prefix bound. Exact-rational replay reconstructs the strict hierarchy curves, DG calibration endpoints and frontier witnesses. The twelve steps in H.6 include the target return-cycle identity, actual closed-class entry, a potential covering unfinished entry, deterministic-stage moment bounds and fixed-seed hard-budget realization.

A separate Lean 4.19.0 file, \texttt{G8CriticalSteps.\allowbreak{}lean}, checks 15 algebraic and finite-path lemmas used in that argument:

\noindent\begin{minipage}{\linewidth}
\par\smallskip\begingroup
\small
\setlength{\tabcolsep}{3pt}
\renewcommand{\arraystretch}{1.10}
\begin{tabularx}{\linewidth}{@{}>{\hsize=1.000000\hsize\linewidth=\hsize\raggedright\arraybackslash}X>{\hsize=1.000000\hsize\linewidth=\hsize\raggedright\arraybackslash}X>{\hsize=1.000000\hsize\linewidth=\hsize\raggedleft\arraybackslash}X@{}}
\toprule
\textbf{Proof step} & \textbf{Checked content} & \textbf{Theorems} \\
\midrule
Killed-kernel bound & Small-set and outside-set inequalities, geometric iteration & 3 \\
Poisson compensation & Finite telescoping, three-square inequality, increment bound & 3 \\
Pathwise cap & One-step guard, kept count, hard cap and excess deletion & 4 \\
Finite-memory loss & Affected-window count, unchanged windows, reward loss and lower-bound algebra & 5 \\
\bottomrule
\end{tabularx}
\endgroup\smallskip\par
{\small Table S6. Selected finite-memory Lean coverage. The countable-chain construction, existence of hitting potentials, conditional expectations and martingale orthogonality are established by the mathematical proof in H.1--H.6. The new file formalizes the listed algebra and finite-path statements.}
\end{minipage}\par

Native compilation returned exit code zero. All 15 axiom reports contain only subsets of \texttt{propext}, \texttt{Classical.\allowbreak{}choice} and \texttt{Quot.\allowbreak{}sound}. The evidence record binds the Lean source, compiler, Mathlib manifest, imported root object and resulting object by SHA-256. This is a separate source closure from the first-order development below.

\subsection{First-order semantics and binary value implementation}

The following development concerns \(m=1,0<a_1=a<B\). Its bounded-LP component separately accepts arbitrary finite serialized dimensions.

The formal Gaussian parameter is \(b_0=1/(2\sigma^2)\). The development defines the normalized mass on every integer, mathematical floors at negative states, hidden-uniform mixtures and complete countable transcript laws. Predictable policies preserve the conditional-state invariant. The cap guard agrees with original path and event masses on legal histories. Its potential \(16b_0c+8b_0|d|\) pays for remaining participation and the pending echo, yielding (7); the pointwise inequalities \((r-\lambda)_+\le r^2/(4\lambda)\) and \((r-\lambda)_+\le(r-1)^2/[4(\lambda-1)]\) give the event conversions.

\noindent\begin{minipage}{\linewidth}
\par\smallskip\begingroup
\small
\setlength{\tabcolsep}{3pt}
\renewcommand{\arraystretch}{1.10}
\begin{tabularx}{\linewidth}{@{}>{\hsize=0.660000\hsize\linewidth=\hsize\raggedright\arraybackslash}X>{\hsize=1.340000\hsize\linewidth=\hsize\raggedright\arraybackslash}X@{}}
\toprule
\textbf{Component} & \textbf{Kernel-checked first-order scope} \\
\midrule
Preparation and transcripts & Complete-word realization, finite expected waits, uniform deadline loss, unrestricted physical/\allowbreak{}source value equality and both-world coordinates \\
Exact compiler & Actual default-root coverage, testing curves, memoized integer value equality, six charged event counts with exponent \(3C-2\), and integer/\allowbreak{}storage bounds \\
Commitments and calibration & Countable-source proof envelope, depth monotonicity, full-depth equality, minimal calibrated depth and physical risk-shift inversion \\
Budget and physical time & Original KL rate and testing inverse, superlinear-deadline leading rate, fixed-ratio square-root frontier and occupation saturation criterion \\
Window and numerical certificates & Two-sided original-DG/\allowbreak{}common-source transfer, full-output rewards, occupation LP, all-precision schedule, five original-frontier precision instances and 56 direct finite-deadline instances \\
Binary value program & Internally generated coefficients and occupation LP, exact bounded-LP value, and accuracy plus polynomial cost for the same heap execution \\
\bottomrule
\end{tabularx}
\endgroup\smallskip\par
{\small Table S7. First-order formal coverage by mathematical object. The compiler uses an explicit finite rational law; the physical connections retain the DG, reduced-slope and zero-start conditions stated in the corresponding results. Sections H.7 and J establish the source-class extension and utility comparisons by mathematical proof and certified arithmetic.}
\end{minipage}\par

Three terminal declarations identify the algorithm's scope. \texttt{binaryCoefficientTable\_\allowbreak{}full\_\allowbreak{}bit\_\allowbreak{}time} fixes positive rational \(b_0,\tau\) and integers \(0<a<B\) before the polynomial constants and every numerical precision \(k\). It constructs a normalized shared finite rational approximation to the DG law and rational approximations to all window full-output KL rewards. \texttt{bounded\_\allowbreak{}rational\_\allowbreak{}LP\_\allowbreak{}binary\_\allowbreak{}polynomial\_\allowbreak{}time} has uniform polynomial constants in the complete serialized input, across all dimensions and rational rows, objectives and stored boxes; it returns the exact attained optimal value or correctly reports infeasibility for the problem including those box inequalities. \texttt{binaryPublicPhi\_\allowbreak{}full\_\allowbreak{}polynomial\_\allowbreak{}bit\_\allowbreak{}time} composes coefficient production, occupation construction, LP evaluation and one \(\tau\) scaling. Its original-frontier error and heap cost concern that same run. Only canonical encodings of the parameters and precision are supplied; coefficient errors and LP optimality are proved internally.

The binary-cell heap model charges finite control, memory operations and arithmetic. Its polynomial execution uses the ellipsoid path, with the earlier Fourier--Motzkin program as a semantic reference. Table S4 verifies five supplied numerical certificates; Section 7.1 measures the exact compiler implementation. Deployment verification calls for a concrete DG sampler, compiler backend and arithmetic libraries.

The frozen R41 source closure contains 1,863 positive modules and 11,558 distinct printed declarations, including definitions. Source, import-object, compiler and foundation hashes bind the native records to Lean 4.19.0 and mathlib commit c44e0c8ee63ca166450922a373c7409c5d26b00b. Reported axioms are subsets of \texttt{propext}, \texttt{Classical.\allowbreak{}choice} and \texttt{Quot.\allowbreak{}sound}. Nineteen negative files reject 44 false mathematical goals; 15 separate controls reject altered saved evidence. Final acceptance rebinds 1,430 inherited and 433 previously compiled current objects, then executes the negative and full-type checks. The evidence map records the source and object bindings for each acceptance check and archive replay.

\section{Certified matrix comparisons for the first-order subfamily}

Sections J.1--J.3 compare the first-order feedback mechanism \(m=1,0<a_1=a<B\) under the stated static or online contracts. Section J.4 proves the prefix lower bound for general finite memory.

\subsection{Finite strategy privacy and utility}

The frozen workload contains 100 binary records and 1000 scheduled prefix outputs. Zeroing one record changes the strategy vector \(Ax\) by a signed column of \(A=15I+J\), whose squared norm is \(15^2+2\cdot15+100=355\). This is the sensitivity used by the matrix mechanism. The workload's realized cap 56 and gap 4 describe the released queries; the internal strategy has its own sensitivity.

For an integer shift vector \(h\) and independent scale-\(\sigma\) DG coordinates, completing the square at integer \(\theta\ge1\) gives
\[
\mathbb E_P\left[\left(\frac{dP}{dQ}\right)^\theta\right]
=\exp\left(\frac{\theta(\theta+1)\|h\|^2}{2\sigma^2}\right).
\tag{J.1}
\]

The translated normalizer is unchanged because \(\theta h\) is integral. The elementary inequality \((1-t^{-1})_+\le\theta^\theta t^\theta/(\theta+1)^{\theta+1}\), applied at \(t=e^{L-\varepsilon}\), yields a testing bound. With \(\theta=6,\sigma=38,\varepsilon=2\), it is \((6^6/7^7)\exp[-12+42\cdot355/(2\cdot38^2)]\).

The implemented source rounds nonzero DG masses down to a 33-bit dyadic grid on \([-230,230]\) and places the residual at zero. It is symmetric and normalized, and its total variation is exactly its excess zero mass. If the certified source distance is \(\eta\), coupling the 100 independent draws gives
\[
\delta_{\rm finite}(2)\le
\frac{6^6}{7^7}\exp\left(-12+\frac{42\cdot355}{2\cdot38^2}\right)
 +(1+e^2)100\eta
\le0.000084068601494238.
\tag{J.2}
\]

All 1000 outputs are public postprocessing of these 100 draws. The consumer reconstructs the 461 positive source masses, inverse matrix, workload and privacy bound from the bound inputs.

Let \(v\) be the exact variance of the finite source. For workload row \(w_t\), the rational-output MSE is \(v\|w_tA^{-1}\|^2\). Exact summation gives mean 16058.915463117209 and maximum marginal 36046.569404175994, to the shown precision. The inverse denominator 1725 is odd, so nearest-integer output rounding has no ties. Symmetry preserves unbiasedness. If the unrounded error has second moment \(v_t\), Cauchy--Schwarz gives rounded MSE at most \(v_t+\sqrt{v_t}+1/4\); Jensen applies the same expression to the mean over outputs. Directed arithmetic yields the upper endpoints in Table 2. Utility is computed directly from the implemented source counts.

The fresh-bit, threshold-information and register counts measure different resources. For the matrix program they are 3300, 15180 and 1000 bits, respectively. The last number bounds 100 private noisy strategy registers on the specified binary data range. The execution audit checks all scheduled outputs against an independent integer implementation. Fifteen process measurements are retained as environment-specific diagnostics. The current fresh replay checks finite certificate generation, the Gaussian primal--dual certificate, finite-program semantics, and both independent finite and execution consumers.

\subsection{The complete ideal-Gaussian reference}

Let \(G=W^TW\) and \(T=1000\). A full-rank \(W\) lets the data-independent additive unbiased Gaussian design be parameterized, on its image, by covariance \(s^2WX^{-1}W^T\) with \(X\succ0\), \(\operatorname{diag}X\le1\). The mean MSE is \(s^2\operatorname{tr}(GX^{-1})/T\). Gaussian components outside the workload image can be projected away. The unit-neighbor Mahalanobis norm is at most \(1/s\). Analytic Gaussian calibration [27] uses
\[
d_s(\varepsilon)=\mathcal N\left(\frac1{2s}-\varepsilon s\right)
-e^\varepsilon\mathcal N\left(-\frac1{2s}-\varepsilon s\right),
\tag{J.3}
\]

where \(\mathcal N\) is the standard normal CDF. Certified integration proves \(d_{1.73435098}(2)>10^{-4}\) and \(d_{1.73435099}(2)<10^{-4}\).

For the feasible upper, exact integer checks prove \(X\succeq I/32\), \(0<X_{ii}\le1\), and \(\|I-XK\|_F\le10^{-7}/32\) for a rational approximate inverse \(K\). Hence \(K-10^{-7}I\preceq X^{-1}\preceq K+10^{-7}I\). Positive leading principal determinants, checked by fraction-free elimination, certify the positive-definiteness obligations.

For the lower, choose positive integer weights \(q_i\), set \(Q=\operatorname{diag}(q_i)\), and supply rational symmetric \(S\succ0\) with \(QGQ-S^2\succ0\). Monotonicity of the positive square root implies \(\operatorname{tr}S\le\operatorname{tr}\sqrt{QGQ}\). For every feasible \(X\), put \(R_X=QXQ\); Hilbert--Schmidt Cauchy--Schwarz gives
\[
\operatorname{tr}(GX^{-1})
=\operatorname{tr}(QGQ R_X^{-1})
\ge\frac{(\operatorname{tr}S)^2}{\sum_iq_i^2}.
\tag{J.4}
\]

Three exact 100-by-100 positive-definiteness checks and the inverse residual yield \(1300.218320640<F_*<1300.262632443\), where \(F_*\) is the normalized shape objective. Multiplying by the appropriate directed squared calibration scales proves (63). The selected feasible mean-optimized design has maximum marginal MSE in \([10188.435943596,10188.530657459]\); the optimization objective here is the mean. Equation (63) therefore bounds the optimal mean error over the stated complete ideal-Gaussian class.

\subsection{Online-prefix utility and the family separation}

Write \(q=\sigma^2\), \(p_q(z)=Z_q^{-1}e^{-z^2/(2q)}\), \(\lambda=41/4\) and \(\delta=10^{-5}\). The first release has no feedback. A neighboring unit shift and the event \(Y_1=k+1\) give payoff \(p_q(k)-\lambda p_q(k+1)\). Since \(Z_q\le1+\sqrt{2\pi q}\), for \(q\in[a,b]\) a positive lower bound is
\[
\frac{e^{-k^2/(2a)}(1-\lambda e^{-(2k+1)/(2b)})}
     {1+\sqrt{2\pi b}}.
\tag{J.5}
\]

For \(k=0\), the first factor is one, also allowing \(0<q\le b\). The intervals \((0,1/5]\), \([1/5,1/2]\), \([1/2,1]\), \([1,3/2]\) and \([3/2,2]\), with \(k=0,2,2,3,5\), cover \((0,2]\). Directed exponential and square-root bounds put all five payoffs above \(1.52\times10^{-5}>\delta\). Thus privacy of the first coordinate, and hence of the complete transcript, requires \(q>2\) for every feedback slope.

The actual variance \(v(q)=\mathbb E_q Z^2\) satisfies \(v'(q)=\operatorname{Var}_q(Z^2)/(2q^2)>0\); dominated convergence justifies differentiation on compact positive \(q\)-intervals. A finite numerator sum gives
\[
v(2)\ge\frac{2e^{-1/4}+8e^{-1}+18e^{-9/4}+32e^{-4}}
 {1+2\sqrt\pi}>1.5366480554>3/2.
\tag{J.6}
\]

For prefix noise, the response coefficient at lag \(\ell\ge1\) is \(b_\ell=[1-(-\gamma)^\ell]/(1+\gamma)\). Consecutive odd/\allowbreak{}even coefficients sum to at least one, so their squares sum to at least \(1/2\); an unmatched odd square is at least \(1/4\). Equation (62) therefore gives
\[
\frac1T\sum_{t=1}^T\mathbb E S_t^2
\ge \frac{v(q)}T\sum_{t=1}^T\frac t4
>\frac{3(T+1)}{16}.
\tag{J.7}
\]

This proves the necessary bound in Proposition 13 over the entire family. The numerical constants are checked by independent ball arithmetic and rational Taylor/\allowbreak{}Machin/\allowbreak{}square-root enclosures.

For the BLT witnesses [7], rational parameters define two decreasing exponential buffers. The Toeplitz strategy uses its maximum signed two-column sensitivity over every legal separation and calibration (J.3). This bound is uniform over common public histories, as required by the adaptive Gaussian factorization theorem [44, Theorem 2.1]. Inversion followed by cumulative summation produces every continuous prefix before public nearest-integer rounding; policies on rounded histories lift to continuous-history policies. Directed integration proves the rounded MSE upper bounds 84.103675364 and 134.862672985 at \(T=512,4096\). Parameter encodings and enclosures accompany the evidence map. Integer prefix and increment transcripts are bijective.

The within-family comparison uses a separate second-moment transfer. Couple a centrally completed finite source with DG so that the innovation difference is conditionally centered and has second moment \(d_v=v-v_{\rm finite}\). At \(\gamma=1/3\), common hidden rounding gives \(\mathbb E(K_t-\widehat K_t)^2\le3d_v/2\). The triangle inequality in the joint time/\allowbreak{}probability \(L^2\) space yields
\[
\left|\sqrt{\operatorname{MSE}_{\rm DG}}-
\sqrt{\operatorname{MSE}_{\rm finite}}\right|
\le\sqrt{\frac{3d_v}{2}\frac{(T+1)(2T+1)}6}.
\tag{J.8}
\]

Finite-state moment propagation and an independent joint-prefix calculation, with this transfer, give the three improvement intervals in Section 7.5. The comparison fixes the certified \(q=9\) design and the depth-one first-success design \(145/16\).

\subsection{Finite-memory prefix lower bounds}

For (2), let \(v=\operatorname{Var}(\eta_t)>0\). Unbiased hidden rounding gives martingale innovations \(\xi_t=K_t+\sum_i(a_i/B)K_{t-i}\), with \(\mathbb E[\xi_t\mid\mathcal F_{t-1}]=0\) and \(\mathbb E\xi_t^2\ge v\). Thus the innovation covariance is diagonal and dominates \(vI\). This retains the actual rounded recurrence and its nonnegative rounding variance.

Let \(Z_T\) be the \(T\)-dimensional lower shift and \(P_T\) the prefix-sum matrix. Define
\[
F_T=I+\sum_{i=1}^m\frac{a_i}{B}Z_T^i,\qquad
W_T=P_TF_T^{-1},\qquad
\|F_T\|_{\rm op}\le1+r.
\tag{J.9}
\]

The matrix \(F_T\) is triangular with diagonal one and is invertible at every finite \(T\). On zero input the prefix vector is \(W_T\xi\). Since \(P_T=W_TF_T\), the Frobenius inequality \(\|P_T\|_F\le\|W_T\|_F\|F_T\|_{\rm op}\) and \(\|P_T\|_F^2=T(T+1)/2\) give
\[
\overline{\operatorname{MSE}}_T
=\frac1T\operatorname{tr}\!\left(W_T\operatorname{Cov}(\xi)W_T^\top\right)
\ge\frac{v}{T}\|W_T\|_F^2
\ge\frac{v(T+1)}{2(1+r)^2}.
\tag{J.10}
\]

This bound holds for every finite \(T\) and every permitted coefficient vector. For fixed source and stable coefficients it gives \(\Omega(T)\) prefix error. Under the privacy request of Proposition 13, the zero-start first release is still DG, so (J.5)--(J.6) require \(v>3/2\); with \(r<1\), (J.10) also recovers the lower bound \(3(T+1)/16\) for general memory.

For \(m=2,a=(1,1),B\ge3\), \(r=2/B\le2/3\), giving \(\overline{\operatorname{MSE}}_T>27(T+1)/100\). At \(T=512,4096\), this exceeds 138.51 and 1106.19, respectively, while the same two-buffer BLT witnesses in J.3 remain below 84.103675364 and 134.862672985. The inequality is uniform if the permitted coefficients or DG scale vary with the deadline, because the first-release variance condition is uniform. It concerns the pure autoregressive update (2) and the complete prefix workload.

\section{First-order calibration and comparison details}

This appendix specializes to \(m=1,0<a_1=a<B\). General finite-memory source transfer and inverse completion are proved in F.4 and G.2.

\subsection{Completion of fixed-budget inverse requests}

For \(C=0\), the inverse is identically zero. Fix \(C\ge1\), rational \(\gamma,\sigma^2\), risk interval \([\delta_{\min},\delta_{\max}]\subset(0,1)\), and requested positive width \(\zeta\). Choose rational \(\eta>0\) with \(\log(1+\eta)<\zeta\), leaving room for displayed logarithm enclosures. Use a rational \(M\ge e^{8C/\sigma^2}\). The moment inequality in (29) follows for every event, and its secant proof uses the same near-optimal event at thresholds \(x<y\), then lets its payoff defect tend to zero.

A first-release unit shift gives \(D_C(1)\ge c_0=1/(1+3\sigma)>0\), since the DG normalizer is smaller than \(1+3\sigma\). A positive rational lower enclosure for \(c_0\) suffices in the following construction. Put \(d_0=\min(\delta_{\min},c_0)/4\), choose rational \(B_0>1\) with \(M/B_0\le d_0/4\), and let \(\mu_0=d_0^2/(8M)\). Choose the informative source-coupling allowance
\[
r_0\le\min\left\{
\frac{d_0}{4(1+B_0)},\frac{d_0^3}{64M(1+B_0)},
\frac{\mu_0}{2},\frac{\mu_0\eta}{8(1+B_0)}\right\}.
\tag{K.1}
\]

A symmetric finite rational source with one-step total variation \(\tau\le r_0/(2C)\) then has informative-graph error \(r=1-(1-\tau)^{2C}\le r_0\). Such a source is effective: a Gaussian tail enclosure tends to zero with the support radius, and directed dyadic approximation of the retained unnormalized weights, keeping the weight at zero exactly one, tends to zero with precision. The finite profile \(F\) satisfies \(|F(\lambda)-D_C(\lambda)|\le(1+\lambda)r\).

Exact bisection between 1 and \(B_0\) finds a point \(b\) with \(d_0\le F(b)\le2d_0\). It terminates because \(F\) is 1-Lipschitz, while the initial endpoints lie strictly above and below this interval. The true-profile secant bound at \(b\) and \(b+d_0/4\), together with (K.1), forces every touching right support at \(b\) to have probability slope \(q_b\ge\mu_0>r\). Also \(F(b)+(1+b)r<\delta_{\min}\). Thus both endpoint safety and inverse conditioning are certified before refinement.

Let \(L\) be the maximum of queried feasible support lines and \(U\) their chord upper. Invert the corrected lower lines as \(\ell(\delta)=\max\{1,\max_i(p_i-r-\delta)/(q_i+r)\}\). Invert \(U+(1+\lambda)r\) for the upper endpoint, setting that endpoint to one when its value at one is already at most \(\delta\). The interval obeys \(1\le\ell\le e^{\varepsilon_C^*}\le u\le b\). Splitting risk space at every affine breakpoint and at the upper clipping boundary makes \(u/\ell\) affine-over-affine with positive denominator. Its maximum on each piece is at an endpoint.

If the vertical sandwich gap is \(E\), then \(u/\ell-1\le[E+2(1+b)r]/(q_b-r)\). Equation (K.1) leaves the strictly positive allowable gap \((q_b-r)\eta-2(1+b)r\ge\mu_0\eta/4\). Dyadic sandwich refinement therefore terminates. Source construction, exact queries, geometry and directed logarithms are all charged; the point oracle has the fixed-budget polynomial bound in Theorem 4.

The accompanying inverse certificates include the zero-epsilon boundary and \(C=3,4\). At \(\gamma=1/2,\sigma=1,C=4\), they give \(14.740484552831<\varepsilon_4^*(10^{-4})\le14.740489644244\). Exact source, query and risk-cell bindings are recorded in the evidence map.

\subsection{Actual-DG commitment and risk-directed witnesses}

At the fixed DG law \(\gamma=1/2,\sigma=1,C=4\), certificates over twenty contiguous cells cover \(I=[10^{-4},1.1\times10^{-4}]\) and give
\[
\begin{array}{c|ccc}
H&1&2&3\\\hline
\inf_{\delta\in I}(\varepsilon_{P_H}-\varepsilon_4^*)&
\ge1.023409037074&\ge0.595242559940&\ge0.094073920355.
\end{array}
\tag{K.2}
\]

Hence \(H_{\rm cal}(4,0.09;I)=4\). At larger allowances, legal physical witnesses and separate necessary bounds give \(H_{\rm cal}(4,0.2;I)=3\) and \(H_{\rm cal}(4,0.7;I)=2\). The sufficient witness includes a capped public preparation, four participations and an echo, with all 34 physical innovations charged. The following records give the shared source, directed inverse endpoints and event coordinates.

For this first-order C4 example, use one radius-eight source with denominator \(2^{80}\). At \(\lambda=2600000\), its transferred profile intervals, in units of \(10^{-6}\), are
\[
\begin{aligned}
P_1 &: [282.449227426,282.615457972],\\
P_2 &: [172.662924345,172.755427709],\\
P_3 &: [107.417496081,107.477705695],
\end{aligned}
\tag{K.2a}
\]

while the exact target is at most 100 in these units. Dividing by \(2600001\) gives the same strict order for testing values. Twenty adjacent risk cells and directed inverse/\allowbreak{}logarithm bounds prove (K.2).

The sufficiency certificates at losses 0.2 and 0.7 use a different lower object: a legal physical event. Prepare origin 2 by common public first-hit search capped at 29 rounds, then use four negative participations and one echo, padding to 34 total releases. Reject the event on preparation failure. At finite threshold 2221685, the transferred event has safe coordinates approximately \(p^-=0.00021892734460638618\) and \(q^+=5.1974012751970526\times10^{-11}\); the consumer uses the exact rational coordinates. The same event across the risk interval gives supremum calibration losses at most 0.196327391279561662 for \(P_3\) and 0.699741708079404244 for \(P_2\). Separate exact-target comparisons give necessary supremum losses at least 0.604926741735487 for \(P_2\) and 1.046484967063540 for \(P_1\). These establish the two minimal depths stated above.

The witness search follows the requested risk. If a sound proof inverse is \(u_H(\delta)\) and the allowed multiplicative threshold loss is \(K\), a failing risk \(\delta_*\) asks for an event satisfying \(p^- -[u_H(\delta_*)/K]q^+\ge\delta_*\). A finite bank of legal events supplies an upper envelope of affine lower inverse lines. Its ratio to the proof upper inverse is checked at joint breakpoints. For a fixed physical program, exact Neyman--Pearson testing of complete public transcripts optimizes the event at that requested threshold. Preparation and source error are charged across its entire physical horizon. Thus proof depth, physical witness choice and numerical resolution can be assessed using one final calibration target.

\textbf{Additional fixed-noise requests.} At \(C=2,\sigma^2=9\), slopes \(1/2,1/3,1/4\) and risks \(10^{-3},10^{-4},10^{-5},10^{-6}\) give twelve requests, evaluated by complete-word integer queries and independent reconstruction of all 48 inverse endpoints. Four informative source draws pay the source-transfer error. Every full-commitment epsilon upper lies below its depth-one lower, with conservative relative gains of 0.233\%--0.623\%. Source uncertainty prevents 12 of the 24 threshold brackets from meeting the predeclared width \(10^{-5}\); their wider intervals still prove the strict calibration improvements and are retained.

At \(\gamma=1/3,\lambda=41/4\), the actual-source profile upper at \(\sigma^2=9\) is below \(9.997374\times10^{-6}\), while the depth-one lower exceeds \(1.058594\times10^{-5}\). The depth-one upper at \(\sigma^2=145/16\) is below \(9.928707\times10^{-6}\); the other two prespecified candidates \(73/8,37/4\) also pass. Separate source second-moment enclosures prove the variance decrease lies strictly between 0.6896551724137\% and 0.6896551724138\%. Each candidate is checked independently using its own source-tail and rounding certificate.

\subsection{The finite encoding and the unbounded-state obstruction}

Consider an already finite legal two-world testing tree with rational probabilities. Each decision node belongs to an information class in which one shared action must be chosen. For weights \(x,y\), first select the hidden world with probabilities \(x/(x+y)\) and \(y/(x+y)\). Give every decision information class one discrete static parameter selecting its action; Lagrange interpolation polynomials on the allowed integer values select the corresponding transition row. Give every distinct public terminal transcript one binary classification parameter. Enter the target state on classification 1 in world \(P\), or classification 0 in world \(Q\). Each allowed parameter assignment is exactly a legal deterministic policy and public event, so maximum reachability equals the original correct-testing value divided by \(x+y\). The finite object therefore connects directly to parametric-chain optimization [15]. The preparation reduction and the shared information classes are required inputs to this encoding.

For the original full integer kernel, a global common positive minorization is impossible. Suppose \(\mathsf K_k(D)\ge\beta\mu(D)\) for every integer \(k\), every set \(D\), some probability law \(\mu\), and \(\beta>0\). At \(k=mB\), \(\mathsf K_{mB}(y)=g_\sigma(y+am)\). Choose an atom \(y_0\) with \(\mu(y_0)>0\); summability gives \(g_\sigma(y_0+am)\to0\), contradicting the assumed fixed positive lower bound. Finite-window and small-set minorization remain available. The contraction-and-tail proof in H.3 supplies the explicit passage from this original noncompact kernel to finite certificates.

\clearpage\typeout{IDEA1_TOTAL_PAGES=\number\numexpr\value{page}-1\relax}
\end{document}